\documentclass[letterpaper,12pt]{report}

\pdfoutput=1

\usepackage[latin1]{inputenc}
\usepackage{geometry}
\usepackage{graphicx}
\usepackage{latexsym}   % Use special symbols.
\usepackage{makeidx}    % Make an index.
\usepackage{cuthesis}   %cuthesis.sty
\usepackage{named}     %named.sty
\usepackage{url}
\usepackage{subfigure}

\newcommand{\xf}[1]{Figure~\ref{#1}}

\newcommand{\lucidL}[1]{{$\mathit{Lucid}$}($L$) }

		{}

\def\myvert{\raise 2.27pt \hbox{\vrule depth 0pt height 8pt width 0.2mm}}
\def\myarrow{\hspace*{0.43mm}%
             \raise 2.29pt\hbox{\vrule depth 0pt height 8pt width 0.16mm}%
             \hspace*{-0.32mm}%
             $\longrightarrow$
             \ %
             }

\makeindex

\author{Miao Song}

\title
{
Dynamic Deformation of Uniform Elastic Two-Layer Objects
}

\degree{Master of Computer Science}
\dept{Computer Science and Software Engineering}
\begin{document}
\begin{abstract}

This thesis presents a two-layer uniform facet elastic object for real-time
simulation based on physics modeling method. 
It describes the elastic object procedural modeling algorithm with particle system from the simplest one-dimensional object,
to more complex two-dimensional and three-dimensional objects.

The double-layered elastic object consists of inner and outer elastic mass spring surfaces
and compressible internal pressure. 
The density of the inner layer can be set different from the density of the outer layer; the motion of the inner layer can be opposite to the motion of the outer layer. These special features, which cannot be achieved by a single layered object, result in improved imitation of a soft body, such as tissue's liquidity non-uniform deformation. The construction of the double-layered elastic object is closer to the real tissue's physical structure.

The inertial behavior of the elastic object is well illustrated in environments with gravity and collisions with walls, ceiling, and floor.
The collision detection is defined by elastic collision penalty method and the motion of the object is guided by
the Ordinary Differential Equation computation.

Users can interact with the modeled objects, deform them, and observe the response to their action in real time.% motion and deformation.

%keywords: deformation, elastic object, liquidity, mass-spring, ordinary differential equation, physical-based modeling, pressure model, soft body, real time. 

\end{abstract}

%\clearpage 
% No heading on the first page
%\thispagestyle{empty}

%\begin{center}
%\textbf{\LARGE Acknoledgements}\\[2ex]
%\textbf{\Large Techniques, Tips, and Traps}\\[2ex]
%\textbf{\large Peter Grogono}\\[2ex]
%\end{center}
\begin{acknowledgments}
%This thesis has been experienced unusual life. But one point I need to declare is it is me to give birth to my baby, no body can take it away, like my dear Deschanel, no one can take her away from me.
This thesis is made possible by these important people in my life:\\
 
I would like to thank Dr. Peter Grogono, my supervisor, for his sure guidance, careful and knowledgeable support, and his patience.
I also want to thank Prof. Jason Lewis for his constant encouragement.
Special thanks to Ms. Catherine LeBel, my immediate superior, and the CSLP (Center for the Study of Learning and Performance) software development team, who I work with, for their able support. Many thanks to Serguei Mokhov for his sturdy belief in the completion of this thesis and for introducing me to many application software tools. I would also like to thank my cherished little treasure, my daughter Deschanel, for her faith in me.
Finally, I must thank my parents, brother, and all of my loved ones for their care and support throughout my studies.  

%For his imaginative, clearheaded thinking on many aspects
%For his patience and professionalism in the composition process

%And for their critical contribution to this project: 
\end{acknowledgments}

%%%%%%%%%%%%%%%%%%%%%%%%%%%%%%%%%%%%%%%%%%%%%%%%%%%%%%%%%%%%%%

%%%%%%%%%%%%%%%%%%%%%%%%%%%%%%%%%%%%%%%%%%%%%%%%%%%%%%%%%%%%%%%%%%

%%%%%%%%%%%%%%%%%%%%%%%%%%%%%%%%%%%%%%%%%%%%%%%%%%%%%%%%%%%%%%%%%%%%%%%%%%%%%%%%%%%%%%%%%%%%
% Start a new page for the table of contents. Use Roman page numbers for
% front matter, then switch to arabic page numbers for the body of the
% text.
%\clearpage
%\pagenumbering{roman}
%\tableofcontents
%\clearpage
%\pagenumbering{arabic}

%%%%%%%%%%%%%%%%%%%%%%%%%%%%%%%%%%%%%%%%%%%%%%%%%%%%%%%%%%%%%%%%%%%%

\chapter{Introduction}
\label{chapt:introduction}
\index{Introduction}

In our real physical world there exist not only rigid bodies\index{rigid bodies} but also soft bodies, such as human and animal's soft parts and tissue, and other non-living soft objects\index{soft bodies}, such as cloth, gel, liquid, and gas. 

Soft body simulation\index{simulation!soft body}, which is also known as deformable object simulation\index{simulation!deformable object}, is a vast research topic and has a long history in computer graphics. It has been used increasingly nowadays to improve the quality and efficiency in the new generation of computer graphics for character animation\index{character animation}, computer games\index{computer games}, and surgical training\index{surgical training}. So far, various elastically deformable models have been developed and used for this purpose. 

In this chapter, we will introduce the concepts about deformable as well as elastic objects.
Moreover, we will explain how important this research is and its present applications. 

%For our research and achievement, we will concentrate on the subset research topic of the elastic simulation, the simulation of dynamic motion of soft tissues. 
\section{Definitions}

\begin{figure}
\hrule\vskip6pt
\begin{center}
\begin{picture}(420,280)
\thicklines
\put(140,230){\framebox(170,40){\texttt{\large Soft Body Deformation}}}
\put(145,135){\framebox(150,40){\texttt{\large Elastic Deformation}}}
\put(-10,135){\framebox(150,40){\texttt{\large Plastic Deformation}}}
\put(300,135){\framebox(160,40){\texttt{\large Fracture Deformation}}}
\put(30,50){\framebox(200,35){\texttt{\large Small Elastic Deformation}}}
\put(250,50){\framebox(200,35){\texttt{\large Large Elastic Deformation}}}
\put(50,-15){\framebox(140,35){\texttt{\large Tissue Animation}}}
\put(270,-15){\framebox(140,35){\texttt{\large Fluid Animation}}}
\thinlines
\put(220,230){\vector(0,-1){53}}
\put(50,205){\line(1,0){330}}
\put(50,205){\vector(0,-1){30}}
\put(380,205){\vector(0,-1){30}}
\put(110,105){\line(1,0){220}}
\put(220,135){\line(0,-1){30}}
\put(110,105){\vector(0,-1){20}}
\put(330,105){\vector(0,-1){20}}
\put(110,50){\vector(0,-1){30}}
\put(330,50){\vector(0,-1){30}}
\end{picture}
\end{center}
\caption{Soft Body Deformation}
\label{fig:SoftBodyDeformation}
\vskip6pt\hrule
\end{figure}
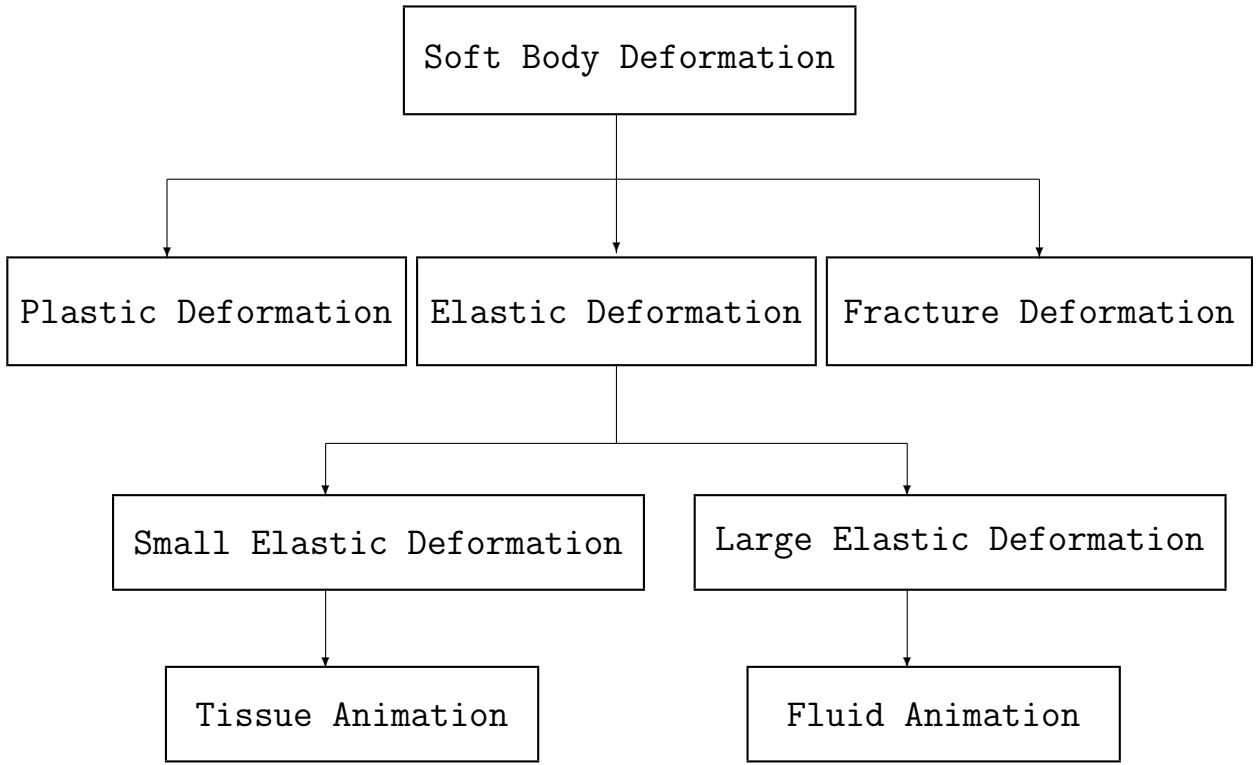

\subsection{Deformable Object}

In engineering mechanics, ``deformable object''\index{deformable object} refers to an object whose shape can be changed due to an applied force, such as tensile (pulling), compressive (pushing), bending, or tearing forces. The deformation can be categorized as the following, depending on the types of material and the forces applied:
\begin{itemize}
\item Elastic deformation\index{deformation!elastic} (small deformation)\index{deformation!small} is reversible. The object shape is temporarily deformed when tension is applied and it returns to its original shape when force is removed. An object made of rubber has a large elastic deformation range; silk cloth material has a moderate elastic deformation range; crystal has almost no elastic deformation range.
\item Plastic deformation\index{deformation!plastic} (moderate deformation)\index{deformation!moderate} is not reversible. The object shape is deformed when tension is applied and its shape is partially returned to its original form when the force is removed. Objects such as silver and gum, which can be stretched at their original length, cannot completely restore their original shapes after deformation.
\item Fracture deformation\index{deformation!fracture} (large deformation)\index{deformation!large} is not reversible, but is different from the plastic deformation. The object is permanently deformed when it is irreversibly bent, torn, or broken apart after the material has reached the end of the elastic deformation ranges. All materials will experience fracture deformation when sufficient force is applied.
\end{itemize}

\subsection{Elastic Object}

Elastic objects\index{elastic!objects} belong to a subset of soft body deformable objects. They are dynamic objects that change shape significantly and keep constant volume in response to collision. They can be bent, stretched, and squeezed. Moreover, they restore their previous shape after deformation. Elastic objects can be divided into two domains: %An example of an elastic object can be a bouncing, deformable ball, soft part of human skin, clay, fluid, or gel modeling.
\begin{itemize}
\item Large elastic deformation\index{deformation!large elastic}, such as fluid deformation, which focuses on flows through space. It tracks velocity and material properties at fixed points in space.
\item Small elastic deformation\index{deformation!small elastic}, such as tissue deformation\index{deformation!tissue}, which uses particle systems\index{particle systems} to identify chunks of matter and track their position\index{position}, acceleration\index{acceleration}, and velocity\index{velocity}.
\end{itemize}

Within this wide research range of soft body simulation, this work has focused on small elastic deformable object simulation, such as tissue animation. Even though there has been many valuable contribution related to this field, there are still many difficulties in accomplishing to realistic and efficient deformable simulation.

%This research area retains challenge and fascination not only because it is a new and novel field,
%but also because it is widely used for surgeon operation training and film animation.  

\section{Animation Techniques}\index{animation! techniques}

This section introduces some basic concepts related to the elastic simulation, such as the subject animation method\index{animation! method}.
Animation relies on persistence of vision and refers to a series motion illusions resulting from the display of static images in rapid-shown succession.
In traditional animation\index{ animation!tradition}, squash and stretch are exaggerated for elastic objects. In order
to be efficient when working with many of single frame images (or simply frames), inbetweening\index{inbetweening} and cel animation\index{cel animation} \cite{TJ84} have been introduced by Disney for manual traditional animation. 

The rate of the animation\index{animation!rate} refers to how many frames are displayed within a given amount of time. If the rate is too low, which is lower than the brain visual retention, the animation becomes jerky because the brain retains the empty frame from the previous image to the next image. 

A frame rate\index{frame rate}, which is the time between two updates of the display, describes the update frequency. In computer games, frames are often discussed in terms of frames per second (fps). The lower bound for smooth animation is between 22 to 30 frames per second. 
%For a hardcore gamer, an fps  of 60 is very good. 

For many years' research, computer-animation has been developed dramatically to replace the amount of manual traditional animation. The techniques of key-framing\index{key-framing}, morphing\index{morphing}, and motion capture\index{motion capture} \cite{HO98} have been widely used.

\begin{itemize}
\item Key-frame animation\index{key-framing}: is based on manual animation. It is a sequence of images of the same object with its local transformations, e.g. values for translation, rotation and scale, computed by interpolating between keyframes.
%\item Procedural method: is the method that automatically generates certain elements of an animation.
\item Morphing\index{morphing}: is a method usually used to estimate and generate a sequence of frames between one object to the other object with same number of vertices. Morphing is a good animation technique when using skeletal
animation would be too complex, e.g. facial animation.
%\item Skeleton animation: is represented by a hierarchy of bones, which are positioned and animated in the world by a translation and a orientation.
%\item Skin animation: is the process of attaching a deformable mesh to a skeletal structure so that the mesh is deformed. smoothly as the skeleton moves
\item Motion capture\index{motion capture}: is the method that attaches sensors on actors bodies and records the data for their movements and apply these data to a computer generated characters.
\end{itemize}

%\section{}
%3D animation appears in two forms:rigid-body and soft-body motion.
%In rigid-body motion,the relative position of each two vertices of the mesh stays .xed and the body moves as one entity.Such motion may be described by six parameters (also known as degrees of freedom):three for the position of the object s center and three for its orientation.The simplicity and compact representation of the rigid motion makes it particularly attractive. However,this model is not strong enough to capture many life-like movements. 
%Soft-body motion does not impose any restrictions on the relation between an object s vertices (as long as they form a valid mesh).It consists of a separate trajectory for each mesh vertex,which allows capturing of smooth and realistic motion.However,the size of .les storing these data is usually very large because each frame is actually a complete 3D object.This is a major factor preventing soft-body motions from becoming popular or 

\section{Elastic Animation}

There are two different methods about elastic animation modeling, which
depends on the predefined simulation or simulation in real time.
 
\paragraph*{Kinematic modeling}\index{kinematic modeling} predefines the positions and velocities of objects.
It does not concern what causes movement and how things get where they are in the
first place and only deals with the actual movement. 
For example, given that a ball's initial speed is 10 kilometers per hour on a perfect smooth plane,
we can use kinematic method to calculate how far it travels in two hours.  

\paragraph*{Dynamic modeling}\index{dynamic modeling} also termed as physically based modeling\index{physically based modeling},
is the study of masses and forces that cause the kinematic quantities,
such as acceleration, velocity, and position, to change as time progresses. 
For example, when we know the ball's initial speed, we need to know how far it travels after
an external force dynamically applied to it. 

For elastic object movement, the dynamic methods calculate how the soft body behaves after external force applied dynamically.
The animator does not need to specify the exact path of an object compared to using the kinematic modeling method.
The system predefines the initial condition of the elastic object, such as position and gravity force.
The animation of the object movement is updated each time step based on the acceleration derived from Newton's Laws of motion.
The dynamic simulation method is more advanced, easier to achieve the realistic motion than kinematic method.
Therefore, we will only represent dynamic simulation of elastic object in this thesis.

%and .It has been developed to computer graphics, film animation, virtual reality, computer games engines, medical imaging tracing, and surgical training.

\section{Applications}\index{applications}

Elastic modeling has been developed and used in several fields, including geometric modeling, computer vision, computational mathematics, physics engines, bio-mechanics, engineering, character animation, and many other fields \cite{gm97}. 

\paragraph*{Character Animation}\index{character animation} There is much advanced animation modeling software, which has advanced features for modeling, texturing, and lighting. However, for modeling the simulation of elastic objects\index{elastic!objects}, 3D artists have to do it manually, frame-by-frame because most of the current 3D software does not provide soft object simulation functionality. Artists have to use not only their drawing skills and intuition, but also posses some knowledge of physics to make the objects behave as if they are in the real world or close. 

The techniques of the non-physically based modeling\index{non-physically based modeling} of the elastic object include modeling the group of control points and changing their property parameters manually frame-by-frame. The virtual objects will not convince audiences because no natural laws of physics are applied. Moreover, key-frame animation is an inefficient way to model elastic objects without functionality provided by software. Hence, most of 3D film animators have to ignore the movement details of soft objects.

The latest version of the most advanced animation tool, Maya\index{maya}, provides the Soft Mod Deformer tool\index{Maya!Soft Mod Deformer tool}, which allows smooth sculpting of a group of objects \cite{sw}.
However, users need to have knowledge about how to use this complex software in order to access this advanced functionality. Moreover, users can only animate elastic object with Kinematic modeling method\index{kinematic modeling method} by setting values through the software interface rather than interact with the object in real time. 

\paragraph*{Computer Games}\index{computer game} Compared to the fancy and lifelike character animation widely used in 3D films, computer games are more concerned about computation efficiency because users interact with the software in real time. As one might notice, the majority of computer games do not portray the characters in detail, even with the mesh and texture modeling. It is not likely that elastic simulation will be widely introduced to computer games because existing elastic models usually require expensive calculation and are inconvenient to use in real time simulation\index{real time simulation}.

%The extreme realism of the game will have players feel every punch and kick by delivering shocking visuals of broken bones, blood stained ripped clothes and real-time facial deformation in fully interactive environments familiar from the movie scenes. real time tessellation [SP?] and deformation

\paragraph*{Surgical Training}\index{surgical training} Surgeons benefit from the rapid development of computer graphics and hardware techniques.
The computer generated visual virtual environment imitates the reality of medical operations and organ construction to fulfill the training purpose. This new application improves surgical outcomes and decreases the research costs. However, the reality and accuracy of the software always require high-end knowledge of physics, mathematic and heavy computation.
It makes it difficult for users to interact with virtual objects in real time.

\section{Thesis Goal}

The elastic object for dynamic simulation, which has been widely used, is the one layer elastic surface with different content within.
The soft objects can be squashed and stretched according to external\index{force!external} and internal forces\index{force!internal} applied on them.
Its computation depends on geometric modeling methods and physical equations.
However, this method is inefficient to imitate the behaviors of real human's tissue because human's or
animal's soft body does not consist of only one layer with either liquid or air inside. \xf{fig:tissue} from a biological research group demonstrates the complexity of human tissue \cite{km}. A tissue is composed of epidermis\index{tissue!epidermis}, dermis\index{tissue!dermis}, fat\index{tissue!fat}, fascia\index{tissue!fascia}, and muscle layer\index{tissue!muscle}. 

\begin{figure}[h]
\hrule\vskip4pt
\begin{center}
 {  \includegraphics[width=2in]{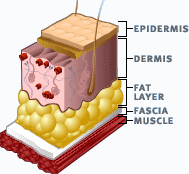}}
\end{center}
\hrule\vskip4pt
\caption{Human Tissue Layers}
\label{fig:tissue}
\end{figure}

\begin{itemize}
\item The epidermis is skin's outermost layer. It is responsible for the skin coloring because it contains the skin's pigment. 
\item The dermis, which is the layer that lies below the epidermis, consists entirely of living cells. It provides the skin elasticity because this layer is composed of bundles of fibers and small blood vessels.
\item The fat, the fascia, and the muscle layer are hypodermis layer of skin. It is an inner layer of and cushion for the skin. This subcutaneous tissue layer varies throughout the body region and hormonal influence. The fat and muscle increase the tension of the skin and protect the bones.  
\end{itemize}

Soft tissues are multi-composite layers; therefore, one layer elastic object is not accurate to model the kind of soft body  exemplified by human tissue.
Moreover, it is difficult to represent the object's inertia caused by the internal material realistically and its liquidity motion based on the various material densities. 

In this thesis, we investigate a more accurate two-layer elastic object\index{elastic!two-layer object}. 
The outer layer of the elastic object represents the epidermis and the dermis layer of a real tissue. 
The inner layer of this object corresponds to the hypodermis layers of skin. 
This two-layer computer generated elastic object provides more accurate modeling method based on the main feature of human tissue. Its deformation is based on the realistic physical consistency of tissues and the laws of established physics.  
The objective of this new model is to be convincing and to have distinct realism to the animated scene by applying proper physics. 
The program should be easy in implementation, convenient to re-use, and gives best elastic body behavior at the minimum cost rather than model the absolute complex object with the exact accurate physical equations. Users should be able to interact with the soft body in real-time and the collision detection and response must be handled correctly.

\section{Organization}

This chapter starts with the introduction of elastic objects, their applications, some basic concepts related to physical based deformable simulation, and the thesis goal. Chapter 2 surveys and analyzes the existing elastic simulation system and its problems. 
Chapter 3 describes the modeling methodology of elastic objects in one-dimension, two-dimension, and three-dimension.
Physically-based modeling and simulation map a natural phenomena to a computer simulation program. There are two basic processes in this mapping: mathematical modeling and numerical solution \cite{ml06}. Chapter 4 introduces mathematical modeling, which describes natural phenomena by mathematical equations. 
Chapter 5 presents the dynamics numerical equation of motion by using ODE (ordinary differential equation)\index{ODE} associated with our elastic simulation system, and discusses the complexity and improvement of the different numerical time integrator of Euler\index{integrator!euler}, Midpoint\index{integrator!midpoint}, and Runge Kutta 4th order\index{integrator!runge kutta 4th order}.
Chapter 6 presents the detailed design and implementation of the simulation system. 
Chapter 7 shows our experimental results with the animation sequences of the elastic object simulation and discusses the effects of changing the simulation parameters.  
Chapter 8 gives the conclusion of our current system, summarizes our contributions based on the existing elastic simulation models, and analyzes the possible development and related work in the future. %achievements of solving those problems with my development.
%Chapter 6 demonstrates the effectiveness of my approach based on methods for modeling by varies testing cases. Moreover, it discusses other elastic object simulation related features. Chapter 7 gives the conclusion of my current system and analyzes the possible development and related work in the future. 

\chapter{Related Work and Background Material}\index{related work}
\label{chapt:related work}
\index{related work}

Research about modeling deformable objects in computer graphics field has been going on for over 40 years and a wide variety techniques have been developed. In this chapter, we will review the existing geometric approaches for modeling elastic objects. These models are all based on physical laws. From the early elastic model, such as particle model, mass-spring  model, finite element model, to recent development such as fluid based model, and pressure model, we briefly introduce their physically-based modeling methods and compare these approaches with their advantages and disadvantages.

\section{Existing Elastic Object Models}

\paragraph*{Particle Model}\index{particle model} has been used by Reeves \cite{TR83} and to model the natural phenomena such as fire, water, liquid as shown in \xf{fig:particle}. Particles move under the force field and constraint without interacting with each other.
 
\begin{figure}[h]
\hrule\vskip4pt
\begin{center}
	{	 \includegraphics[width=3in]{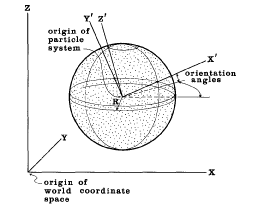}}
\caption{Particle System}
\label{fig:particle}
\end{center}
\hrule\vskip4pt
\end{figure}

%In another particel model, presented by Wyvill et al \cite{gw86}, the energy implicit functions to model the inter-particle connection are used with a large number of particles. (arranged themselves into volumes or surfaces?)
The advantage of this particle model is that the method is easy to implement. It is the earliest model to imitate the natural phenomena.

The disadvantage is that all the particles are independent and there are no forces connecting the particles. Therefore, for some phenomena, such as springs and mass, the method cannot achieve.

\paragraph*{Deformable Surface}\index{deformable surface} was introduced first time by Terzopoulos \textit{et al.} \cite{tpb87}, using finite difference for the integration of energy-based Lagrange equations based on Hooke's law. 

It was very successful in creating and animating surfaces. However, this method requires not only the discretization of the surfaces into spline patches, but also the specification of local connectivity for spring mass systems. These involve a significant amount of manual preprocessing before the surface model can be used.

\paragraph*{Linear Mass Spring System}\index{linear mass spring system} has been widely used for modeling elastic objects as shown in \xf{fig:massspring}. It is actually derived from the particle model; however, it simplifies the modeling of the inter-particle connection by using flexible springs. Three dimensional systems contain a finite set of masses connected by springs, which are assumed to obey Hooke's Law. 

%the cloth spring. Later, spring-mass based models\cite{gl99}.

\begin{figure}[h]
\hrule\vskip4pt
\begin{center}
	{	 \includegraphics[width=3in]{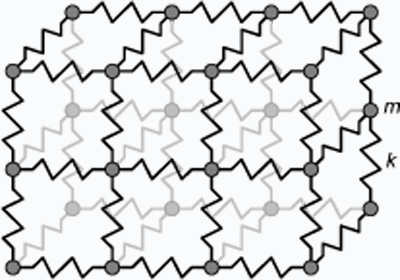}}
\caption{Mass-spring Model}
\label{fig:massspring}
\end{center}
\hrule\vskip4pt
\end{figure}

This method was first introduced by Terzopoulos \cite{TPF89} to describe melting effects. Particles, which are connected by springs, have an associated temperature as one of their properties. The stiffness of the spring is dependent on the temperature of the linked two particles. Increased temperature decreases the spring stiffness. When the temperature reaches the melting point, the stiffness becomes zero.

The advantages of mass-spring model are that it is easy to construct and display the simulation dynamically. 

The disadvantages are that such system restricts to only the objects with small elastic deformation with approximation of the true physics, not for the objects that require large elastic deformation, such as fluid. This method also has difficulties to simulate the separation and fusion of a constant volume object. Moreover, the spring stiffness is problematic. If the spring is too weak, for the closed shape model with only simple springs to model the surface will be very easy to collapse. If we try to avoid the collapse, we need to model with spring stiffer, and then we will have difficulty to choose the elasticity because the object is nearly rigid. Another disadvantage is that the mass spring system has less stability and requires the numerical integrator to take small time steps \cite{BW92} than FEM model. 

%\paragraph*{Angular Springs With Stiff Distance Constraints} is similar to linear mass spring system; however, the force is applied based on angles between springs rather than distance between particles. In this system, mass distances are held constant by stiff springs.

%\paragraph*{Discreet Force Field} presents a vector field of forces using a three dimensional array. Each particle in the system contributes repulsive and / or attractive forces to nearby force `elements' in the array. Later all of the particles have the force of their discrete force 'element' applied to them.

\paragraph*{Finite Element Method}\index{FEM model} known as FEM Model \cite{gm97}, is the most accurate physical model compared to other models. It treats deformable object as a continuum, which means the solid bodies with mass and energy distributed all over the object. This continuum model is derived from equations of continuum mechanics. The whole model can be considered as the equilibrium of a general object subjected to external forces. The deformation of the elastic object is a function of these forces and the material property. The object will stop deformation and reach the equilibrium state when the potential energy is minimized. The applied forces must be converted to the associated force vectors and the mass and stiffness are computed by numerically integrating over the object at each time step, so the re-evaluation of the object deformation is necessary and requires heavy pre-processing time \cite{gm97}. 

%based techniques (Fedkiw, 2003): 

The advantage of FEM model is that it gives more realistic deformation result than mass-spring system because the physics are more accurate.  

The disadvantage is that the system lacks efficiency. Because the energy equation will be used, the FEM is only efficient for the small deformation of the elastic object, such as application to the plastic material, which has a small deformation range. Alternatively, the object has less control elements needed to be computed, as in cloth deformation. If we need to simulate the human soft body parts or facial animation, the deformation rate is very high. It will be very difficult and sometimes impossible to carry out the integration procedure over the entire body. Therefore, it has been limited to apply in real-time system because of the heavy computational effort (usually it is done off-line). Moreover, the implementation is complicated.

\paragraph*{Fluid Based Model}\index{fluid based model} \cite{ND02} consists of two components: an elastic surface and a compressible fluid as shown in \xf{fig:fluid}. The surface is represented as a mass spring system. The fluid is modeled using finite difference approximations to the Navier-Stokes equations of fluid flow. \xf{fig:fluid} describes how this model simulates the fluid flows down a sink simulated. The inner layer is modeled by a particle system, which is similar to real water molecules. Using the numerical methods, the motion of each particle can be computed. In this example, the motion of the each particle is at the center of the basin, and points down to the sink. 

\begin{figure}[h]
\hrule\vskip4pt
\begin{center}
	{ \includegraphics[width=3in]{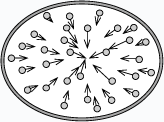}}
\caption{Fluid-Based Soft-Object Model}
\label{fig:fluid}
\end{center}
\hrule\vskip4pt
\end{figure}

The fluid based model uses physically based modeling and it produces realistic fluid animation. It illustrates the behavior of fluid in environments with gravity and collisions with planes.

The disadvantage of this model is the heavy computation for both elastic surface and density inside fluid. It also provides a solution to the constant volume problem.

\paragraph*{Pressure Model}\index{pressure model} was introduced by M. Matyka \cite{mm041, mm042, mm042}. It simulates an elastic deformable object with a internal pressure based on the ideal gas law as shown in \xf{fig:pressure}. The object volume is calculated approximately by bounding box, shaped as sphere, cube, or ellipsoid. Another method to determine the object volume is based on Gauss's Theorem.

\begin{figure}[h]
\hrule\vskip4pt
\begin{center}
	{	 \includegraphics[width=1.5in]{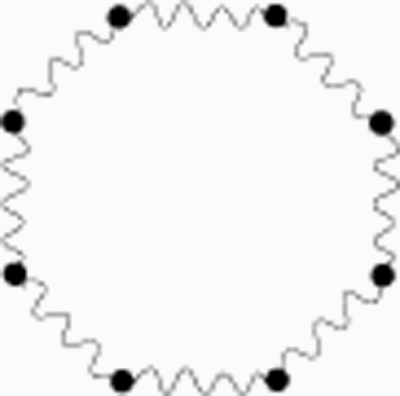}}
\caption{Pressure Soft Body Model}
\label{fig:pressure}
\end{center}
\hrule\vskip4pt
\end{figure}

Advantage of this model is that it gives very convincing effects about elastic property in real time simulation. The object behaves like a balloon filled only with air. 

However, it can not imitate more interesting effects, such as human tissue. It can not achieve the effect of semi-liquid deformable object because the air pressure density is uniform inside of the object, which is different from liquid with non-uniform density. It is not accurate for describing the inertia of the semi-liquid object.

%\paragraph*{SPAM} Smooth-Particle Applied Mechanics
%can model both fluids and solids using free-moving particles.
%\section{Existing Physical Engine}

%Professional game physics engine Havok

%Virtual reality simulation engine Vortex

\section{Summary of The Existing Models}
Previous work on deformable object animation uses physically-based methods with local and global deformations applied directly to the geometric models. Based on the survey of the existing elastic models, we conclude their usage as the two types:
\begin{itemize}
\item Interactive models are used when speed and low latency are most important and physical accuracy is secondary. Typical examples include mass-spring models and spline surfaces used as deformable models. These can satisfy the character animation with exaggerated unrealistic deformation.
\item Accurate models are chosen when physical accuracy is important in order to accomplish the surgical training purpose which requires the accurate tissue modeling. The continuum simulation model, for instance, the most accurate model, FEM, is not ideal for simulation requiring real time interaction and the object undergoing large deformation. 
\end{itemize}

%When two flexible objects collide, they exert reaction forces on each other resulting in the deformation of both objects. When one object self collides, it may deform and result n self-intersection. 

In short, elastic object simulation is a dilemma of demanding accuracy and interactivity.

\chapter{Procedural Modeling Methodology}
\label{chapt:procedural modeling methodology}
\index{procedural modeling methodology}

\section{Graphics Objects Modeling Methods}

\paragraph*{Polygonal Methods}\index{graphics objects modeling methods!polygonal} create geometric objects that can be described by their convex planar polygonal surfaces. These methods are easy to describe the shapes of mathematical objects rendered on graphics system. However, they are difficult to describe physical objects, such as cloud and fire, and their complex behaviors combined with physical laws \cite{ea03}.

\paragraph*{Procedural Methods}\index{graphics objects modeling methods!procedural} build natural phenomena, 3D models and textures in an algorithmic manner and generate polygons only during the rendering process. The details of the object modeling can be controlled upon vary requests. Meanwhile, these methods are easy to combine computer graphics with physical laws \cite{ea03, wk07}.

\section{Procedural Methods}
We use procedural modeling methods in our elastic object simulation system. One of the possible approaches to procedural modeling, a particle system, is designed to model elastic objects as described in this section. This particle system is also capable of describing the complex behaviors of elastic objects based on physical laws, such as solving differential equations of thousands of particles on those elastic objects. Another approach is language-based procedural method \cite{ea03}, which generates complex objects with simple programs. 

In order to model an elastic object, we need to study the following basic data structures, which are varied in one-dimensional, two-dimensional, and three-dimensional modeling methods.

\paragraph*{Particles}\index{particle} are objects that have mass, position, velocity, and forces applied on them, but have no spatial extent. Moreover, the differential position and velocity change are two properties for these computation of each particle.
\paragraph*{Springs}\index{spring} are massless with natural length not equal to zero. They are the linkage of particles. There must be at least one spring connects with two particles paired by modeling algorithm.
\paragraph*{Faces}\index{face} are the data type that consists of springs as the edges and particles as the vertices.

\section{1D}
The techniques used in an one-dimensional object are presented here, which are applied subsequently to models in two and three dimensions. An one-dimensional object with one end fixed as shown in \xf{fig:OneD1}. The other end is interacted by users with mouse as in \xf{fig:OneD2}. The interacted force is restricted to one dimension.

%the spring by dragging it or squeeze it. The force of a single spring is restricted to one dimensional. The dangling end strives apart at certain distance from the fixed end because spring generates opposite force when pulled apart or pushed together along with a spring damping term which models the loss energy. A rubber rope is a good example of 1D elastic objects.\cite{go01}

\begin{figure*}[h]
\hrule\vskip4pt
\begin{center}
	\subfigure[The Initial Spring System is at Rest]
	{\label{fig:OneD1}
	 \includegraphics[width=1.6in]{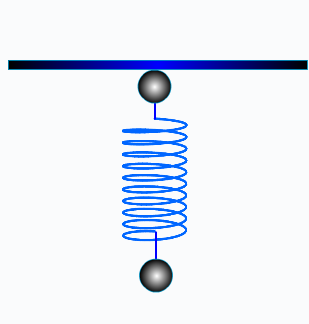}}
	 \hspace{.3in}
	\subfigure[The Compressed Force is Applied]
  {\label{fig:OneD2}
	 \includegraphics[width=1.6in]{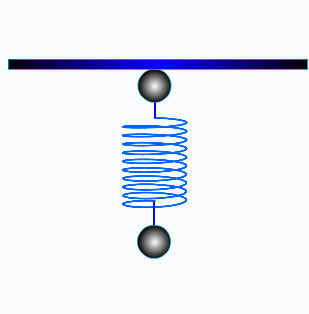}}
	  \hspace{.3in}
	\subfigure[The Stretched Force is Applied]
  {\label{fig:OneD3}
	 \includegraphics[width=1.6in]{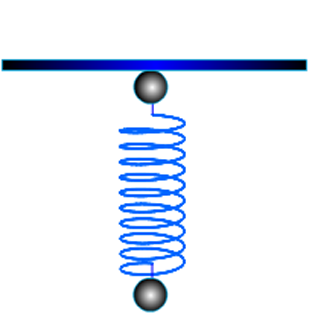}}
\caption{One Dimensional Elastic Object}
\end{center}
\hrule\vskip4pt
\end{figure*}

\subsection{Geometric Data Type}
\paragraph*{Particle} There are two mass particles $P_0$ and $P_1$ on a single spring $SP_1$.
\paragraph*{Spring} In one-dimensional object, only one type of the spring, structural spring, is introduced. Structural spring, is used to model the object shape, connected by the two mass particles in this case. In \xf{fig:OneD1}, the spring is at the initial state of equilibrium. The natural length of the spring is $l$ and the force $f$ is zero. In \xf{fig:OneD2}, the spring is compressed by an external mouse force. The current spring length $l'<l$ and the spring force restores the elastic object to its equilibrium position $f>0$. When the spring is stretched out as shown in \xf{fig:OneD3}, the current spring length $l'>l$ and the force of the spring $f<0$.

\subsection{Modeling Algorithm}

\begin{itemize}
\item Step1: Create two particles $P_0$ and $P_1$ with positions $(x_0, y_0)$ and $(x_1, y_1)$ shown in \xf{fig:1dobj}. %Fig.\ref{fig:1dobj}.
\item Step2: Create a spring $S_1$ with these two particles as two ends $Sp_1$ and $Sp_2$.
\end{itemize}

\begin{figure}[h]
\hrule\vskip4pt
\begin{center}
\includegraphics[width=3.5in]{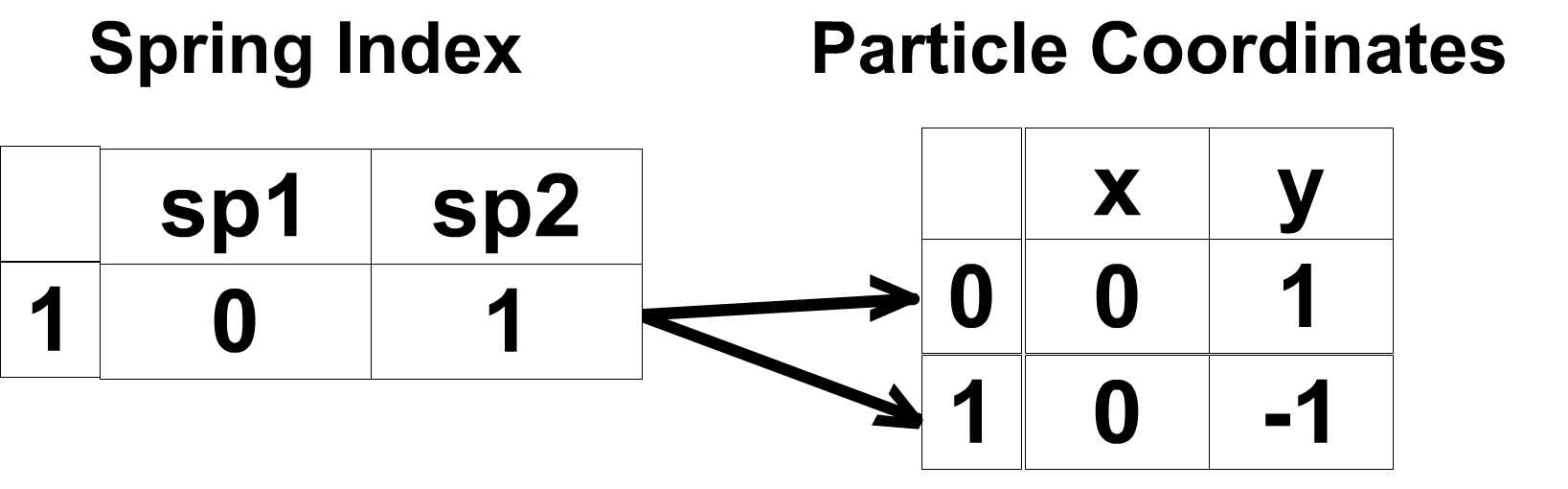}
\caption{Data Structure for One-dimensional Object Representation}
\label{fig:1dobj}
\end{center}
\hrule\vskip4pt
\end{figure}

%For each particle, its strucures The velocity is ${\bf V}_{1\left(x_1,y_1\right)}$ and ${\bf V}_{2\left(x_2,y_2\right)}$. with mass value $m_1$ and $m_2$.
%\item Step 2: The particle $P_{1}$ is fixed, so its gravity force has no effect. The particle $P_{2}$ is connected with a spring, and its gravity force is ${\bf F}^{g}$.
%\item Step 3: Calculate all forces accumulated on the first mass particle, and calculate all the forces accumulated on the second mass particle.
%\item Step 4: Integration the object's motion by calculating the derivative velocity and its new position for each particle.
%\item Step 5: Loop through Step 3.
%\end{itemize}

%\subsection{Result and Summary}
%We have modeled a one-dimensional elastic object with one spring connected by two particles. Furthermore, we have shown the forces to model a simple elastic object.

\section{2D}
In this section, we extend the one-dimensional elastic object to two dimensions. We create two separated elastic circles, inner circle and outer circle. Both of them consist of the same modeling structure as one-dimensional mass particles  and springs. Then, the two concentric circles, inner and outer, are connected by various springs to become one two-layered elastic object. However, the distinct features in two-dimensional object have more types of springs presented and the air pressure inside the two-layer close shape is calculated. The spring surface prevents infinite expansion of the air; meanwhile the internal pressure avoids the surface collapsing.

\subsection{Geometric Data Type}

Two-dimensional object is made of three types of primitives, mass particles, springs, and indexed triangular faces.

\paragraph*{Particles} are defined based on their coordinates related to $x$ and $y$ axes. Consider a unit circle with twelve particles as an example shown in \xf{fig:TwoD11}. The spatial position for each particle $P_i$ is $(x_i,y_i)$ can be defined by the two equations:

\begin{figure}[h]
\hrule\vskip4pt
\begin{center}
	{	 \includegraphics[width=2in]{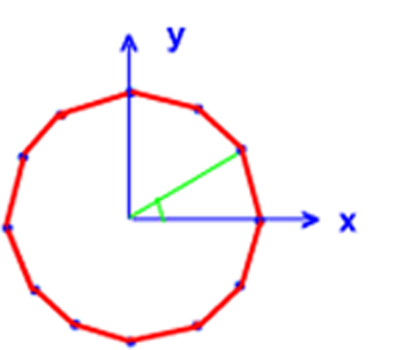}}
\caption{Two-dimensional Elastic Object with Single Layer}
\label{fig:TwoD11}
\end{center}
\hrule\vskip4pt
\end{figure}

\noindent
$x(\theta) = \cos (\theta+\Delta\theta)$

\noindent
$y(\theta) = \sin (\theta+\Delta\theta)$

\noindent
where

\noindent
$0^0 \leq \theta \leq 360^0$

\noindent
$\Delta\theta$ is a small angle stepping along $\theta$
%\begin{equation}
%x = Radius \cdot \cos \left( \frac{i \cdot 2 \cdot \pi}{numParticles} \right)
%\end{equation}
%\begin{equation}
%y = Radius \cdot \sin \left( \frac{i \cdot2  \cdot \pi}{numParticles} \right)
%\end{equation}

\paragraph*{Springs} In additional to the structural spring, which also exists in one-dimensional object, there are two other types of springs, radius spring and shear spring.

\begin{figure*}[h]
\hrule\vskip4pt
\begin{center}
	\subfigure[Structural Springs]
	{\label{fig:TwoDS}
	 \includegraphics[width=1.8in]{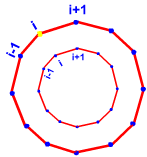}}
	 \hspace{.6in}
	\subfigure[Radius Springs]
  {\label{fig:TwoDR}
	 \includegraphics[width=1.8in]{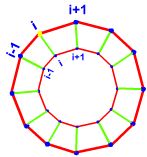}}

	 % \hspace{.3in}
	\subfigure[Shear Left Springs]
  {\label{fig:TwoDSL}
	 \includegraphics[width=1.8in]{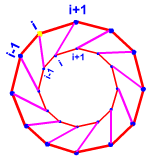}}
	  \hspace{.6in}
	\subfigure[Shear Right Springs]
  {\label{fig:TwoDSR}
	 \includegraphics[width=1.8in]{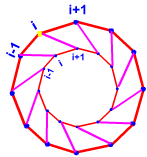}}
\caption{Four Types of Springs on a Two-dimensional Object}
\end{center}
\hrule\vskip4pt
\end{figure*}

\begin{diamonds}
\item	Structural springs\index{spring!structural}: give the basic structure of inner circle and outer circle to prevent neighboring particles from getting too close to one another as shown in \xf{fig:TwoDS}. Linkage of each structural spring i is to connect with two particles${P}_{i}$ and ${P}_{i+1}$ or ${P}_{i-1}$ and ${P}_{i}$.

\item Radius springs\index{spring!radius}: are the springs connected from particles on inner circle to the particle on the outer circle as part of the circle radius in order to prevent the bending of the surface as shown in \xf{fig:TwoDR}. Linkage of each radius spring $i$ is to connect with particle ${P}^{inner}_{i}$ and the particle ${P}^{outer}_{i}$.

\item Shear springs\index{spring!shear}: are springs connected from particles on inner circle to their diagonal neighbors on outer circle in order to avoid the surface fold over . Linkage of each left shear spring\index{spring!left shear} i is to connect with particle ${P}^{outer}_{i}$ diagonally and the particle ${P}^{inner}_{i-1}$; connect with particle ${P}^{outer}_{i+1}$ diagonally and the particle ${P}^{inner}_{i}$ and so on as shown in \xf{fig:TwoDSL}. Linkage of each right shear spring\index{spring!right shear} i is to connect with particle ${P}^{outer}_{i-1}$ diagonally and the particle ${P}^{inner}_{i}$; connect with particle ${P}^{outer}_{i}$ diagonally and the particle ${P}^{inner}_{i+1}$ and so on as shown in \xf{fig:TwoDSR}.
\end{diamonds}

%They prevent the object from distorting. shear spring resists stretching and thus prevents shearing.

%Bend spring revents the object from folding over. Every node connected to its second neighbor in every direction. Bend spring resists contracting and thus prevents bending.

%\begin{figure*}[h]
%\begin{center}
%	{\label{fig:TwoD111}
 %		\includegraphics[width=2in]{images/TwoD111.png}}
%\caption{Three Types of Springs on a Two-dimensional Object}
%\end{center}
%\end{figure*}

\paragraph*{Faces} are the data structure for the only purpose of drawing and displaying a filled object to a two-dimensional object. The triangular facets\index{face!triangular} can be drawn separately and  visualized as a filled circle as shown in \xf{fig:TwoD1}.

\begin{figure}[h]
\hrule\vskip4pt
\begin{center}
	{ \includegraphics[width=2in]{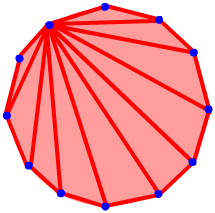}}
\caption{Two-dimensional Elastic Object Facets}
\label{fig:TwoD1}
\end{center}
\hrule\vskip4pt
\end{figure}

\subsection{Modeling Algorithm}

\begin{itemize}

\item	Step 1: Define the number of particles as $n=12$ in our example. Then, the step size is $\Delta \theta= \frac{360^0}{12} =30$ degrees.

\item Step 2: Define the group of particles' position on inner circle and the ones on outer circle as shown in \xf{fig:2dobj}. The first particle $P_0$ is at $(\cos\theta, \sin\theta)$ where $\theta=0^0$, the second particle is at  $(\cos\theta, \sin\theta)$  where $\theta=\theta +\Delta\theta=30^0$... By multiplying the inner and outer coordinates with a different radius value, for example, $Radius_{inner}=1.5$, and $Radius_{outer}=2$ to create two concentric circles.

\item Step 3: Add the structural springs $S_0$, $S_1$, ...$S_{11}$ to the inner circle according to the spring index of inner particles as shown in \xf{fig:2dobj}. The same method is applied to outer structural springs on outer circle. The last spring, $S_{11}$ in our example, is composed of two particles $P_{11}$ and $P_0$ as two ends in order to make a close shape.

\begin{figure}[h]
\hrule\vskip4pt
\begin{center}
	{	 \includegraphics[width=4in]{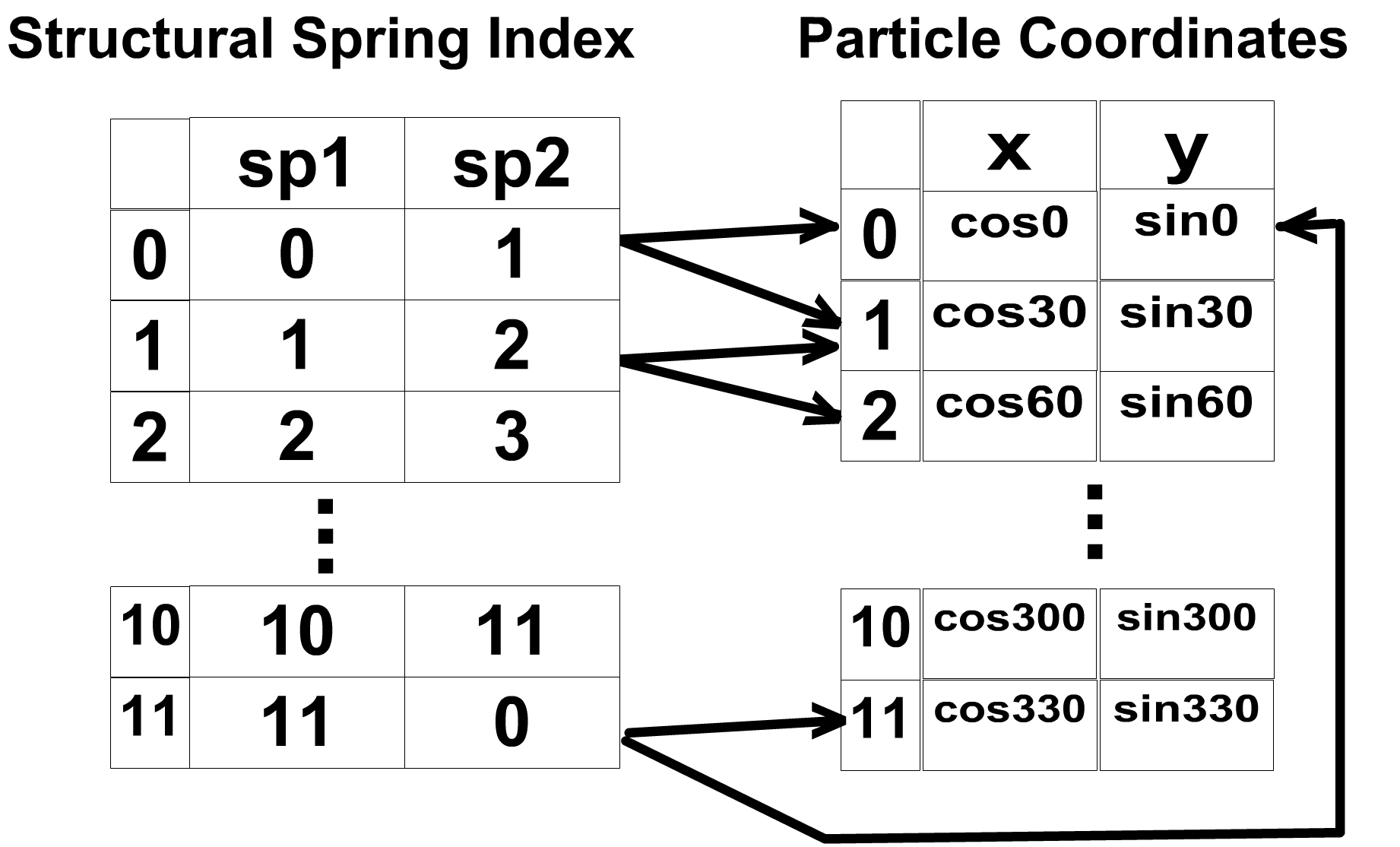}}
\caption{Data Structure for Two-dimensional Object Representation 1}
\label{fig:2dobj}
\end{center}
\hrule\vskip4pt
\end{figure}

\item Step 4: Loop through the number of structural springs $n=12$ to add the same number of radius springs according to the linkage of the inner and outer particles as shown in \xf{fig:2dobj2}.

\begin{figure}[h]
\hrule\vskip4pt
\begin{center}
	{	 \includegraphics[width=3in]{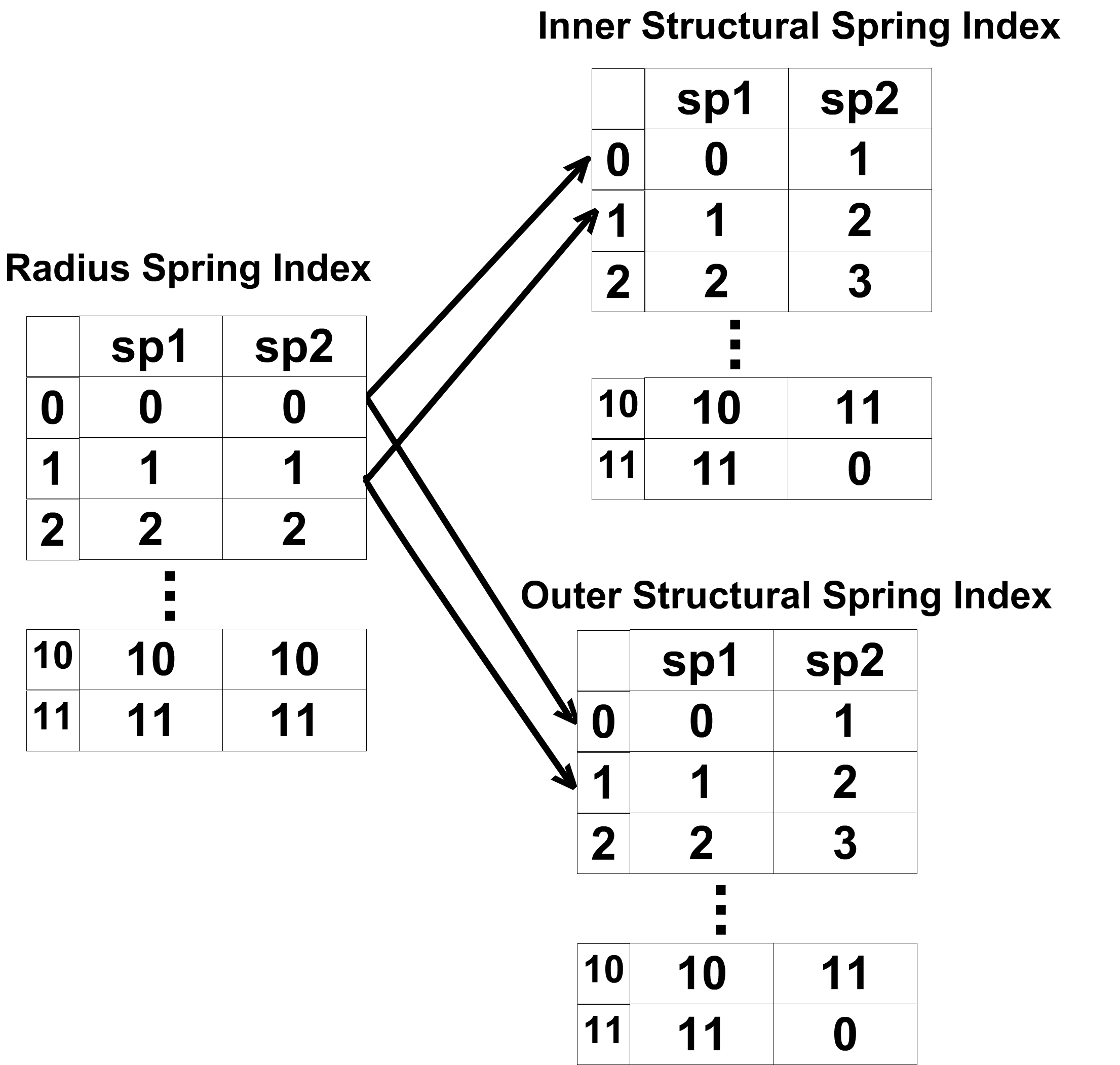}}
\caption{Data Structure for Two-dimensional Object Representation 2}
\label{fig:2dobj2}
\end{center}
\hrule\vskip4pt
\end{figure}

\item Step 5: Loop through the number of structural springs to add the same number of shear left springs and shear right springs according to the linkage of the inner and outer particles as shown in \xf{fig:2dobj3}.

\begin{figure}[h]
\hrule\vskip4pt
\begin{center}
	{	 \includegraphics[width=5in]{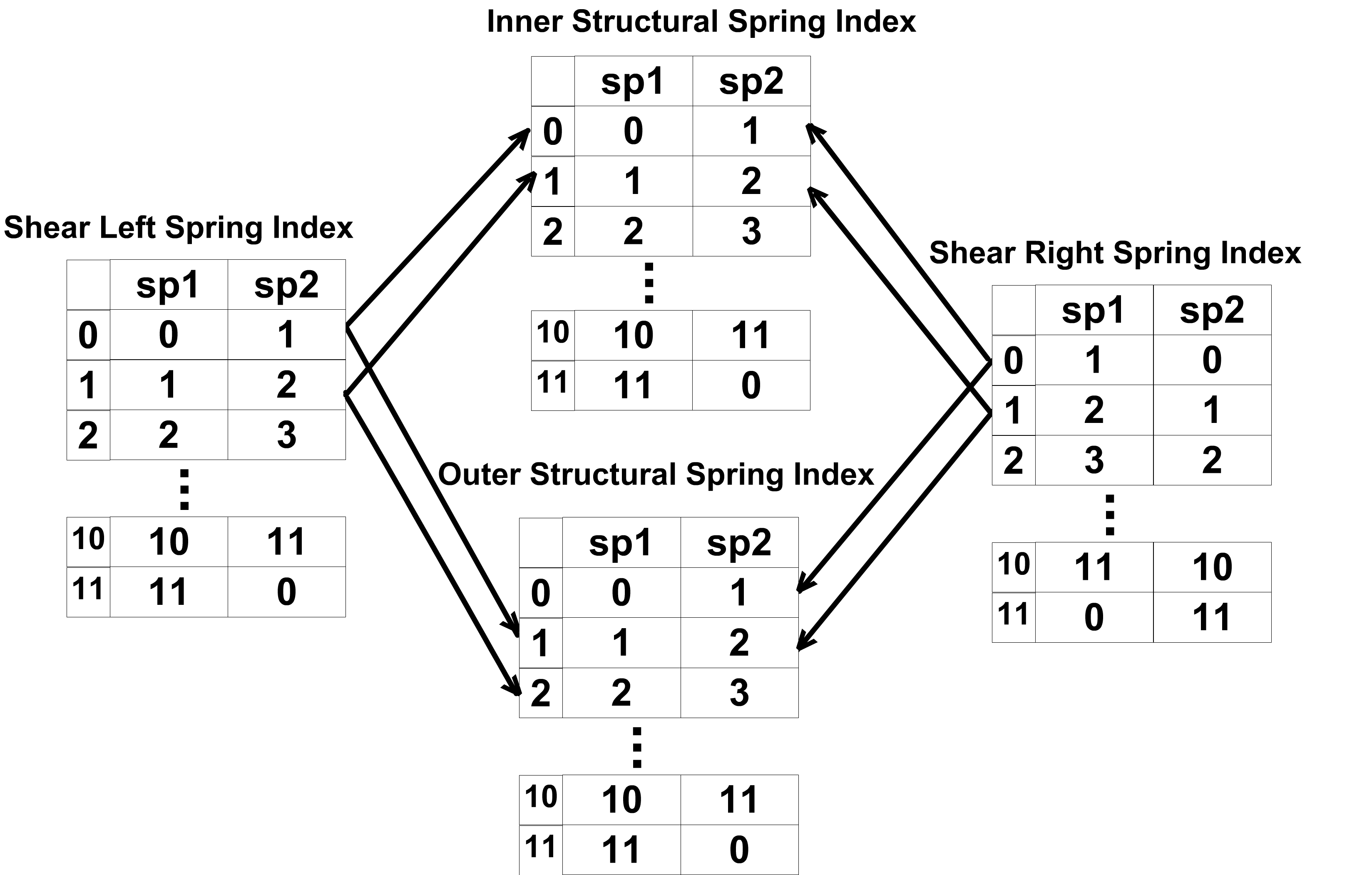}}
\caption{Data Structure for Two-dimensional Object Representation 3}
\label{fig:2dobj3}
\end{center}
\end{figure}
\hrule\vskip4pt
\end{itemize}

%\subsection{Result and Summary}
%I have modeled a two-dimensional double layered elastic object based on the knowledge from one-dimensional object. The modeling of different springs has been introduced. Moreover, the new force, air pressure force and approximated calculation of the object volume have been discussed also.
\section{3D}
In this section, a more complicated three-dimensional elastic object is extended from the two-dimensional object. In the two-dimensional model, the structural springs' index is the most important key data structure to link up all the particles and reference about the index of the particles. This spring linkage method will still work for the model based on the non-uniform sphere geometric modeling method. However, in the other geometric modeling method, the uniform sphere modeling, the faces' index is the key data structure of the linkage to other data structure, such as particles and springs. The reason is because in the later geometric modeling method, each facet on the object is used for subdivision of other facets in each iteration. Compared with a two-dimensional object, the three-dimensional object consists the same types of primitives, such as particles, springs, and faces, but extended to $z$ axis.
\subsection{Non-Uniform Sphere}
\subsubsection{Geometric Data Type}
% For any particle $i$, the mass, position, velocity, the derived new position, the derived new velocity, and the the forces act on it will represented as the following:

%\noindent
%$mass_{i} = \left(m_x, m_y, m_z\right)$

%\noindent
%$position_{i} = \left(r_x, r_y, r_z\right)$

%\noindent
%$velocity_{i} = \left(v_x, v_y, v_z\right)$

%\noindent
%$dposition_{i} = \left(dr_x, dr_y, dr_z\right)$

%\noindent
%$dvelocity_{i} = \left(dv_x, dv_y, dv_z\right)$

%\noindent
%$force_{i} = \left(f_x, f_y, f_z\right)$

One of the simplest non-uniform modeling methods\index{non-uniform modeling} to generate an approximate facet sphere uses Polar to Cartesian Coordinates method. Consider $\theta$ the angle on $xy$-plane (around $z$-axis), known as the Azimuthal Coordinate. The angle $\phi$ is from $z$-axis, known as the Polar Coordinate. If we fix $\theta$ and draw curves as we change $\phi$, we get circles of constant longitude; if we fix $\phi$ and vary $\theta$, we obtain circles of constant latitude \cite{ea03}.

\begin{figure}[h]
\hrule\vskip4pt
\begin{center}
	{ \includegraphics[width=2in]{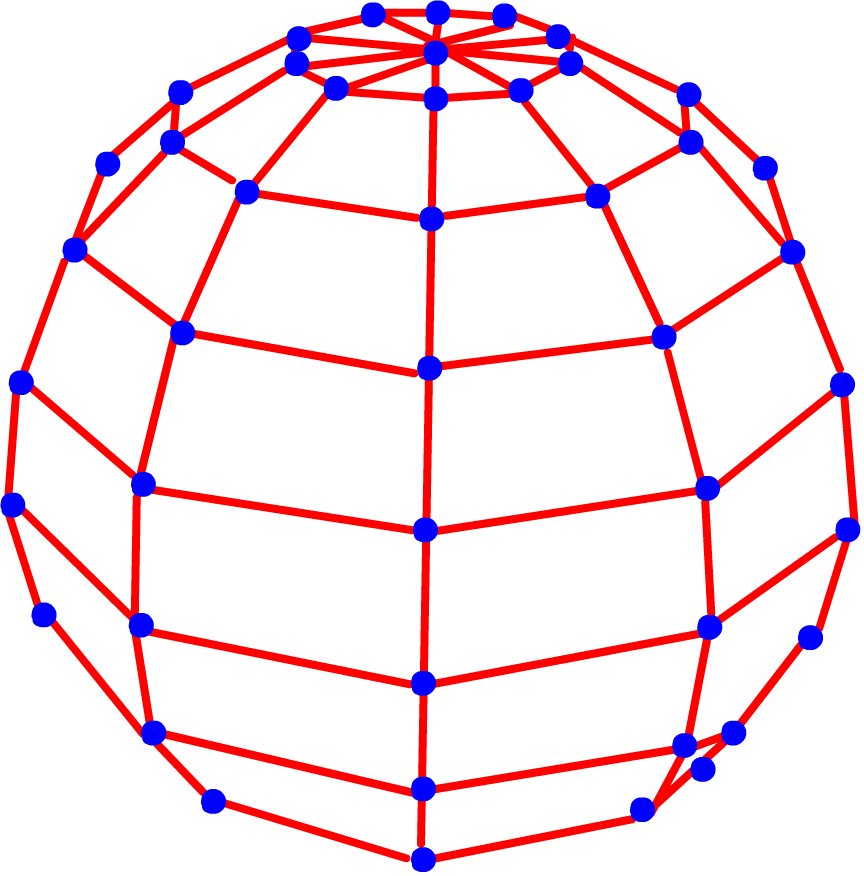}}
\caption{Polar Cartesian Coordinates Non-uniform Sphere Generation}
\label{fig:ThreeD}
\end{center}
\hrule\vskip4pt
\end{figure}

\paragraph*{Particles} The spherical coordinates for a particle $i$ can be defined by the three equations:

\noindent
$x(\theta, \phi) = \cos\theta \sin\phi$

\noindent
$y(\theta, \phi) = \cos\theta \cos\phi$

\noindent$z(\theta, \phi)= \sin\phi$

\noindent
where

\noindent
$0^0 <= \theta <= 360^0$

\noindent
$-90^0 <= \phi <= 90^0$

\noindent
By stepping $\theta$ and $\phi$ in small angles $\Delta\theta$ and $\Delta\phi$ between their bounds as the number of slices and stacks, the particles are:

\noindent
$P_0(x_0, y_0, z_0)$=
$(\sin\theta\cos\phi,
\cos\theta\cos\phi,
\sin\phi)$

\noindent
$P_1(x_1, y_1, z_1)$=
$(\sin\theta\cos(\phi+\Delta\phi) , \cos\theta\cos(\phi+\Delta\phi),
\sin(\phi+\Delta\phi))$

\noindent
$P_2(x_2, y_2, z_2)$=
$(\sin(\theta+\Delta\theta)\cos\phi,
\cos(\theta+\Delta\theta)\cos\phi,
\sin\phi)$

\noindent
$P_3(x_3, y_3, z_3)$=
$(\sin(\theta+\Delta\theta)\cos(\phi+\Delta\phi) , \cos(\theta+\Delta\theta)\cos(\phi+\Delta\phi),
\sin(\phi+\Delta\phi))$

\noindent
$\cdots$

%c=M_PI/180;//degrees to radians, M_PI =3.14159...
%for(phi=-80;phi<=80;phi+=20)
%{glBegin(GL_QUAD_STRIP);
%for(theta=-180;theta<=180;theta+=20)
%{
%x=sin(c*theta)*cos(c*phi);
%y=cos(c*theta)*cos(c*phi);
%z=sin(c*phi);
%glVertex3d(x,y,z);
%x=sin(c*theta)*cos(c*(phi+20));
%y=cos(c*theta)*cos(c*(phi+20));
%z=sin(c*(phi+20));
%glVertex3d(x,y,z);
%}
%glEnd();
%}

%glBegin(GL_TRIANGLE_FAN);
%glVertex3d(x,y,z);
%c=M_PI/180;
%z=sin(c*80);
%for(theta=-180;theta<=180;theta+=20)
%{
%x=sin(c*theta)*cos(c*80);
%y=cos(c*theta)*cos(c*80);
%z=sin(c*phi);
%glVertex3d(x,y,z);
%}
%glEnd();
%x=y=0;
%z=-1
%glBegin(GL_TRIANGLE_FAN);
%glVertex3d(x,y,z);
%c=M_PI/180;
%z=-sin(c*80);
%for(theta=-180;theta<=180;theta+=20)
%{
%x=sin(c*theta)*cos(c*80);
%y=cos(c*theta)*cos(c*80);
%z=sin(c*phi);
%glVertex3d(x,y,z);
%}
%glEnd();

\noindent
However, at the North and South Pole areas, we can only use triangles to present because all lines of latitude are converged.

\noindent
The particle at the North Pole area can be presented as:

\noindent
$P(x, y, z)$=
$(\sin(\theta+\Delta\theta)\cos{90}^{0},
\cos(\theta+\Delta\theta)\cos{90}^{0},
\sin{90}^{0})$

\noindent
The particle at the South Pole area can be presented as:

\noindent
$P(x, y, z)$=
$(\sin(\theta+\Delta\theta)\cos{90}^{0},
\cos(\theta+\Delta\theta)\cos{90}^{0},
-\sin{90}^{0})$

%$\theta$ is longitude (eastward positive, westward negative);
%$\phi$ is latitude (northward positive, southward negative).

\paragraph*{Springs} There are also three types of springs in three-dimensional objects as we described in two-dimensional objects, such as structural, radius, and shear springs.

\begin{figure}[h]
\hrule\vskip4pt
\begin{center}
	{	\includegraphics[width=3in]{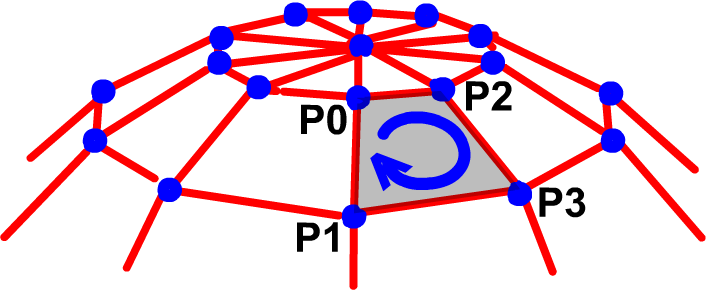}}
\caption{Quadrilaterals and Triangles on Non-uniform Sphere}
\label{fig:3dobjnu1}
\end{center}
\hrule\vskip4pt
\end{figure}

\begin{itemize}
\item Structural spring is still the basic data structure
to form the shapes of inner and outer spheres. Four particles define four springs as the proper order. Taking the first four particles $P_0$, $P_1$, $P_2$, and $P_3$ as an example, the first four springs's are $S_0=P_0P_2$, $S_1=P_2P_3$, $S_2=P_3P_1$, $S_3=P_1P_0$ shown in \xf{fig:3dobjnu1}. The structural springs on two poles are also defined by the particles on poles as proper order.
\item Radius and shear springs, which connect inner and outer layers, follow the same methods as in two-dimensional object.
\end{itemize}

\paragraph*{Faces}
Any quadrilateral-facet\index{face!quadrilateral} on the body of sphere can be represented by four springs:
$S_i$, $S_{i+1}$, $S_{i+2}$, and $S_{i+3}$. Any triangular-facet on the poles can be represented by three springs: $S_j$, $S_{j+1}$, and $S_{j+2}$.

\subsubsection{Modeling Algorithm}

\begin{itemize}
\item	Step 1: Define the number of slices and stacks of a sphere, $n_{slice}=10$ and $n_{stack} = 10$ in our example. Then, the step size is $\Delta \theta= \frac{360^0}{10} =36^0$ and $\Delta \phi= \frac{180^0}{10} =18^0$.
\item Step 2: Define the group of particles' position on inner circle and the ones on outer circle shown in \xf{fig:3dobjnu}. %The first particle $P_0$ is at $(\sin\theta\cos\phi, \cos\theta\cos\phi, \sin\phi)$ where $\theta=0$, the second particle is at  $(\cos\theta, \sin\theta)$  where $\theta=\theta +\Delta\theta=30$...
By multiplying the inner and outer coordinates with a different radius value, for example, $Radius_{inner}=1.5$, and $Radius_{outer}=2$ to create two concentric spheres.

\begin{figure}[h]
\hrule\vskip4pt
\begin{center}
	{	 \includegraphics[width=5.5in]{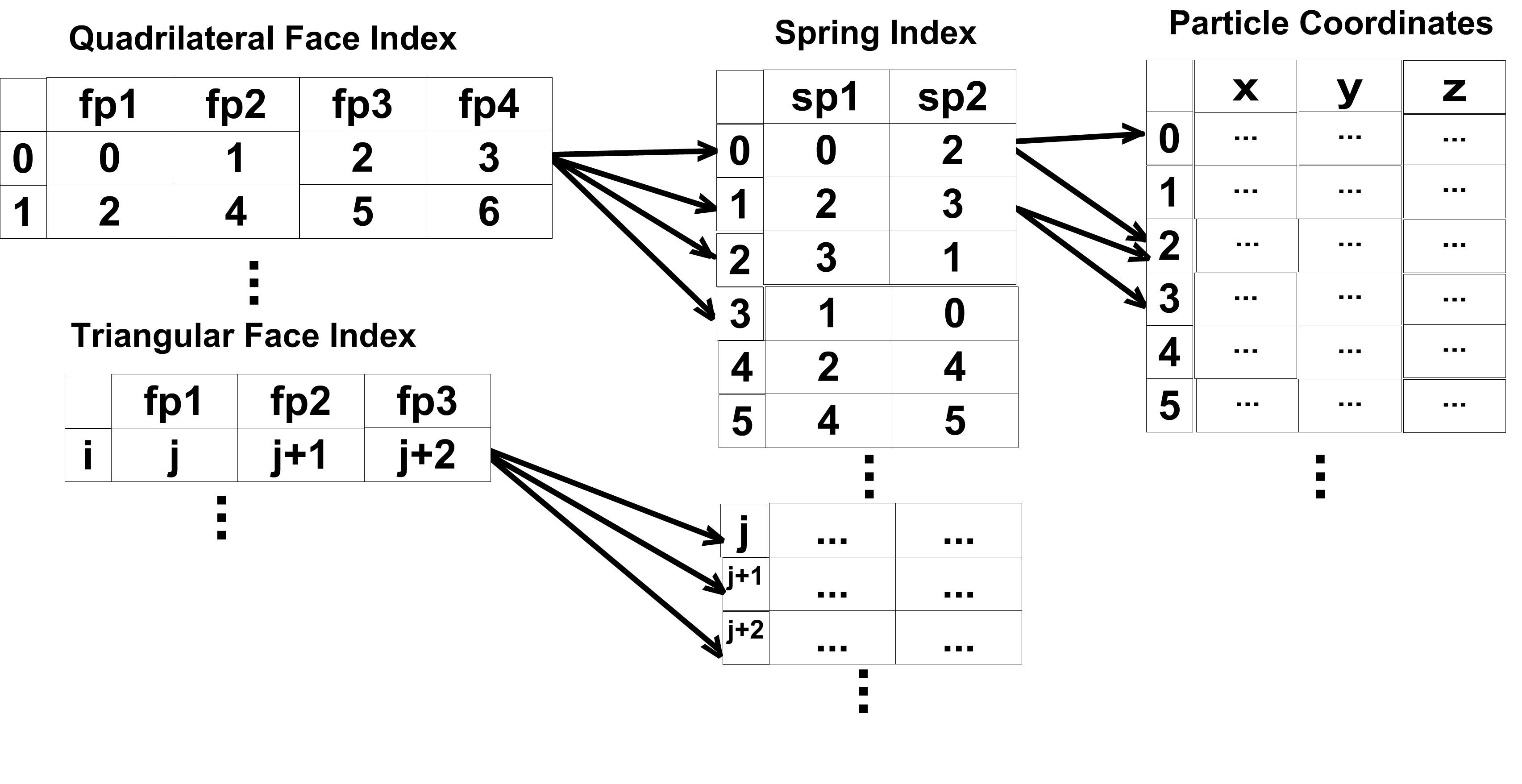}}
\caption{Data Structure for Three-dimensional Non-uniform Object Representation}
\label{fig:3dobjnu}
\end{center}
\hrule\vskip4pt
\end{figure}

\item Step 3: Add the structural springs $S_0$, $S_1$, ... to the inner circle according to the spring index of inner particles as shown in \xf{fig:3dobjnu}. The same method is applied to outer structural springs on outer circle. %The last spring, $S_{11}$ in our example, is composed of two particle $P_{11}$ and $P_0$ as two ends in order to make a close shape.
\item Step 4: Loop through the number of structural springs to add the same number of radius springs and shear springs according to the linkage of the inner and outer particles as described in two-dimensional object modeling method, shown in \xf{fig:2dobj2} and \xf{fig:2dobj3}.
\end{itemize}

\subsection{Uniform Sphere}
An important drawback of the non-uniform sphere model is that the faces vary in both shape (some are triangles and some are quadrilaterals) and size.
\subsubsection{Geometric Data Type}
Surface refinement is a simple way for uniform modeling. It is started with a kernel polyhedron, which is a regular polyhedron with faces that are equilateral triangles. We have used an octahedron with bisecting each face at the same time recursively. This method is a powerful technique for generating approximations to curves and surfaces of a sphere to any desired level of accuracy.

\begin{figure*}[h]
\hrule\vskip4pt
\begin{center}
	\subfigure[The Initial Octahedron Shape]
	{\label{fig:iteration0} \includegraphics[width=1.5in]{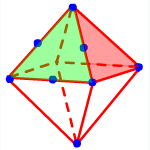}}
	 \hspace{.2in}
	\subfigure[The Unit Facet Sphere Object With One Iteration]
  {\label{fig:iteration1} \includegraphics[width=1.5in]{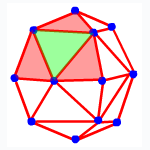}}
	 \hspace{.2in}
	\subfigure[The Unit Facet Sphere Object With Second Iteration]
  {\label{fig:iteration2} \includegraphics[width=1.5in]{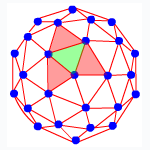}}
\caption{Uniform Sphere Generation}
\end{center}
\hrule\vskip4pt
\end{figure*}

\paragraph*{Particles} The algorithm starts with a regular octahedron shown in \xf{fig:iteration0}. The shape is composed of eight equilateral triangles, determined by six vertices, $P_0(0,0,1)$, $P_1(0,0,-1)$, $P_2(-1,-1,0)$, $P_3(1,-1,0)$, $P_4(1,1,0)$, and $P_5(-1,1,0)$. The vertices of the kernel polyhedron are known to lie on the surface of a unit sphere of radius $r=1$. We fix the two vertices $P_0$ and $P_1$ on $z$ axis and normalize the other five vertices by multiplying $\frac{1}{\sqrt{2}}$ in order to make them lie on the unit sphere, centered at the origin. The six vertices after normalization are $P_0(0,0,1)$, $P_1(0,0,-1)$, $P_2(-0.7,-0.7,0)$, $P_3(0.7,-0.7,0)$, $P_4(0.7,0.7,0)$, and $P_5(-0.7,0.7,0)$.

\paragraph*{Faces} We talk about faces before talking about springs because the face is the key data structure for recursive subdivision and its index is referenced by spring index. The first eight triangular faces defined by the six particles are $f_0 =P_0P_3P_4$, $f_1 =P_0P_4P_5$, $f_2 =P_0P_5P_2$, $f_3 =P_0P_2P_3$, $f_4 =P_1P_4P_3$, $f_5 =P_1P_5P_4$, $f_6 =P_1P_2P_5$, $f_7 =P_1P_3P_2$.
\paragraph*{Springs} Each face is composed of three springs. Therefore, the first twelve springs on the octahedron are $S_0=P_0P_3$, $S_1=P_3P_4$, $S_2=P_4P_0$, $S_3=P_4P_5$, $S_4=P_5P_0$, $S_5=P_5P_2$, $S_6=P_2P_0$, $S_7=P_2P_3$, $S_8=P_1P_4$, $S_9=P_1P_3$, $S_{10}=P_1P_5$, and $S_{11}=P_2P_1$.

\paragraph*{Subdivision} We can subdivide a single triangular face of the kernel polyhedron by projecting the midpoints $pa$, $pb$, $pc$ of its three edges onto the surface of the sphere as shown in \xf{fig:3dobju1}.

\begin{figure}[h]
\hrule\vskip4pt
\begin{center}
	{	 \includegraphics[width=5.5in]{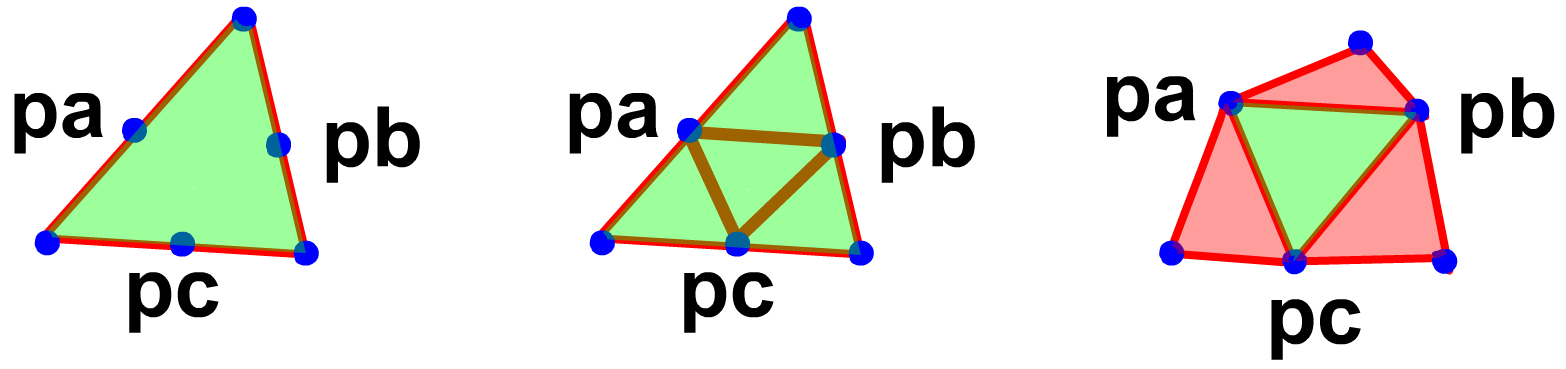}}
\caption{Subdivision of A Triangle By Bisecting Sides}
\label{fig:3dobju1}
\end{center}
\hrule\vskip4pt
\end{figure}

This face is split into four faces by bisecting each edge. The four new triangles are still in the same plane as the original triangle. We move the new vertices $pa$, $pb$, $pc$ to the unit sphere by normalizing each new vertices. The number of particles increases by a factor of 2. The number of springs increases by a factor of 3. The number of facets increases by a factor of 4. We subdivide another 7 triangles with the same method. After subdividing the octahedron once, the number of particles are 12, the number of triangular faces is 32, and the number of springs is 36. We can repeat the subdivision process $n$ times to generate successively closer approximations to the sphere.%Four new triangles are defined using the three original points and the three projected points. After the randomly particle and face generated, each particle will be normalized as the distance to the radius so that it will reside on the sphere.

\subsubsection{Modeling Algorithm}
\begin {itemize}
\item Step 1: Define a collection of particles to create a closed equilateral triangles shape of the elastic object. Define an octahedron object as the initial object with 6 particles, 8 triangular faces, and 12 structural springs as shown in \xf{fig:3dobjui}.

\begin{figure}[h]
\hrule\vskip4pt
\begin{center}
	{	 \includegraphics[width=5.5in]{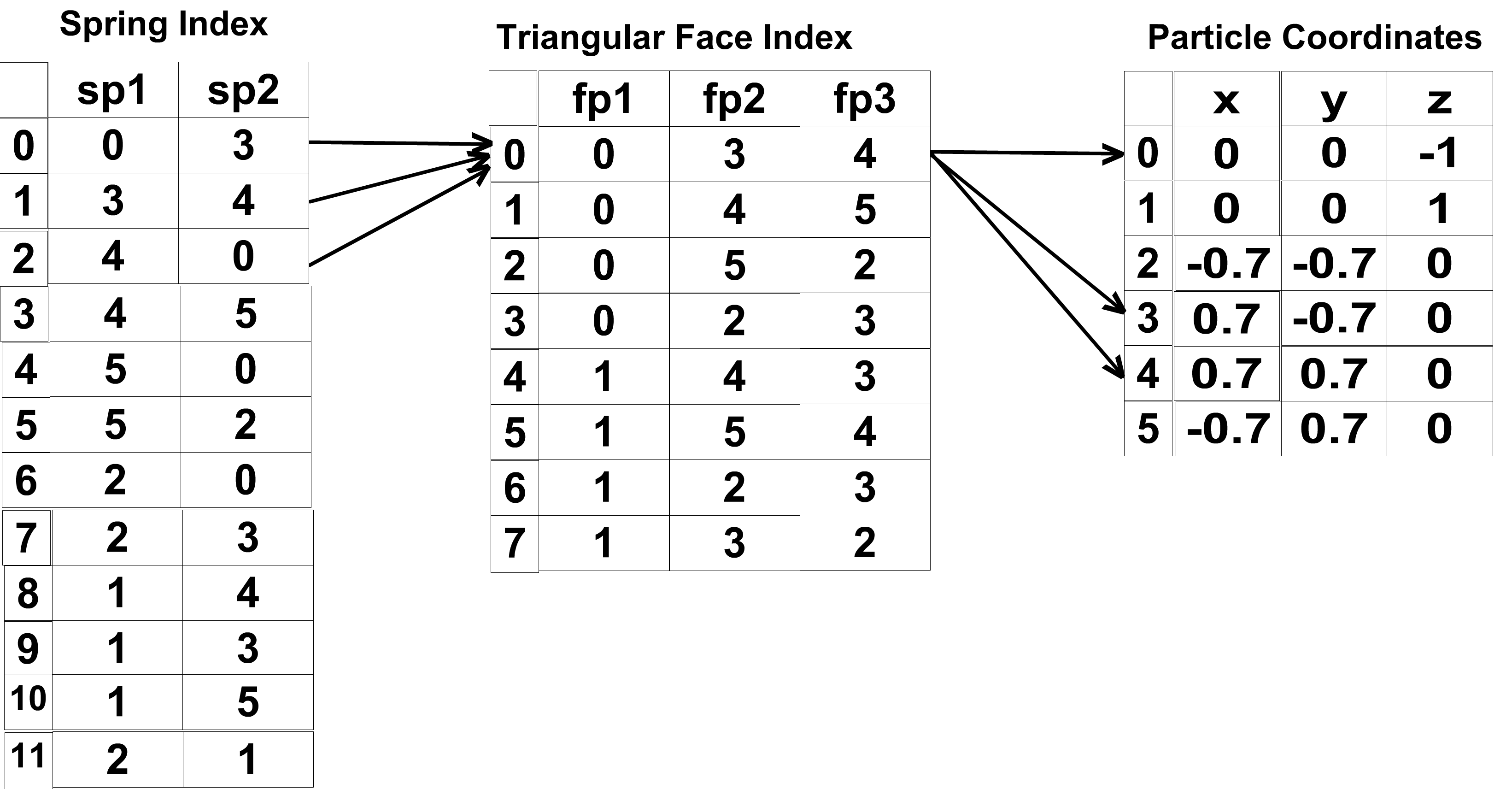}}
\caption{Data Structure for Three-dimensional Uniform Object Representation without Subdivision}
\label{fig:3dobjui}
\end{center}
\hrule\vskip4pt
\end{figure}

\begin{figure}[h]
\hrule\vskip4pt
\begin{center}
	{	 \includegraphics[width=6in]{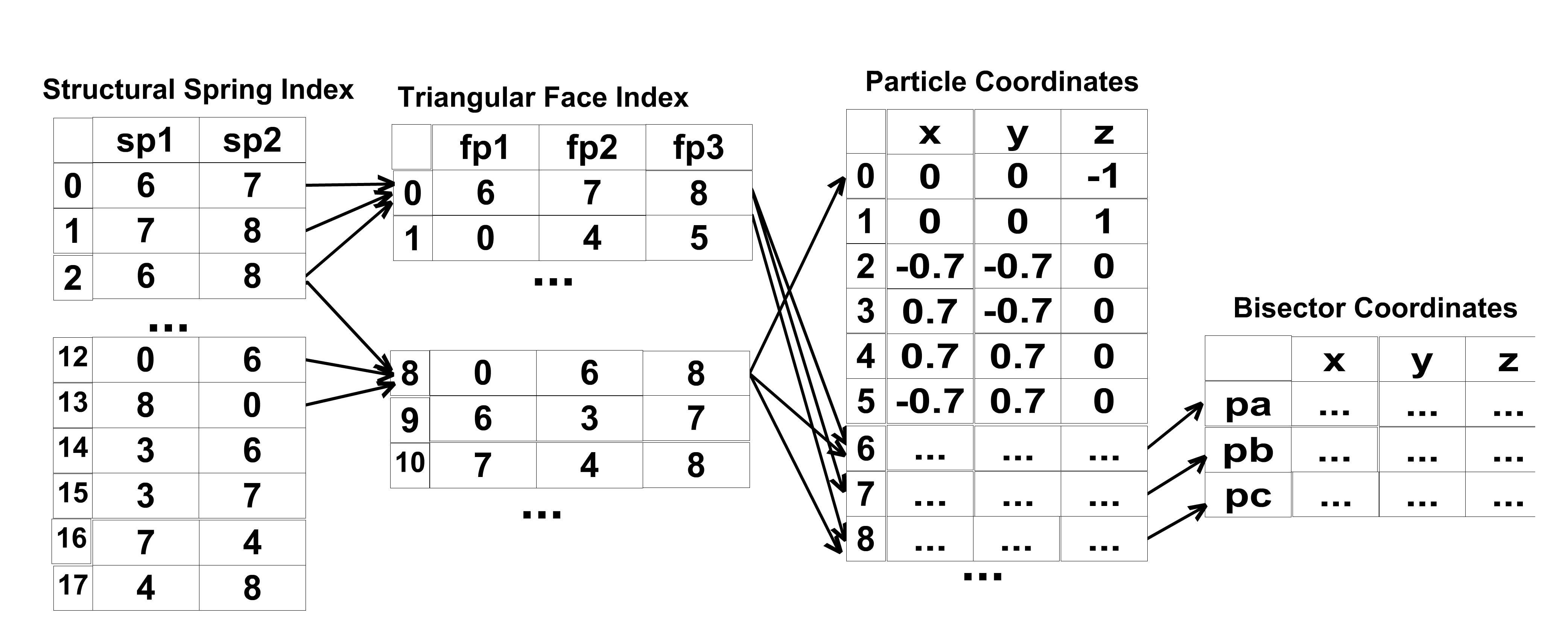}}
\caption{Data Structure for Three-dimensional Uniform Object Representation with the Number of Subdivision n=1}
\label{fig:3dobjui1}
\end{center}
\hrule\vskip4pt
\end{figure}
%\item Step 2: All the particles have the default gravity force $\vec{F}^{g}$ points down to the opposite of y axis.
\item Step 2: Connect the particles with the structural springs according to the edge order of the octahedron to make an inner layer of the three-dimensional object. Each particle is separated equidistantly from its neighbors.
\item Step 3: Check if there is need of subdivision to approach a more spherical object.
\item Step 4: In the first subdivision, the object becomes 12 particles, 32 triangles, and 36 structural springs. The \xf{fig:3dobjui1} shows how new data can be inserted into a collection of particles, faces, and springs after  subdivision.Use the first triangle as a concrete example. In the initial octahedron shape, find the midpoint of each edge on $F_0$. Normalize the coordinates of these three new particles to make them lie on the sphere. Push these three new particles to the particle container. The first face of the initial octahedron has three pointers that point to particles $P_0$, $P_3$, and $P_4$ as shown in \xf{fig:3dobjui}. After subdividing this triangle, it becomes four smaller triangles connected by the bisectors $pa$, $pb$, and $pc$. The three new triangles are pushed onto container $Faces$. The middle triangle replaces the original big triangle because the original triangle does not exist anymore. The pointers on each face point to the correspondent particles as indicated in \xf{fig:3dobjui1}. New structural springs are added to Spring container correspondent to new faces only if there is no such spring has existed yet. The subdivision of the remaining faces follows the same method. More faces are approaching the object to a unit facet sphere.
\item Step 5: Repeat Step 1 to 5 to create an outer layer with larger radius value than the inner layer.
\item Step 6: Loop through the number of structural springs to add the same number of radius springs according to the linkage of the inner and outer particles.
\item Step 7: Loop through the number of structural springs to add the same number of shear left springs and shear right springs according to the linkage of the inner and outer particles.
\end{itemize}

\subsection{Comparison of Non-uniform and Uniform Methods}

The advantages of both methods are that they can be used to describe complex behaviors combined with physical laws, such as elasticity.
% Moreover, the details of object modeling can be controlled upon requests initiated by the user.
Additionally, the level of detail (LoD) of the object can be adjusted depending on the proximity of the
object on the display to the human's eye.

The disadvantage of the non-uniform sphere modeling method is that the facets of the sphere do not have approximately equal size. The facets become smaller at the poles and bigger at the ``equator''. Therefore, the springs are shorter at the poles and longer at the equator. The normal of each spring varies from equator and the ones on the poles. Consequently, this non-uniform modeling method increases errors in force computations for each particle.

The disadvantage of the uniform facet sphere generation algorithm is that it can not generate surfaces of arbitrary resolution. It can be shown that at all levels of recursion, particles at the kernel points are connected to four springs if the kernel object is an octahedron (as shown in \xf{fig:iteration1}). In other cases, all the particles at the kernel points are connected to five springs if the kernel object is an icosahedron (20 faces); all the particles at the kernel points are connected to three springs if the kernel object is a tetrahedron. All particles at recursively generated points are connected to six springs. This will result the irregular surface stiffness and might cause the non-spherical shape because the same pressure will displace the regions of a surface about the kernel points further than the rest of the surface.

To solve this problem, the sum of the spring forces accumulated at a particle can be normalized by multiplying a factor of $\frac{6}{n_{springs}}$, where $n_{springs}$ is the number of springs connected to this particle. For example, if $particle_a$ is a kernel point, which is connected to four springs and the sum of the spring forces is $f_a$; and if $particle_b$, which is the point generated from subdivision, connects to six springs and the sum of the six spring forces is $f_b$. $f_a$ is multiplied by a factor of $\frac{6}{4}$ and $f_b$ is multiplied by a factor of $\frac{6}{6}$.

Our simulation system ignores the described drawback resulting from the uniform sphere modeling method. We find a set of air pressure and spring stiffness parameter values at which the simulation is stable by trial and error. Thus, the difference of the forces for every particle either connected to four springs or six springs is not addressed in this work.

\chapter{Physical Based Modeling Methodology}
\label{chapt:physical based modeling methodology}
\index{Physical Based Modeling Methodology}

A one-dimensional object model includes gravity force ${\bf F}^{g}$, user applied force ${\bf F}^{a}$, and collision force ${\bf F}^{c}$ as external forces; linear structural spring force ${\bf F}^{h}$ and spring damping force ${\bf F}^{d}$ as internal forces.

\begin{equation}
{\bf F} = {\bf F}^{g} + {\bf F}^{h} + {\bf F}^{d} + {\bf F}^{a} + {\bf F}^{c}
\end{equation}

A two-dimensional object model is considered as a closed shape with air pressure inside. Then, the air pressure ${\bf F}^{p}$ is a new internal force exist in two-dimensional object in addition to the common forces in one-dimensional object. Accumulation of forces on a three-dimensional object is similar to forces applied on the two-dimensional one. The only difference is that all forces on three-dimensional objects are extended to axis z.

\begin{equation}
{\bf F} = {\bf F}^{g} + {\bf F}^{h} + {\bf F}^{d} + {\bf F}^{a} + {\bf F}^{p} + {\bf F}^{c}
\end{equation}

\section{Gravity Force}
Gravity force is a constant force at which the earth attracts objects based on their masses. In most cases, the particle system does not include gravitation, but, in our system, particle gravities represent object's density. Users can set particle gravities to a non-zero value. $\textit{g}$ is a constant scalar of the gravitational field. %${\bf g}$ is a constant vector of the gravitational field which points to opposite of $y$ axis.

\begin{equation}
{\bf F}^{g}= m\textit{g}
\end{equation}

%A gravitational force for a one-dimensional or two-dimensional object is
%\[{\bf F}^{g}=
%\left[\begin{array}{c}
%{0} \\
%{-m{\bf g}} \end{array}\right]\]

%A gravitational force for a three-dimensional object is
%\[{\bf F}^{g}=
%\left[\begin{array}{c}
%{0} \\
%{-m{\bf g}} \\
%{0} \end{array}\right]\]

\section{Spring Hooke's Force}
Spring force is a linear force exerted by a compressed or
stretched spring upon two connected particles. The particles
which compress or stretch a spring are always acted upon
by this spring force which restores them to their
equilibrium positions. It is calculated as following according
to Hooke's law, which describes the opposing force exerted by a spring.

\begin{equation}
{\bf F}_{12}^{h} = -\left(||{\bf r}_{2}-{\bf r}_{1}||-r_{l}\right) \, {k}_{s}
\end{equation}

\noindent
where

\noindent
${\bf r}_{1}$ is the first particle position,

\noindent
${\bf r}_{2}$ is the second particle position,

\noindent
${r}_{l}$ is default length of the  resting spring between the two particles,

\noindent
${k}_{s}$ is the stiffness of the spring,

\noindent
when $||{\bf r}_{2}-{\bf r}_{1}||-r_{l}=0$, the spring is resting,

\noindent
when $||{\bf r}_{2}- {\bf r}_{1}||-r_{l}>0$, the spring is extending,

\noindent
when $||{\bf r}_{2}-{\bf r}_{1}||-r_{l}<0$, the spring is contracting.

We have discussed the type of structural spring in one-dimensional object. In two and three dimensional object model, the same method applies on the other three types of spring, such as radius springs, shear left springs, and shear right springs with different spring stiffness and spring damping factor. So, the total Hooke's spring force is:

\begin{eqnarray}
{\bf F}_{total} ^{h} = { \bf F}_{structure}^{h} + {\bf F}_{radius}^{h} +	{\bf F}_{shear left}^{h} + {\bf F}_{shear right}^{h}
\end{eqnarray}

%This force should be contributed to the two particles connected with this spring by adding the force to one particle and subtracting the force from the other particle.

%\begin{eqnarray*}
%\vec{\bf F}^{total spring} = { \bf F}_{structure}^{h} + {\bf F}_{structure}^{d} + {\bf F}_{radium}^{h} + {\bf F}_{radium}^{d}\\
% 											 + {\bf F}_{shear left}^{h} + {\bf F}_{shear left}^{d} + {\bf F}_{shear right}^{h} + {\bf F}_{shear right}^{d}
%\end{eqnarray*}

\section{Spring Damping Force}
Spring damping force is also called viscous damping. It is opposite force of the Hook spring force in order to simulate the natural damping and resist the motion. It is also opposite to the velocity of the moving mass particle and is proportional to the velocity because the spring is not completely elastic and it absorbs some of the energy and tends to decrease the velocity of the mass particle attached to it. It is needed to simulate the natural damping due to the forces of friction. More importantly, it is useful to enhance numerical stability and is required for the model to be physically correct \cite{brar97}.

%a force with magnitude proportional to that of the velocity of the object but opposite in direction to it.
\begin{equation}
{\bf F}_{12}^{d} =
\left({\bf v}_{2}-{\bf v}_{1}\right)\cdot
\left(\frac{{\bf r}_{2}-{\bf r}_{1}}{||{\bf r}_{2}-{\bf r}_1||}\right) \, {k}_{d}
\end{equation}
\noindent
where

\noindent
$\left(\frac{{\bf r}_{2}-{\bf r}_{1}}{||{\bf r}_{2}-{\bf r}_1||}\right)$ is the direction of the spring,

\noindent
${\bf v}_{1}$ and ${\bf v}_{2}$ is the velocity of the two masses,

\noindent
${k}_{d}$ is spring damping coefficient.

\noindent
When the two endpoints moving away from each other, the force imparted from the damper will act against that motion; when the two endpoints moving toward each other, the damper will act against the squeeze motion. The damper will always acts against the motion. The total spring damping force is:

\begin{eqnarray}
{\bf F}_{total} ^{d} = { \bf F}_{structure}^{d} + {\bf F}_{radius}^{d} +	{\bf F}_{shear left}^{d} + {\bf F}_{shear right}^{d}
\end{eqnarray}

%This damping force should be contributed to the two particles connected with this spring by adding the force to one particle and subtracting the force from the other particle.
\section{Drag Force}
Drag force is the force when users interact with the elastic object through mouse. At the moment users click the mouse, the simulation system finds the nearest particle $i$ to the current position of the mouse.
If users drag this particle $i$, the drag force contributes to force of this nearest particle. The forces applied on rest particles are effected by the new user applied force, which is passed through by springs.

We consider one end of the string connects to the mouse position and the other end of the string connects to the nearest particle on the object. This string is elastic, so it has all the spring's properties, such as hook spring force and damping force. The drag force can be presented as following:

\begin{equation}
{\bf F}^{a}=-\left(||{\bf r}_{m}-{\bf r}_{i}||-r_{lm}\right)\, {k}_{sm}+ \left({\bf v}_{m}-{\bf v}_{i}\right)\cdot\left(\frac{{\bf r}_{m}-{\bf r}_{i}}{||{\bf r}_{m}-{\bf r}_{i}||}\right)\, {k}_{dm}
\end{equation}

\noindent
where

\noindent
${\bf r}_{m}$ is the mouse position,

\noindent
${\bf r}_{i}$ is the particle position nearest to mouse,

\noindent
${\bf r}_{lm}$ is default length of the  resting mouse spring,

\noindent
${k}_{sm}$ is the stiffness of the mouse spring,

\noindent
${\bf v}_{m}$ is the velocity of the mouse represented as a mass

\noindent
${\bf v}_{i}$ is the mass for the nearest particle,

\noindent
${k}_{dm}$ is spring damping coefficient for the mouse spring.

\noindent
${\bf F}^{a}$ is a momentary force for interacting with the elastic simulation system. This force is accumulated to the current forces already applied on this nearest particle.

In a one-dimensional object simulation system, the nearest particle to mouse is either $P_0$ or $P_1$. In a two-dimensional and three-dimensional object system, the drag force is only applied on the outer layer of the double layered object when user interacts the object with mouse. %The nearest particle from the outer layer to the mouse position is found when mouse is pressed, and the drag force for rest of particles on inner layer and outer layer are passed through by the four different types of springs.

\section{Air Pressure Force}
In order to describe an elastic object more accurately, especially soft body of human beings and animals, the calculation only about the elastic force on the object's surface is not enough. We add the flow pressure force inside of the elastic object to make the object wobbly looking when it is deformed. %The internal force will be applied on one direction and relieved in another direction.

The pressure force will be calculated for every spring, then update each particle's direction. The pressure vector is always acting in a direction of normal vectors to the surface, so the shape will not deform completely. If pressure is simulated without also simulating the mass-spring system, the object will explode.

\begin{figure*}[h]
\hrule\vskip4pt
\begin{center}
	\subfigure[The External Gravity Force Is Applied Producing A Pressure Wave]
	{\label{fig:TwoD3}
 		\includegraphics[width=2in]{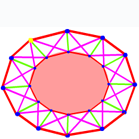}}
 		\hspace{.6in}
	\subfigure[The Object Restores Its Shape With Internal Air Pressure Force]
	{\label{fig:TwoD2}
 		\includegraphics[width=2in]{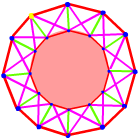}}
\caption{Double-layered Two-dimensional Elastic Object Filled With Air}
\end{center}
\hrule\vskip4pt
\end{figure*}

In \xf{fig:TwoD3}, the object is deformed from bottom because of the gravity force. If there is no internal air pressure, the object will collapse unless the springs are hard enough to avoid the failure. With the very hard springs, it is difficult to simulate the reality of the elasticity. The real elastic object restores its shape as described in \xf{fig:TwoD2}. The simplified version of the Ideal Gas Law \cite{mm041}, also known as Clausius Clapeyron Equation, is used to describe such effect:

\begin{equation}
P V = N R T
%P V = NRT \iff P = \frac{NRT}{V}
\end{equation}

\noindent
where

\noindent
$P$ is the pressure value,

\noindent
$V$ is the volume of the object,

\noindent
$N$ is number of mols,

\noindent
$R$ is the gas constant,

\noindent
$T$ is the gas temperature. Therefore, the pressure force is:

\begin{equation}
{\bf F}^p = P \, {\bf n}
\end{equation}

\noindent
where

\noindent
${\bf F}^p$ is the pressure force vector,

\noindent
${\bf n}$ is the normal vector to the springs on the object.

\begin{equation}
P = \frac{NRT}{V}
\end{equation}

\subsection{Volume}
In order to find an estimate pressure inside of the object which will be applied to particles later, we need to calculate the volume of the object. The approximation of the volume is calculated with Gauss' Theorem \cite{ek07}: % for closed shape object. %considered to be enough because we only care about the volume in general change.

\begin{equation}
V = \int\int\int_v f(x,y,z) \, dx \, dy \, dz \Longleftrightarrow      V = \int\int\int_v f(x,y,z)\, dV
\label{volume}
\end{equation}

\noindent
where triple integrals of a function $f(x,y,z)$ define a volume integral of an elastic sphere. Moreover, triple integrals can be transformed into surface double integrals over the boundary surface of a region if the three-dimensional object is closed shape by divergence theorem \cite{ek07}:

\begin{equation}
V = \int\int\int_V \Delta {\bf F}\, dV \Longleftrightarrow     S = \int\int_S {\bf F} \, dS
\end{equation}

\noindent
where

\noindent
${\bf F}$ is a vector field,

\noindent
$V$ is the object volume,

\noindent
$S$ is the object surface.

\noindent
Double integrals over a plane region may be transformed into line integrals by Green's Theorem in the Plane:

\begin{equation}
\int\int_S \Delta {\bf F}\, dx\, dy \Longleftrightarrow     \int_L {\bf F} \, dL
\end{equation}

\noindent
where $L$ is the object edge and $dL$ is the length of the edge. 

Therefore, a triple integrals function $f(x,y,z)$ shown in Eq.\ref{volume}, which defines a volume integral of an elastic sphere, can be transformed to line integrals as shown in Eq.\ref{volumetoline}. 

\begin{equation}
V \approx \int_L {\bf F} \, dL
\label{volumetoline}
\end{equation}

We assume on the line, the vector field ${\bf F} = (x, 0)$, the simplified integration of body volume is \cite{mm041, ek07}:

\begin{equation}
\int_L {\bf F} \, dL=  \displaystyle\sum_{i=0}^{i=NUMS-1}\frac{1}{2}({\bf x_{1}}-{\bf x_{2}})\, {\bf n_x}\, dL
\end{equation}

\noindent
where

\noindent
$V$ is the volume of the object,

\noindent
$({\bf x_{1}}-{\bf x_{2}})$ is the absolute difference of the line (represents spring here) of the start and end particles at axis x,

\noindent
${\bf n_x}$ is the normal vector to this line (spring) at axis x,

\noindent
$dL$ is the line's (spring's) length.

%The internal air pressure force prevents the object from collapsing when it is compressed by external forces. Moreover, it simulates the object in reality by demonstrating the derivate calculation of the force and volume.
\subsection{Normals}

%The normal force is the support force exerted upon an object which is in contact with another stable object. For example, if a book is resting upon a surface, then the surface is exerting an upward force upon the book in order to support the weight of the book. On occasions, a normal force is exerted horizontally between two objects which are in contact with each other.

Normals are unit vectors perpendicular to specified data structure, such as particle (vertex psuedo-normals), spring (line), and face (polygonal facets) on the object. %They are used in rendering scenes techniques to calculate the light sources are reflected to the camera view. Normal calculation is also useful in mapping texture techniques.
\begin{itemize}
\item Particle normal, or vertex psuedo-normals, does not exist for vertices; however, it can be considered as the average of the normals of the subtended neighbor particles. To calculate the particle normal is to sum up the normals for each face adjoining this particle, and then to normalize the sum.

%// Calculating vertex normals using a vertex list and polygon faces (vertex indices);

% loop through your vertex list {

%       clear this_vertex_normal;    // initialized to [0,0,0]

 %      loop through your face lists {

 %                for each face vertex, check if its
 %                the same as the vertex as in our
 %                current outer vertex loop.                 If it is, {                            this_vertex_normal += this_face.normal;
 %                }
 %      }

 %      this_vertex_normal.Normalize();
 %}

\item Spring (line) normal in two dimension is based on the two particles  $P_1, P_2$ connected on the spring. It is perpendicular to the spring itself.

\item Face (plane) normal in three dimension is determined by right-hand rule, which is perpendicular to its surface based on the any pair of springs on the surface. The normal for a triangle surface composed with three particles $P_1, P_2, P_3$ is computed as the vector cross product of the springs $P_2-P_1$ and $P_2-P_3$.
\end{itemize}

The usage of normal calculation method in our elastic object simulation system is for analysis of the direction of the pressure force inside of the object. Only spring normal is calculated here because all the internal and external forces will apply on each spring, and the spring will define the particles' force which connect onto it.
\subsubsection{2D Normals}

For the single spring $Spring_{12}$, the Cartesian coordinates for particle $P_1$ is $(x_1, y_1)$; the Cartesian coordinates for particle $P_2$ is $(x_2, y_2)$. The normal to this spring is the spring rotate $90^0$ at axis z according to the space position. So, we can get the components of the normal in x axis and y axis as following

\[\left[\begin{array}{c}
{x'} \\
{y'} \end{array}\right]= \left[\begin{array}{cc}
\cos{90^{0}} & -\sin{90^{0}} \\
\sin{90^{0}} & \cos{90^{0}} \end{array} \right]
\left[\begin{array}{c}
{x_2-x_1} \\
{y_2-y_1} \end{array} \right]\]

\[\left[\begin{array}{c}
{x'} \\
{y'} \end{array}\right]= \left[\begin{array}{c}
-\left({y_2-y_1}\right)\\
{x_2-x_1} \end{array} \right]\]

\subsubsection{3D Normals}

The calculation of the 3D normals of springs is important because it will define the direction of the internal air pressure force, either compress the elastic object or expand its volume. In theory, in three-dimensional simulation, the normal of a spring in the space position is represented as an average of the normals of faces connected to it. However, in our elastic object simulation system, we use a simplified estimated normal based on the normal algorithm of the two-dimensional calculation discussed above. Instead of rotating a line $90^0$ at z axis to get its normal vector in two-dimension, the estimated algorithm is rotating a line $90^0$ at z axis, y axis, and x axis to get its normal in three-dimension. 

\[\left[\begin{array}{c}
{x'} \\
{y'} \\
{z'} \\
{1} \end{array}\right]=
\left[\begin{array}{cccc}
\cos{90^{0}} & -\sin{90^{0}} & 0 & 0 \\
\sin{90^{0}} & \cos{90^{0}} & 0 & 0\\
0 & 0 & 1 & 0 \\
0 & 0 & 0 & 1 \end{array} \right]
\left[\begin{array}{cccc}
\cos{90^{0}} & 0 & -\sin{90^{0}} & 0 \\
0 & 1 & 0 & 0 \\
\sin{90^{0}} & 0 & \cos{90^{0}} & 0\\
0 & 0 & 0 & 1 \end{array} \right]\]
\[\left[\begin{array}{cccc}
1 & 0 & 0 & 0 \\
0 & \cos{90^{0}} & \sin{90^{0}} & 0 \\
0 & -\sin{90^{0}} & \cos{90^{0}} & 0 \\
0 & 0 & 0 & 1 \end{array} \right]
\left[\begin{array}{c}
{x_2-x_1} \\
{y_2-y_1} \\
{z_2-z_1} \\
{1} \end{array} \right]\]

\noindent
Therefore,

\[\left[\begin{array}{c}
{x'} \\
{y'} \\
{z'} \\
{1} \end{array}\right]= \left[\begin{array}{c}
{z_2-z_1} \\
{y_2-y_1} \\
-\left({x_2-x_1}\right)\\
{1} \end{array} \right]\]

\noindent
We use a vector $(1,0,0)$ as an example to prove this algorithm, the normal vector for this vector is:

\[\left[\begin{array}{c}
{x'} \\
{y'} \\
{z'} \\
{1} \end{array}\right]= \left[\begin{array}{c}
{0-0} \\
{0-0} \\
-\left({1-0}\right)\\
{1} \end{array} \right]=\left[\begin{array}{c}
{0}\\
{0}\\
{-1}\\
{1} \end{array} \right]\]

\noindent
This result is reasonably correct and believable despite of the fact that if vector $(1,0,0)$ lies in the $xz$-plane or lies in the $xy$-plane.

\noindent
However, it shows that this estimation algorithm has the limitation for some cases, for example vector $(0,1,0)$, the normal vector is:

\[\left[\begin{array}{c}
{x'} \\
{y'} \\
{z'} \\
{1} \end{array}\right]= \left[\begin{array}{c}
{0-0} \\
{1-0} \\
-\left({0-0}\right)\\
{1} \end{array} \right]=\left[\begin{array}{c}
{0}\\
{1}\\
{0}\\
{1} \end{array} \right]\]

\noindent
This result shows the normal vector is the vector itself, which is obviously wrong. However, with this estimated algorithm, the simulation result appears enough realistic;
%\footnote{In our sphere only up to two springs can be parallel and aligned with $y$, making the error effect negligible and corrected by the other connected springs.}
 moreover, it requires less
computational effort\footnote{Our estimation only takes 3 additions vs. 12 multiplications and 6 additions for two cross products and three more additions
and divisions for the averaging.}.

\section{Collision Force}

If an object continues traveling under a force without colliding with other objects, it will be very difficult to describe objects' motion and elastic response in reality. Collision force is the force to make object bounce away from the fixed interacting plane when elastic object collision happens.There are two steps to describe the collision effects: detection and reaction. Detect the elastic object if particles hit anything; adjust their position by computing the impulse.

\subsection{Collision Detection}

Collision Detection is a geometric problem of determining if a moving object intersected with other objects at some point between an initial and final configuration. In our elastic object simulation system, we are concerned with the problem of determining if any of $n$ particles collide with any of $m$ solid planes.

\paragraph*{Perfect Elastic Collision}

\begin{figure}[h]
\hrule\vskip4pt
\begin{center}
{\includegraphics[width=2in]{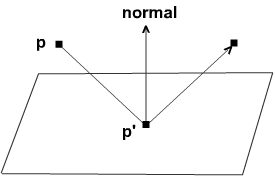}}
\caption{Particle Inelastic Collision and Impact}
\label{fig:collision}
\end{center}
\hrule\vskip4pt
\end{figure}

We take one particle collides with a plane shown in \xf{fig:collision} as an example. We can detect this collision by inserting the particle position into the plane equation:

\begin{equation}
P(x,y,z) = ax + by + cz + d
\end{equation}

\noindent
If $P(x,y,z)>0$, the particle is within the plane. If $P(x,y,z)=0$, the particle collides with the plane. If $P(x,y,z)<0$, the particle penetrates the plane. At each time step, looping through all the particles on the object, each particle is checked if it is outside of the interacting plane.

When the particle $i$ collides with the plane, if there is a perfect elastic collision as in \xf{fig:collision}, the particle does not lose its energy, so its speed does not change. However, its direction after the collision is in the direction of a perfect reflection.% \cite{ea03}.

\begin{equation}
{\bf F}^c = 2((P-P')\cdot \, {\bf n} ) \, {\bf n}-(P-P')
\end{equation}

\noindent
where

\noindent
${\bf F}^c$ is the the direction of a perfect reflection

\noindent
${\bf n}$ is the normal at the point of collision $P'$ and the previous position of particle $P$

\noindent
$P-P'$ is the vector from the particle to the surface.

\paragraph*{Damped Elastic Collision}
If there is a damped elastic collision, the particle cannot penetrate the surface, and it cannot bounce from the surface because of the force being applied to it, then we need to apply the damped elastic collision method. The particle loses some of its energy when it collides with another object. The coefficient of restitution of a particle is the friction of the normal velocity retained after the collision. Therefore, the angle of reflection is computed as for the inelastic collision, and the normal component of the velocity is reduced by the coefficient of restitution.

% an artificial collision spring, which has elasticity and damping properties as an ordinary spring, is generated between this particle and the plane. This artificial collision spring will push the colliding particle backward according to the normal force, then all the other particles will be applied with this new distortion force. Eventually, the elastic object will stop its motion because of friction loss is calculated. This is the simplest case of collision detection.

%A particle can be treated as a line segment from its previous position to its current position:
%If we are colliding against static objects, then we just need to test if the line segment intersects the object. Colliding against moving objects requires some additional modifications that we will also look at.

%$F_friction = \mu F_{Normal}$ where $\mu$ is the friction coefficient

%General goals: given two objects with current and previous orientations specified, determine if, where, and when the two objects intersect - Alternative: given two objects with only current orientations, determine if they intersect - Sometimes, we need to find all intersections. Other times, we just want the first one. Sometimes, we just need to know if the two objects intersect and dont need the actual intersection data.

\subsection{Collision Response}
Collision Response is a physics problem of determining the forces of the collision.
In elastic collision, elastic object should bounce away from the colliding plane and some energy is lost in the collision response as described in the penalty method.
%When collision happens, must perform some action to prevent the object penetrating even deeper. Object should bounce away from the colliding object. Some energy is usually lost during collision. Several ways to handle collision response, we will use the penalty method.

%When collision happens, put an artificial collision spring at the point of collision, which will push the object backwards and away from the colliding object. Collision springs have elasticity and damping, just like ordinary springs.
\begin{equation}
{\bf F}^c = -e \, {\bf F}^c
\end{equation}

\noindent
where $e$ is elasticity of the collision and $0.0 \leq e \leq 1.0$. At $e=0$, the particle does not bounce at all; $e=1$, the particle bounces with no friction.

%\noindent
%$f$ is sliding friction
%N: plane normal
%V0: initial projectile velocity
%V1: final projectile velocity
%subscript (n): component parallel to n
%subscript (p): component perpendicular to n
%V0n = V0.projectOn(N)
%V0p = V0 - V0n
%V1 = -V0n*e + V0p*f
%if no sliding friction:
%V1 = -V0n*e + V0p*e = (-V0n + V0p)*e

In an one-dimensional object, the boundaries are the walls and floors. In a two-dimensional and three-dimensional object, the particles on the outer layer still follow the same method and same pre-defined boundary as the one-dimensional object. However, for the particles on the inner layer, the boundary is constrained to the outer layer instead of the wall and floor.

%\section{Friction force}
%The friction force is the force exerted by a surface as an object moves across it or makes an effort to move across it. The friction force opposes the motion of the object. For example, if a book moves across the surface of a desk, then the desk exerts a friction force in the opposite direction of its motion. Friction results from the two surfaces being pressed together closely, causing intermolecular attractive forces between molecules of different surfaces. As such, friction depends upon the nature of the two surfaces and upon the degree to which they are pressed together. The friction force can be calculated using the equation:
%$F_friction = \mu F_{Normal}$ where $\mu$ is the friction coefficient

\section{Force Accumulation Algorithm}

The following algorithm describes how different forces are accumulated and applied to an elastic object. For a one-dimensional object, some steps will be skipped, for example, there are no other types of spring computations except structural springs because other types of springs only apply on two-layer 2D or 3D objects. Moreover, there are no pressure force accumulation and volume computation because these steps are only available for closed shape objects.

\begin{itemize}
\item Step 1: Loop through the number of particles to assign particles with mass value $m$ and compute gravity force ${\bf F}^g$. Gravity force, which is independently on each particle, either depends on a constant force, or one or more of particle position, particle velocity, and time \cite{aw99}. If the object is one-dimensional, the mass of each particle can be different. If the object is two or three dimensional, the mass of the particles on inner or outer layer can also be set differently.
%\item Step 9: Split object into regions, integrate density and weight in each region to modify mass. The current system considers the particle weights are all same. In the future model, the density and weight of the particles can be vary in order to get the non-unit motion.
\item Step 2: Loop through the number of the structural springs to accumulate the structural spring force. %Spring force is n-ary force,
\item Step 3: Loop through the number of the radius springs to accumulate the radius spring force.
\item Step 4: Loop through the number of the shear springs to accumulate the shear spring force.
%\item Step 2 Changing the spring constants for the different types of springs of the system. Also, different density effects can be realized. The consequence is that the force of each particle varies from the different compression of springs.
\item Step 5: Initialize density as gas, liquid, or rubber inside of the body and introduce some simple physics to describe it. In the current system, only air pressure material is considered and only pressure equation will be used for this extra force computation.
%\item Step 6 Calculate all forces accumulated on the each inner and outer layer particle except air pressure force.
\item Step 6: Calculate volume of the inner layer and outer layer of the elastic object.
%\item Step 14: Calculate volume of the outer layer of the elastic object.
\item Step 7: Calculate the normals of springs on each triangular face to define the pressure force direction.
\item Step 8: Calculate the force from the internal air pressure by multiplying the force value by normal vector of the spring.
\item Step 9: Accumulate pressure force to each particle.
\item Step 10: If users apply the drag force, compute the user applied force and accumulate this force to the dragged particle.
\item Step 11: Integrate the object's momentum motion by calculating the derived velocity and its new position for each particle. This step will be explained in next chapter.
\item Step 12: Resolve collision detection and response and define the updated position.
\end{itemize}
\chapter{Numerical Integration Methodology}
\label{chapt: numerical integration methodology}
\index{Numerical Integration Methodology}

Assume, after the elastic object simulation system creates an elastic object based on the methodology described in Chapter 3 with its initial force state in \xf{fig:dt0} as described in Chapter 4, the system starts the simulation. The simulation system is updated a finite number of times. The object is at the state in \xf {fig:dt100} after 50 discrete time steps.

\begin{figure*}[h]
\hrule\vskip4pt
\begin{center}
\subfigure[Elastic Object at the Initial Step]
	{\label{fig:dt0}
	 \includegraphics[width=2in]{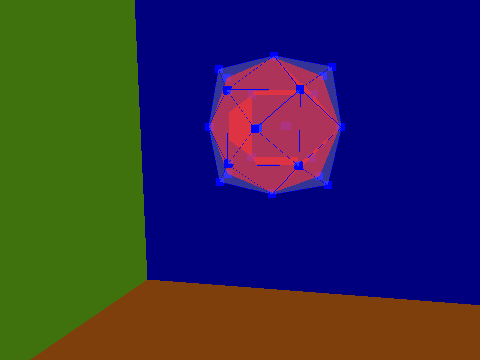}}
	 \hspace{.6in}
	 	\subfigure[Elastic Object at the Step 50]
	{\label{fig:dt100}
	 \includegraphics[width=2in]{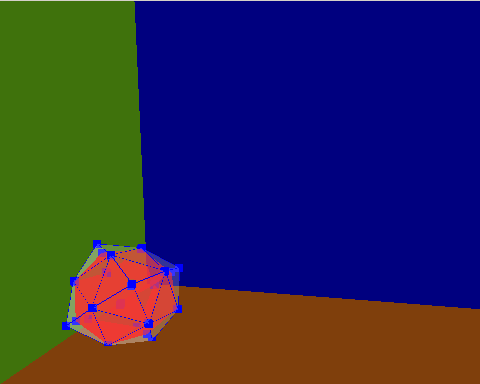}}
\caption{Elastic Object at Different Time States}
\end{center}
\hrule\vskip4pt
\end{figure*}

In each update, the accumulated impact forces on the object tell it how to change the velocity for next step and result in a re-computation of the forces. The dynamic force applied on this object may be the collision force when the object reaches the boundary; or, the mouse dragging force when user interacts with the object. Overall, the shape deformation, a mapping of the positions of every particle in the original object to those in
the deformed body of this elastic object, is also computed in real time. Therefore, it is important to study differential equations, which govern dynamics and geometric representation of objects \cite{ml06} and tell us how the velocity and displacement of the particles are integrated dynamically from the knowledge of force applied onto them.

\section{Differential Equations}
Differential equations describe a relation between a function and one or more of its derivatives. The order of the equation is the order of the highest derivative it contains. The elastic object simulation system is associated with initial value problems because it always seeks the particles' velocity and position at next time step $t+h$ from their initial state at time $t$. We will concentrate on ODE (ordinary differential equation), where all derivatives are with respect to single independent variable, often representing time, such as position and velocity, during the derivate of the state at discrete time steps \cite{ea03}.

\begin{equation}
%\dot{y} = {\bf A}(y,t) %= \frac{\partial x}{\partial t}$$
{y}' = {\bf A}(y,t)
\label{eq:dfe}
\end{equation}

\noindent
where

\noindent
${\bf A}$ is a function of $y$ and $t$,

\noindent
$y$ is a vector, which is the state of the system,

\noindent
${y}'$ is a vector, which is $y$'s time derivative.

\noindent
Suppose that we integrate the Eq.\ref{eq:dfe} over a short time $h$

\begin{equation}
\label{eq:dfeh}
%\int_t^{t+h}\dot{y} \, dt= y(t+h) - y(t)=\int_t^{t+h}{\bf A}(y,t) \,dt
y(t+h) - y(t)=\int_t^{t+h}{\bf A}(y,t) \,dt
\end{equation}

\noindent
where

\noindent
$h$ is the small stepsize of time,

\noindent
$y(t)$ is the initial state at the start point $t$,

\noindent
$y(t+h)$ is the value we need to find over time thereafter.

\noindent
Thus

\begin{equation}
\label{eq:dfeh1}
y(t+h)\approx  y(t) + h{\bf A}(y(t),t)
\end{equation}

\subsection{Explicit Euler Integrator}

The simplest ODE integration method is Explicit Euler Integration method or Forward Euler method. It evaluates the forces at time $t$, compute derivatives ${\bf A}$ at the state of $t$ by multiplying the interval $h$, and add it to the current state $t$. %Then re-determine the slope of the curve and taking the next step along the new tangent.
Consider a Taylor series expansion as in Eq.\ref{eqn:Taylor}:

\begin{equation}
\label{eqn:Taylor}
%y({t_0}+{\partial t}) = y({t_0}) + {\partial t}\dot{y}({t_0})+ \frac{{{\partial t}^2}{\ddot{y}}({t_0})}{2!}+\ddots+\frac{{{\partial t}^n}\left(\frac{\partial^{n}y}{\partial{t}^n}\right)}{n!}
%\frac{{{h}^4}}{4!}\dot{\dot{\dot{\dot{{y}}}}}({t})+
%%y({t}+{h}) = y({t}) + {h}\dot{y}({t})+ \frac{{{h}^2}}{2!}{\dot{\dot y}}({t})+
%%\frac{{{h}^{3}}}{3!}{\dot{\dot{\dot y}}({t})}+
%%\cdots+\frac{{{h}^n}}{n!}\left(\frac{\partial^{n}y}{\partial{t}^n}\right)+\cdots
y({t}+{h}) = y({t}) + {h}{y}'({t})+ \frac{{{h}^2}}{2!}{y}''({t})+
\frac{{{h}^{3}}}{3!}{y}'''({t})+
\cdots+\frac{{{h}^n}}{n!}\left(\frac{\partial^{n}y}{\partial{t}^n}\right)+\cdots
\end{equation}
%\begin{equation}
%\label{eq:Taylor}
%x({t}+{\partial t}) = x({t}) + {\partial t}\dot{x}({t})+ \frac{{{\partial t}^2}{\ddot{x}}({t})}{2!}+\ddots+\frac{{{\partial t}^n}\left(\frac{\partial^{n}x}{\partial{t}^n}\right)}{n!}
%\end{equation

\noindent
Euler method retains only first derivative:

\begin{equation}
%y({t}+{h}) = y({t}) + {h}\dot{y}({t}) + O(h^2)
y({t}+{h}) = y({t}) + {h}{y}'({t}) + O(h^2)
\label{eq:euler}
\end{equation}

\begin{figure}[h]
\hrule\vskip4pt
\begin{center}
	{	 \includegraphics[width=3.5in]{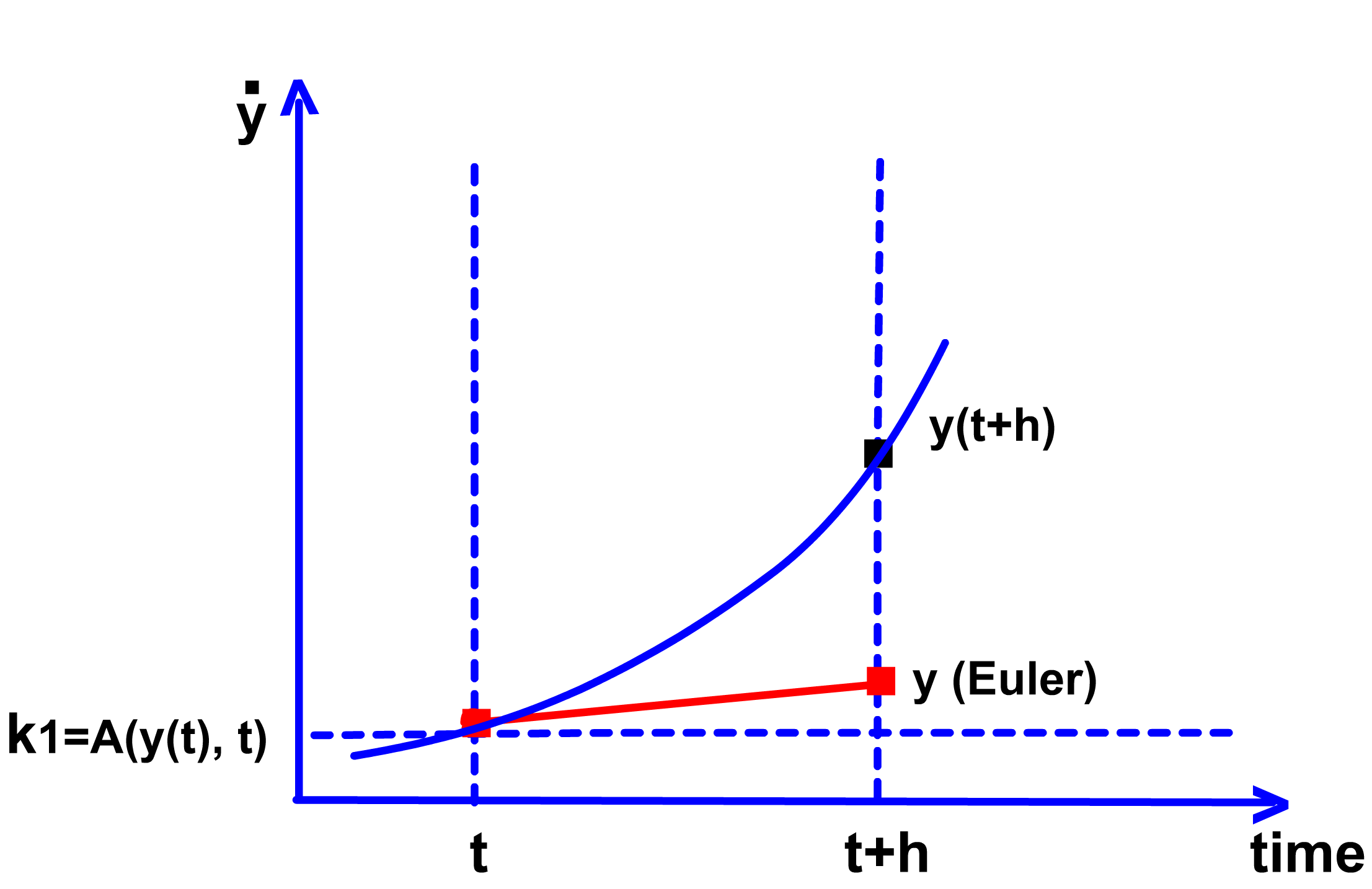}}
\caption{Euler Integrator}
\label{fig:euler1}
\end{center}
\hrule\vskip4pt
\end{figure}

\noindent
We split the series into elements, which we will later use
in a re-usable manner throughout integrator framework, where

\noindent
$k_0$ which represents the first term in Eq.\ref{eq:euler}, is the initial state

\begin{equation}
k_0 = y({t})
\end{equation}

\noindent
$k_1$ which represents the second term in Eq.\ref{eq:euler}, is the function to find the simplest estimation, the Euler slope of the interval.

\begin{equation}
%k_1 = \dot{y}({t}) = {\bf A}(y(t),t)
k_1 = {y}'({t}) = {\bf A}(y(t),t)
\end{equation}

\noindent
Thus

\begin{equation}
y({t}+{h}) = k_0 + {{h}}k_1
\end{equation}

We can apply this method iteratively to compute further values at state $t+ 2h$, $t+3h$,.... \cite{wr03} This method is easy to implement; however, it is a low accuracy prototype ODE. In \xf{fig:euler1}, we can see Euler method only calculates the derivative, also called slope, at the beginning of the interval and adds it to the value at the initial state; therefore, it is asymmetric and not stable. % according to the beginning and the end of the interval.
.
%\section{Implicit methods}

%formula $y_{n+1} = y_n {h}f(y_n,t_n)$

%Implicit method is also called backward Euler method, which is based on the explicit methods, sometimes, the ODE can become non accurate and do not provide a very good result. So, implicit method can be applied to solve the stability problem. This method can use the large step size. However, it is more expensive in time and space computation and difficult

%For spring mass:
%$v(t+\partial t) = (v(t)-\partial t*SpringConstant/Mass*x(t) + \partial t*gravity) / (1 + \partial t*dt*SpringConstant/SpringMassMass) $

%$x(t+\partial t) = x + \partial t*v(t+\partial t) $

%If I set the spring coefficient value very high because I want very stiff springs, the little Euler integrator cannot handle it.
\paragraph*{Pseudocode for Euler Method}
\begin{center}
\begin{prog}
\verb"Line 1: define A(y(t), t)"
\verb"Line 2: initial values y0 and t0"
\verb"Line 3: stepsize h and number of steps n"
\verb"Line 4: for i from 1 to n do"
\verb"Line 5: k1 = A(y(t), t)"
\verb"Line 6: y = y + hk1"
\verb"Line 7: t = t + h"
\end{prog}
\end{center}
\subsection{Midpoint Integrator}
Compared to the Euler method, the one-sided estimate algorithm, midpoint integrator is a symmetric estimate method with a higher per-step accuracy. It computes the derivative at the center of the interval first, then computes the end of the interval.

%Midpoint Integration is more symmetric integration by taking midpoint of the interval of Euler Method to make the real steps of the interval.

\noindent
The midpoint integrator, just like others, is based on the
Taylor's series. It retains only first three derivative term:

%\begin{eqnarray*}
\begin{equation}
\label{eq:midpoint}
%y({t}+{h}) = y({t}) + {h}\dot{y}({t})+ \frac{{{h}^2}}{2!}\dot{\dot{{y}}}({t})+ O({h}^3)
y({t}+{h}) = y({t}) + {h}{y}'({t})+ \frac{{{h}^2}}{2!}{y}'({t})+ O({h}^3)
\end{equation}

\begin{figure}[h]
\hrule\vskip4pt
\begin{center}
	{	 \includegraphics[width=4in]{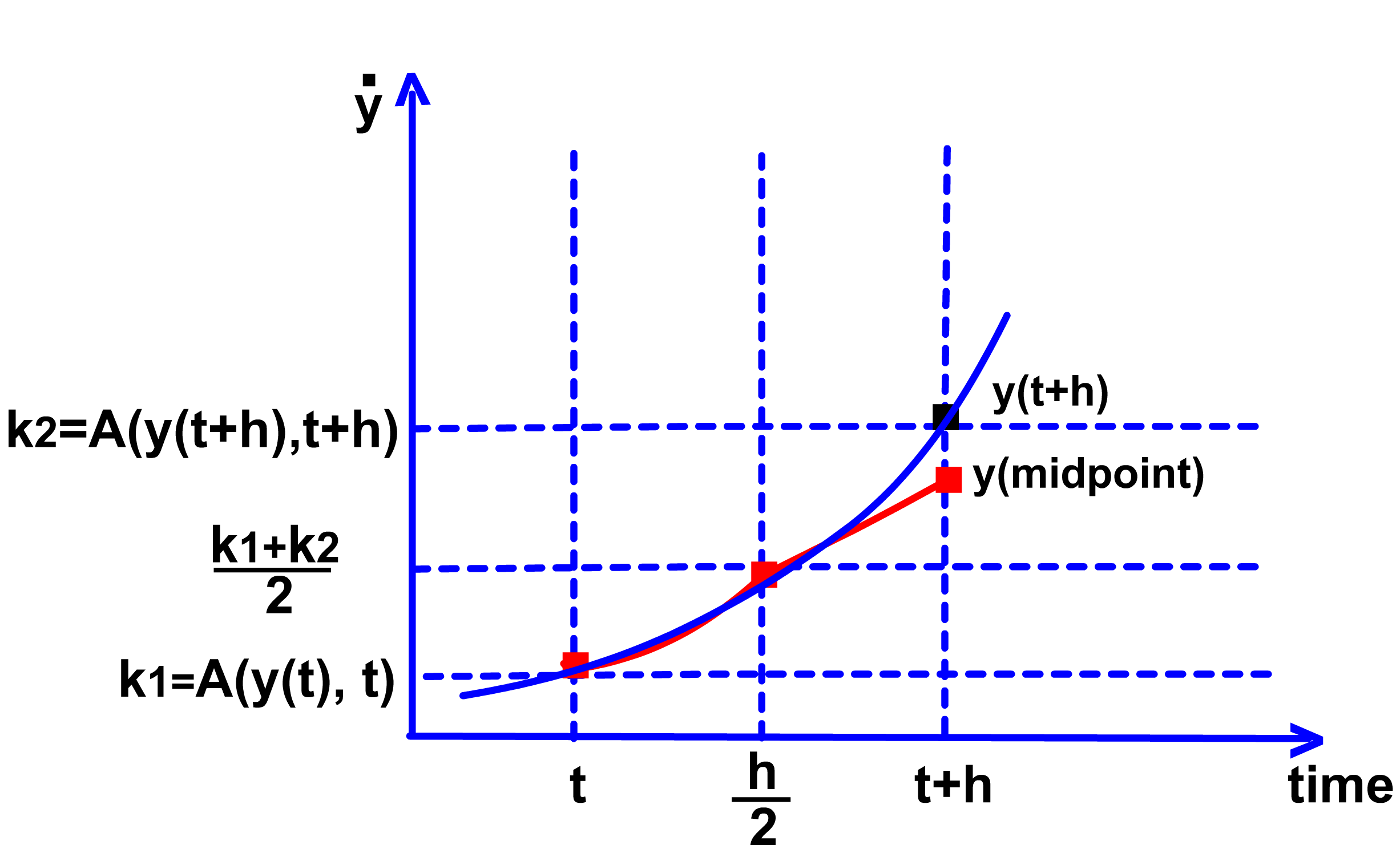}}
\caption{Midpoint Integrator}
\label{fig:midpoint}
\end{center}
\hrule\vskip4pt
\end{figure}

\noindent
We split the series into elements again for explanation of the method, where

\noindent
$k_0$, which represents the first term in Eq.\ref{eq:midpoint}, is the initial state at time $t$.

\begin{equation}
k_0 = y({t})
\end{equation}

\noindent
$k_1$ which represents the second term in Eq.\ref{eq:midpoint}, is the function to find the the simplest Euler slope of the interval at time $t$.

\begin{equation}
%k_1 = \dot{y}({t}) = {\bf A}(y(t),t)
k_1 = {y}'({t}) = {\bf A}(y(t),t)
\end{equation}

\noindent
$k_2$ is the function to find the the simplest Euler slope of the interval at time $t+h$.

\begin{equation}
\label{eq:k2}
%k_2 = \dot{y}({t+h}) = {\bf A}(y(t+h),t+h)
k_2 = {y}'({t+h}) = {\bf A}(y(t+h),t+h)
\end{equation}

\noindent
Since the unknown $(y+h)$ appears on the right side of Eq.\ref{eq:k2}, in
${\bf A}(y(t+h),t+h)$ as one of the arguments of function ${\bf A}$, we can use
the value obtained using the Euler method in Eq.\ref{eq:euler}.

\begin{equation}
\label{eq:k2}
{\bf A}(y(t+h),t+h)\approx {\bf A}(y(t) + h{\bf}(y(t),t), t+h)={\bf A}(y(t) + hk1, t+h)
\end{equation}

\noindent
The midpoint integration technique obtains a more accurate estimate of the slope than Euler's technique. The following equation computes the integrand at the middle of the interval of $t$ and $t+h$ shown in \xf{fig:midpoint}. Thus,

\begin{equation}
\label{eq:k1k2}
y({t}+{h}) = k_0 + {{h}}\frac{k_1+k_2}{2}
%= k_0 + \frac{{h}}{2}({{\bf A}(y(t),t)+{\bf A}(y(t+h),t+h)})
%k_2 = {h}\ddot{y}({t}) = {h}{\bf A}(y+\frac{k_1}{2},\frac{t}{2})
\end{equation}

Compared to Euler Method, Midpoint Method, also called the Runge-Kutta method of order 2, goes from $t$ to $t+h$, we must evaluate function ${\bf A}$ twice. By using Taylor's theorem to evaluate the per-step error, we would find that it is now $O(h^3)$. Therefore, this method is more stable than Euler Method with same step size.
\paragraph*{Pseudocode for Midpoint Method}
\begin{center}
\begin{prog}
\verb"Line 1: define A(y(t), t)"
\verb"Line 2: initial values y0 and t0"
\verb"Line 3: stepsize h and number of steps n"
\verb"Line 4: for i from 1 to n do"
\verb"Line 5: k1 = A(y(t), t)"
\verb"Line 6: k2 = A(y(t+h), t+h)= A(y+hk1, t+h)"
\verb"Line 7: y = y + h/2(k1+k2)"
\verb"Line 8: t = t + h"
\end{prog}
\end{center}
\subsection{Runge Kutta Fourth Order Integrator}

Runge Kutta Fourth Order integrator evaluates the derivative four times. It is the most accurate integrator that we describe compared to Euler and Midpoint.

\begin{figure}[h]
\hrule\vskip4pt
\begin{center}
	{	 \includegraphics[width=4in]{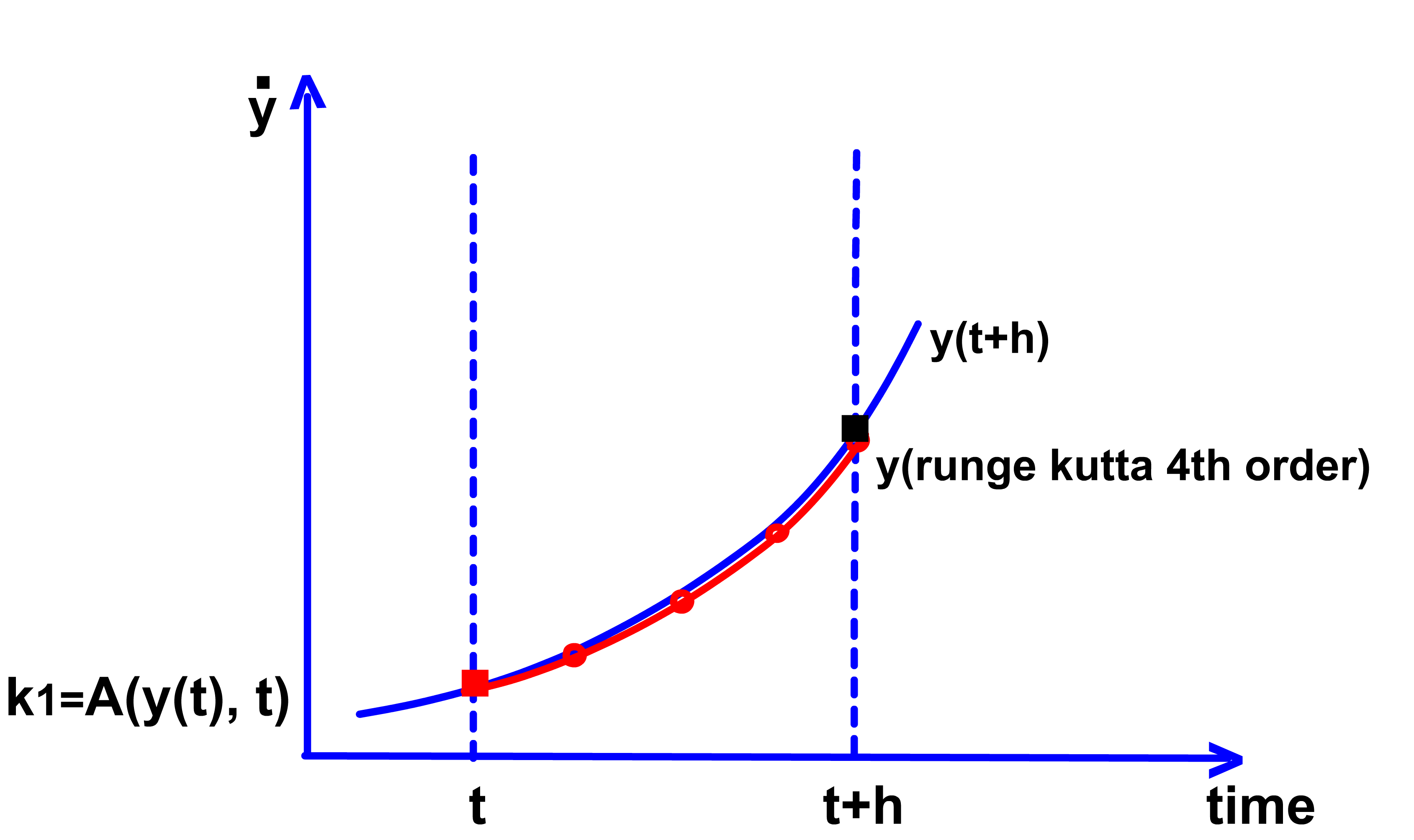}}
\caption{Runge Kutta 4th Order Integrator}
\label{fig:rk4}
\end{center}
\hrule\vskip4pt
\end{figure}

\noindent
The Runge Kutta Fourth integrator, is also based on the
Taylor's series. It retains only first five derivative term with a local truncation error $O(h^5)$:

\begin{equation}
\label{eq:rk4}
%y({t}+{h}) = y({t}) + {h}\dot{y}({t})+ \frac{{{h}^2}}{2!}\dot{\dot{{y}}}({t})+
%\frac{{{h}^3}}{3!}\dot{\dot{\dot{{y}}}}({t})+\frac{{{h}^4}}{4!}\dot{\dot{\dot{\dot{{y}}}}}({t})+O({h}^5)
y({t}+{h}) = y({t}) + {h}{y}{'}({t})+ \frac{{{h}^2}}{2!}{y}{''}({t})+
\frac{{{h}^3}}{3!}{y}{'''}({t})+\frac{{{h}^4}}{4!}{y}{''''}({t})+O({h}^5)
\end{equation}

\begin{equation}
k_0 = y({t})
\end{equation}

\begin{equation}
%{k}_{1} = \dot{y}({t}) = {\bf A}(y(t),t)
{k}_{1} = {y}'({t}) = {\bf A}(y(t),t)
\end{equation}

\begin{equation}
{k}_{2} = {\bf A}(y(t)+{h}\frac{{k}_{1}}{2},t+\frac{h}{2})
\end{equation}

\begin{equation}
{k}_{3}  = {\bf A}(y(t)+{h}\frac{{k}_{2}}{2},t+\frac{h}{2})
\end{equation}

\begin{equation}
{k}_{4}  = {\bf A}(y(t)+{h}{k}_{3}, t+h)
\end{equation}

\begin{equation}
y({t}+{h}) = {k}_{0} + \frac{1}{6}{h}({k}_{1}
+ 2\, {k}_{2} + 2\, {k}_{3} + {k}_{4})
\end{equation}

where

$k_0$ is the initial state

%$k_1$ is the function to find the simple Euler slope
$k_1$ is the slope at the left end of interval,

%$k_2$ is the function to find the better estimate slope at the middle of the interval $k_1$
$k_2$ is the slope at the middle point using the Euler formula to go from $t$ to $t+ \frac{h}{2}$,

%$k_3$ is again to the function to find the better estimate slope at the middle of the interval $k_3$

$k_3$ is the second approximation to the slope at the midpoint,

%$k_4$ is the function to use endpoint slope

$k_4$ is the slope at $t+h$ using the Euler formula and the slope $k_3$ to go from $t$ to $t+h$.
\paragraph*{Pseudocode for Runge Kutta Fourth Order Method}
\begin{center}
\begin{prog}
\verb"Line 1: define A(y(t), t)"
\verb"Line 2: initial values y0 and t0"
\verb"Line 3: stepsize h and number of steps n"
\verb"Line 4: for i from 1 to n do"
\verb"Line 5: k1 = A(y(t), t)"
\verb"Line 6: k2 = A(y+h/2(k1), t+h/2)"
\verb"Line 7: k3 = A(y+h/2(k2), t+h/2)"
\verb"Line 8: k4 = A(y+hk3, t+h)"
\verb"Line 9: y = y + h/6(k1+2*k2+2*k3+k4)"
\verb"Line 10: t = t + h"
\end{prog}
\end{center}

\section{Newton's Laws}

After the force accumulation on the object, it is important to find the acceleration {\bf a} in order to define the motion of objects in their next time step. The physical law that governs the motion of objects is the Newton's Second law. It states that the force ${\bf F}$ is proportional to the time rate of change of its linear momentum. Momentum is the product of mass $m$ and velocity {\bf v}.
%Then, the classical equation \ref{equ:newton2} can be rewritten as:
%
\begin{equation}
{\bf F} \approx {m}\frac{\Delta{\bf v}}{\Delta t}
\label{equ:newton22}
\end{equation}

% \begin{equation}
% %{\dot {\bf v}}={\bf a}  %= \frac{\partial {v}}{\partial t}
%  {\bf v}'={\bf a}
% \end{equation}
%
% \begin{equation}
% %{\dot {\bf r}}={\bf v}  %={\frac{\partial {p}}{\partial {t}}}
%  {\bf r}'={\bf v}
% \end{equation}

%force ${\bf F}$ equals mass $m$ multiplied by its acceleration ${\bf a}$ \cite{wr03}.

%\begin{equation}
%{\bf F} = {m}{\bf a} %\Rightarrow {\bf a}=\frac{\bf F}{m}
%\label{equ:newton2}
%\end{equation}

\paragraph*{Velocity} {\bf v} is the integral of acceleration {\bf a} with respect to the time $t$. Therefore, integrating the acceleration gives us the new velocity {\bf v}.
%is the derivative of the velocity, which is differentiated with respect of time, the partial velocity and partial time in first order ODE.
\begin{equation}
%{\dot {\bf v}}={\bf a}  %= \frac{\partial {v}}{\partial t}
%  {\bf v}'={\bf a}
  {\bf v}=\int{\bf a} {dt}
\end{equation}
% \noindent

\paragraph*{Position} {\bf r} is the integral of velocity {\bf v} with respect to the time $t$. Therefore, integrating the velocity gives us the new position {\bf r}.
%  %the derivative of the position ${\bf r}$ of the particle,
% %which is differentiated with respect of time, the partial position and partial time in first order ODE
\begin{equation}
%{\dot {\bf r}}={\bf v}  %={\frac{\partial {p}}{\partial {t}}}
%  {\bf r}'={\bf v}
 {\bf r}=\int{\bf v} {dt}
\end{equation}

\noindent
Let's take one particle on the object as an example and understand how the different integrators work.
%%%%%%%%%%%%%%%%%%%%%%%%%%%%%%%%%%%%     get the acceleration a = F/m     %%%%%%%%%%%%%%%%%%%%%%%%%%%%%%%%%%%%%%%%%%%%%%%%
%\item
%\begin{equation}
%\end{equation}
%%%%%%%%%%%%%%%%%%%%%%%%%%%%%%%%%%%%%%%%%  v = at    %%%%%%%%%%%%%%%%%%%%%%%%%%%%%%%%%%%%%%%%%%%%%
%\item
%%%%%%%%%%%%%%%%%%%%%%%%%%%%%%%%%%%%     x = vt   %%%%%%%%%%%%%%%%%%%%%%%%%%%%%%%%%%%%%%%%%%%%%%%%%%%%%%
%\item
%\begin{equation}
%\end{equation}
%%%%%%%%%%%%%%%%%%%%%%%%%%%%%%%%%%%%    a = v 1st  %%%%%%%%%%%%%%%%%%%%%%%%%%%%%%%%%%%%%%%%%%%%%%%%%%%%%%
%\item
\subsection{Newton's Laws in Euler Integrator}

Based on the Euler Integrator method shown in Eq.\ref{eq:euler}, the new velocity and position of a particle can be integrated follows.

\paragraph*{Velocity} can be represented as the following equation:
%\paragraph*{Acceleration} is the derivative of the velocity, which is differentiated with respect of time, the partial velocity and partial time in first order ODE.
%\begin{equation}
%{\dot {\bf v}}={\bf a}  %= \frac{\partial {v}}{\partial t}
%\end{equation}
%\noindent
%Therefore,velocity is the integral of acceleration with respect of the time. If we integrate the acceleration vector over time, it gives us the new velocity. The velocity vector has been changed over this time.

\begin{equation}
%v(t+h) = v_{k0} + v_{k1}
%v(t+h) = v(t) + {h}\dot{v}({t})
{\bf v}(t+h) \approx {\bf v}(t) + {h}{\bf v}'({t})
\label{eq:veuler}
\end{equation}

% \noindent
% where

\noindent
$v_{k0}$ represents the first term in Eq.\ref{eq:veuler}, which is the initial velocity at time $t$

\begin{equation}
{v_{k0}} = {\bf v}(t)
\end{equation}

\noindent
$v_{k1}$ represents the second term in Eq.\ref{eq:veuler}, which is the function to compute the derivative velocity in the period $h$

\begin{equation}
%v_{k1} = \int_t^{t+h}\dot{\bf a}\Delta t = {{\bf a}(t)}h
%v_{k1} = {h}\dot{v}({t}) = {{\bf a}(t)}h
v_{k1} = {h}{\bf v}'({t}) = {{\bf a}(t)}h
\end{equation}
%%%%%%%%%%%%%%%%%%%%%%%%%%%%%%%%%%%%    a = x 2nd   %%%%%%%%%%%%%%%%%%%%%%%%%%%%%%%%%%%%%%%%%%%%%%%%%%%%%%
%\item
%%%%%%%%%%%%%%%%%%%%%%%%%%%%%%%%%%%%    a = x/t 2nd   %%%%%%%%%%%%%%%%%%%%%%%%%%%%%%%%%%%%%%%%%%%%%%%%%%%%%%
%\item
%The acceleration is the second derivative of position, which is partial position and partial time in second order ODE:
%\begin{equation}
%{\bf a} = \dot{{\dot {\bf r}}}% = {\frac{\partial ^{2} {p}}{\partial {t}^{2}}}
%\end{equation}
%%%%%%%%%%%%%%%%%%%%%%%%%%%%%%%%%%%%%%%%% v = x/t		 %%%%%%%%%%%%%%%%%%%%%%%%%%%%%%%%%%%%%%%%%
%\item
%\paragraph*{Velocity} is handled similarly. It is the derivative of the position ${\bf r}$ of the particle,
%which is differentiated with respect of time, the partial position and partial time in first order ODE
%\begin{equation}
%{\dot {\bf r}}={\bf v}  %={\frac{\partial {p}}{\partial {t}}}
%\end{equation}
%If we integrate the velocity vector over time, it gives us how the position vector changed over this time.
%The position $p$ and the velocity ${\bf v}$ space is called phase space, which can be concatenated to form a 6-vector:
%\begin{equation}
%[ {\dot{{\bf r}}_{x}},  {\dot{{\bf r}}_{y}}, {\dot{\bf r}_{z}}, {\dot{\bf v}_{x}}, {\dot{\bf v}_{y}}, {\dot{\bf v}_{z}} ] = [{\bf v}_{x}, {\bf v}_{y}, {\bf v}_{z}, \frac{{\bf f}_x}{m}, \frac{{\bf f}_y}{m}, \frac{{\bf f}_z}{m}]
%\end{equation}
%%%%%%%%%%%%%%%%%%%%%%%%%%%%%%%%%%%%%%%%%%%%%  v(t+dt) = vt + (F/m) dt 	 %%%%%%%%%%%%%%%%%%%%%%%%%%%%%%%%%%%%%%%%%
%\item
%%%%%%%%%%%%%%%%%%%%%%%%%%%%%%%%%%%%%%%%%%%%%%%  x(t) = x(t_0) + v(t)dt   %%%%%%%%%%%%%%%%%%%%%%%%%%%%%%%%%%%%%%%%%%%%
%\item

\paragraph*{Position} can be represented as the following equation
\begin{equation}
%r(t+h) = r_{k0} + r_{k1}
%r(t+h) = r(t) + {h}\dot{r}({t})
{\bf r}(t+h) \approx {\bf r}(t) + {h}{\bf r}'({t})
\end{equation}

% \noindent
% where

\noindent
$r_{k0}$ is the initial position at time $t$

\begin{equation}
{r_{k0}} = {\bf r}(t)
\end{equation}

\noindent
$r_{k1}$ is the function to find the travel position in the period $h$

\begin{equation}
%r_{k1} = \int_t^{t+h}\dot{\bf v}\Delta t = {{\bf v}(t)}h
%r_{k1} = {h}\dot{r}({t}) = {{\bf v}(t)}h
r_{k1} = {h}{\bf r}'({t}) = {{\bf v}(t)}h
\end{equation}

%\end{itemize}

%\end{eqnarray*}

%$$y({t}+{h}) = y({t}) + {h}\dot{y}({t})+ \frac{{h^2}}{2}\ddot{y}({t})+ O(h^3)$$
%$$k0 = y({t})$$
%$$k1 = {h}\dot{y}({t}) = {h}f(y,t)$$
%$$k2 = {h}\ddot{y}({t}) = {h}f(y+k1/2,t/2)$$
%$$y({t}+{h}) = k0 + k2$$

\subsection{Newton's Laws in Midpoint Integrator}

We apply the midpoint algorithm theory on the Newton's law in order to
achieve higher accuracy in the the relationship between the velocity and the position according the Eq.\ref{eq:midpoint}.
%We simply plug in the elements $k_1...k_n$ into the Newton's law equation.

%$$v_{k_1} = a(t)h$$
%$$p_{k_1} = v(t)h$$
%$$v_{k_2} = v(t) +{v_{k_1}}h$$
%$$p_{k_2} = p(t)+{p_{k_1}}h$$
%$$v(t) = v(t) + \frac{v_{k_1}+v_{k_2}}{2}h$$
%$$p(t) = p(t) + \frac{p_{k_2}+p_{k_2}}{2}h$$
\paragraph*{Velocity} can be represented as the following equation
% is the integral of acceleration with respect of the time. If we integrate the acceleration vector over time, it gives us the new velocity. The velocity vector has been changed over this time.
\begin{equation}
%v({t}+{h}) = v({t}) + {h}\dot{v}({t})+ \frac{{{h}^2}}{2!}\dot{\dot{{v}}}({t})
%v(t+h) = v_{k0} + \frac {v_{k1}+v_{k2}}{2}
{\bf v}({t}+{h}) \approx {\bf v}({t}) + {h}{\bf v}'({t})+ \frac{{{h}^2}}{2!}{{\bf v}}''({t})
\end{equation}

% \noindent
% where

\noindent
$v_{k0}$ is the initial velocity at state $t$

\begin{equation}
{v_{k0}} = {\bf v}(t)
\end{equation}

\noindent
$v_{k1}$ is the function to compute the derivative velocity in the period $h$

\begin{equation}
%v_{k1} = \int_t^{t+h}\dot{\bf a}\Delta t = {{\bf a}(t)}h
%v_{k1} = \dot{v}({t}) = {{\bf a}(t)}h
v_{k1} = {\bf v}'({t}) = {{\bf a}(t)}h
\end{equation}

\noindent
$v_{k2}$ is the function to compute the derivative velocity in the period $t+h$

\begin{equation}
%v_{k2} = \int_t^{t+h}\dot{\dot{\bf a}}\Delta t = v_{k0} + v_{k1} = {\bf v}(t) +{{\bf a}(t)}h
%v_{k2} = \dot{v}({t+h}) = {\bf v}(t) +{{\bf a}(t)}h
v_{k2} = {\bf v}'({t+h}) = {\bf v}(t) +{{\bf a}(t)}h
\end{equation}
Therefore, the new velocity of a particle is
\begin{equation}
{\bf v}(t+h) = v_{k0} + \frac {v_{k1}+v_{k2}}{2}
\end{equation}
%%%%%%%%%%%%%%%%%%%%%%%%%%%%%%%%%%%%%%%%%%%%%%%  x(t) = x(t_0) + v(t)dt   %%%%%%%%%%%%%%%%%%%%%%%%%%%%%%%%%%%%%%%%%%%%
%\item
%If we integrate the velocity vector over time, it gives us how the position vector changed over this time.
\paragraph*{Position} can be represented as the following equation

%\begin{equation}
%r(t+h) = r_{k0} + \frac {r_{k1}+ r_{k2}}{2}
%\end{equation}
\begin{equation}
%r({t}+{h}) = r({t}) + {h}\dot{r}({t})+ \frac{{{h}^2}}{2!}\dot{\dot{{r}}}({t})
%v(t+h) = v_{k0} + \frac {v_{k1}+v_{k2}}{2}
{\bf r}({t}+{h}) \approx {\bf r}({t}) + {h}{\bf r}'({t})+ \frac{{{h}^2}}{2!}{\bf r}''({t})
\end{equation}
% \noindent
% where

\noindent
$r_{k0}$ is the initial position at state $t$

\begin{equation}
{r_{k0}} = {\bf r}(t)
\end{equation}

\noindent
$r_{k1}$ is the function to find the travel position in the period $h$

\begin{equation}
%r_{k1} = \int_t^{t+h}\dot{\bf v}\Delta t = {{\bf v}(t)}h
%r_{k1} = \dot{r}({t}) = {{\bf v}(t)}h
r_{k1} = {\bf r}'({t}) = {{\bf v}(t)}h
\end{equation}

\noindent
$r_{k2}$ is the function to find the travel position in the period $t+h$

\begin{equation}
%r_{k2} = \int_t^{t+h}\dot{\dot {\bf v}}\Delta t = r_{k0} + r_{k1} = {r}(t) +{{\bf v}(t)}h
%r_{k2} = \dot{r}({t+h}) = {\bf r}(t) +{{\bf v}(t)}h
r_{k2} = {\bf r}'({t+h}) = {\bf r}(t) +{{\bf v}(t)}h
\end{equation}
Therefore, the new position of a particle is
\begin{equation}
{\bf r}(t+h) = r_{k0} + \frac {r_{k1}+r_{k2}}{2}
\end{equation}

\subsection{Newton's Laws in the Runge Kutta Fourth Order Integrator}

\noindent
%Velocity is also the integral of acceleration with respect of the time. If we integrate the acceleration vector over time, it gives us the new velocity. The velocity vector has been changed over this time.
Based on the Runge Kutta Fourth Order method we have shown in Eq.\ref{eq:midpoint}, the new velocity and position of a particle can be integrated as following.

\paragraph*{Velocity} can be represented as the following equation
\begin{equation}
%v(t+h) = v_{k0} + \frac {v_{k1}+2v_{k2}+2v_{k3}+v_{k4}}{6}
{\bf v}({t}+{h}) \approx {\bf v}({t}) + {h}{\bf v}{'}({t})+ \frac{{{h}^2}}{2!}{\bf v}{''}({t})+
\frac{{{h}^3}}{3!}{\bf v}{'''}({t})+\frac{{{h}^4}}{4!}{\bf v}{''''}({t})
\end{equation}

% \noindent
% where

\noindent
$v_{k0}$ is the initial velocity at time $t$

\begin{equation}
{v_{k0}} = {\bf v}(t)
\end{equation}

\noindent
$v_{k1}$ is the function to compute the derivative velocity in the period $h$

\begin{equation}
%v_{k1} = \int_t^{t+h}\dot{\bf a}\Delta t = {{\bf a}(t)}h
v_{k1} = {{\bf a}(t)}h
\end{equation}

\noindent
$v_{k2}$ is the function to compute the derivative velocity of the Euler integration in the period $h/2$ based on the previous step

\begin{equation}
%v_{k2} = \int_t^{t+\frac{h}{2}}\dot {\dot {{\bf a}}}\Delta t = v_{k0} + \frac{v_{k1}}{2} = {\bf v}(t) + \frac{{{\bf a}(t)}h}{2}
v_{k2} = v_{k0} + \frac{v_{k1}}{2}
\end{equation}

\noindent
$v_{k3}$ is the function to compute the derivative velocity of the second approximation based on the $v_{k2}$ in the period $h/2$

\begin{equation}
%v_{k3} = \int_t^{t+\frac{h}{2}}\dot {\dot {\dot{\bf a}}}\Delta t = v_{k0} + \frac{v_{k2}}{2} %= {\bf v}(t) +{{\bf a}(t)}
v_{k3} =  v_{k0} + \frac{v_{k2}}{2}
\end{equation}

\noindent
$v_{k4}$ is the function to compute the final resulting velocity change of $v_{k3}$ from $v_{k0}$

\begin{equation}
%v_{k4} = \int_t^{t+h}\dot{\dot{\dot {\dot {{\bf a}}}}}\Delta t = v_{k0} + v_{k3} %= {\bf v}(t) +{{\bf a}(t)}h
v_{k4} =  v_{k0} + v_{k3}
\end{equation}
Therefore, the new velocity of the particle is
\begin{equation}
{\bf v}({t}+{h}) = v_{k0} + \frac{1}{6}{h}({v_{k1}}
+ 2\, {v_{k2}} + 2\, {v_{k3}} + {v_{k4}})
\end{equation}

%%%%%%%%%%%%%%%%%%%%%%%%%%%%%%%%%%%%%%%%%%%%%%%  x(t) = x(t_0) + v(t)dt   %%%%%%%%%%%%%%%%%%%%%%%%%%%%%%%%%%%%%%%%%%%%
%\item
\noindent
If we integrate the velocity vector over time, it gives us how the position vector changed over this time.

\paragraph*{Position} can be represented as the following equation
\begin{equation}
%r(t+h) = r_{k0} + \frac {r_{k1}+2r_{k2}+2r_{k3}+r_{k4}}{6}
{\bf r}({t}+{h}) \approx {\bf r}({t}) + {h}{\bf r}{'}({t})+ \frac{{{h}^2}}{2!}{\bf r}{''}({t})+
\frac{{{h}^3}}{3!}{\bf r}{'''}({t})+\frac{{{h}^4}}{4!}{\bf r}{''''}({t})
\end{equation}

% \noindent
% where

\noindent
$r_{k0}$ is the initial position at time $t$

\begin{equation}
{r_{k0}} = {\bf r}(t)
\end{equation}

\noindent
$r_{k1}$ is the function to find the travel position in the period $h$

\begin{equation}
r_{k1} = {{\bf v}(t)}h
\end{equation}

\noindent
$r_{k2}$ is the function to find the travel position of the Euler integration in the period $h/2$ based on the previous step

\begin{equation}
r_{k2} = r_{k0} + \frac{r_{k1}}{2} = {\bf r}(t) +\frac{{{\bf v}(t)}h}{2}
\end{equation}

\noindent
$r_{k3}$ is the function to find the travel position of the second approximation based on the $r_{k2}$ in the period $h/2$

\begin{equation}
r_{k3} = r_{k0} + \frac{r_{k2}}{2}
\end{equation}

\noindent
$r_{k4}$ is the function to find the travel position change of $r_{k3}$ from $r_{k0}$

\begin{equation}
r_{k4} = r_{k0} + {r_{k3}}% = {p}(t) +\frac{{{\bf v}(t)}h}{2}
\end{equation}

Therefore, the new position of the particle is
\begin{equation}
{\bf r}({t}+{h}) = r_{k0} + \frac{1}{6}{h}({r_{k1}}
+ 2\, {r_{k2}} + 2\, {r_{k3}} + {r_{k4}})
\end{equation}
%$$v_{k1} = a(t)h$$
%$$p_{k1} = v(t)h$$
%$$v_{k2} = (v(t)+\frac{v_{k1}}{2})h$$
%$$p_{k2} = (p(t)+\frac{p_{k1}}{2})h$$
%$$v_{k3} = (v(t)+\frac{v_{k2}}{2})h$$
%$$p_{k3} = (p(t)+\frac{p_{k2}}{2})h$$
%$$v_{k4} = (v(t)+v_{k3})h$$
%$$p_{k4} = (p(t)+p_{k3})h$$
%$$p(t+\partial t) = p(t) + \frac{{p_{k1}} + 2{p_{k2}} + 2{p_{k3}} + {p_{k4}}}{6}$$
%$$v(t+\partial t) = v(t) + \frac{{v_{k1}} + 2{v_{k2}} + 2{v_{k3}} + {v_{k4}}}{6}$$

\section{Comparison of Three Integrators}

\subsection{Efficiency}
For a given step size, Euler is more efficient because it requires only one derivative evaluation per step. Mid Point requires about twice as much computation than the Euler integrator because Mid Point uses two steps to calculate velocity and position at the next time. Runge Kutta Fourth Order requires about four times as much computation as Euler integrator because it use four steps to calculate the velocity and position at the next time step \cite{wr03}.
%\begin{equation}
%RK 4 = 2 \cdot Midpoint = 4 \cdot Euler
%\end{equation}
For some configuration, if speed is the priority, Euler integration is convenient to use, but at the expense of accuracy and stability.

%there may be need the Euler method because of the need of smaller simulation step. However, RK 4 will allow a much larger step size than Euler.

\subsection{Accuracy}
Smaller time steps means more stability and accuracy. But also means more computation.
If a given step size is $h$, error of Euler method is $O({h}^2)$ as a first-order method, error of midpoint is $O({h}^3)$, and error of RK 4 is $O({h}^5)$ \cite{wr03}.
\begin{itemize}
\item The Euler method is based on keep the first two terms of the Taylor series expansion
\begin{equation}
y({t}+{h}) = y({t}) + {h}y'({t}) + O(h^2)
%y({t_0}+{\partial t}) = y({t_0}) + {\partial t}\dot{y_0}({t})+O({\partial t}^2)
%y({t_0}+{\partial t}) = y({t_0}) + {\partial}\dot{y}({t})+ \frac{{{\partial t}^2}\dot{\dot{y}}({t})}{2!}+\ddots+\frac{{{\partial t}^n}\left(\frac{\partial^{n}y}{\partial{t}^n}\right)}{n!}
\end{equation}
\item An improved method which involves the second derivative is Midpoint method as following
\begin{equation}
%y({t_0}+{\partial t}) = y({t_0}) + {\partial t}\dot{y_0}({t})+ \frac{{{\partial t}^2}}{2}\ddot{y_0}({t})+ O({\partial t}^3)
y({t}+{h}) = y({t}) + {h}y'({t})+ \frac{{{h}^2}}{2!}y''({t})+ O({h}^3)
\end{equation}

\item An improved method which involves the four derivative is Runge Kutta method as following
\begin{equation}
%y({t_0}+{\partial t}) = y({t_0}) + {\partial t}\dot{y_0}({t})+ \frac{{{\partial t}^2}}{2}\ddot{y_0}({t})+ O({\partial t}^5)
y({t}+{h}) = y({t}) + {h}{y}{'}({t})+ \frac{{{h}^2}}{2!}{y}{''}({t})+
\frac{{{h}^3}}{3!}{y}{'''}({t})+\frac{{{h}^4}}{4!}{y}{''''}({t})+O({h}^5)
\end{equation}
\end{itemize}
\subsection{Stability}

With smaller step time value, such as 10 ms, the system integrated by any of the three methods is stable. However, if we give the system a higher step time value, such as 50 or 100 ms, with same mass, damping coefficient, gravity acceleration, the elastic object under Euler system will explode after a short period because its numerical instability causes the mass to oscillate out of control; midpoint and Runge Kutta Fourth Order integrator are more stable \cite{wr03}.

%The numerical solution involves computing an efficient and accurate solution of the mathematical equations. Finite precision of numbers, limited computational power and memory forces us to approximate the mathematical model with simple procedures.

\chapter{Design and Implementation}
\label{chapt:design and implementation}
\index{Design and Implementation}
%In Chapter 3, we have described the elastic object modeling methods, their elastic properties and related forces accumulated on them.
%In Chapter 4, we have introduced the different integrators and explained how they integrate forces and define the new position and orientation for the object in next system update.
In this chapter, we will present the detailed design of the two-layer elastic object physical based simulation system and its implementation.

\section{Elastic Object Simulation System Design}

In this section, an overview of the framework and the algorithm for the elastic simulation system is given.

\subsection{Domain Analysis-Based Modeling}

\begin{figure}[!h]
\hrule\vskip4pt
\begin{center}
	{\includegraphics[width=3.5in]{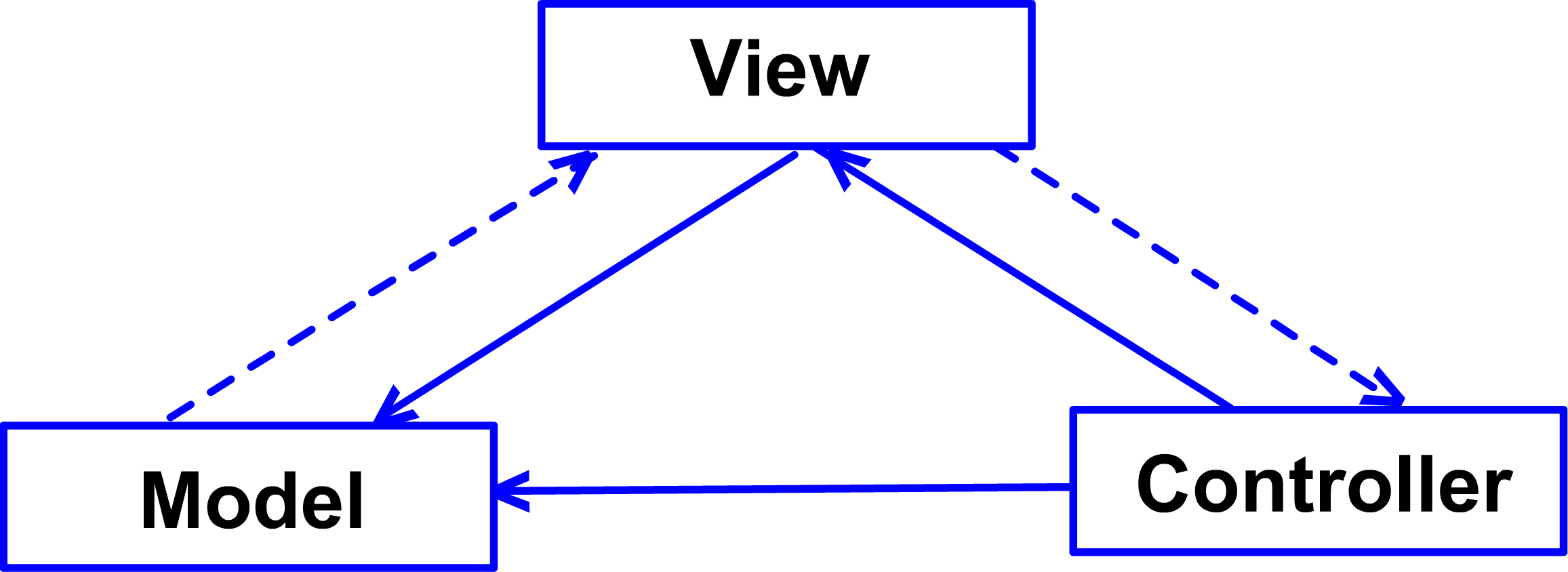}}
\caption{Model-View-Controller}
\label{fig:MVC}
\end{center}
\hrule\vskip4pt
\end{figure}
This elastic object simulation system has been designed and implemented according to the well known architectural pattern, Model-View-Controller\cite{wk07}.
This pattern is ideal for real time simulation because it simplifies the dynamic tasks handling by separating data (Model) from user interface (View).
Thus, the user's interaction with the software does not impact the data handling; the data can be reorganized without changing the user interface. The communication between the Model and the View is done through another component: Controller.
In our current simulation system, the application has been split into these three separated components:
\begin{itemize}
\item Model is an application of object modeling. It stores the geometric modeling methods of the elastic objects and the data of the objects themselves, such as one-dimensional, two-dimensional, and three-dimensional elastic objects and their associated data structure, such as vector, particles, springs, and faces.
\item View is the screen presentation to render the Model and a user interface for dynamical simulation. The view in my system is the GLUT window which displays the elastic object and allows the user to use mouse and keyboard to interact with the elastic object.
\item Controller handles the processes and responds from the user interaction and invokes the changes to the model. When the user interacts with the elastic object through the GLUT window by dragging it with mouse, the controller handles the new dragging force from the user interface, integrates the new force to find out the change of the acceleration and velocity, and where the object should move to in next display update. This is done through the
series of registered GLUT callback functions that process the input from the user.
\end{itemize}
%1. Apply forces
%2. Integrate forces for new velocity
%3. Integrate velocity for new position
%4. Do collision detection
%4a. Apply impulse of collisions, compute new velocity, possibly tweak positions.
\section{Elastic Object Simulation System Implementation}
The system is implemented using OpenGL and the C++ programming language with object oriented programming paradigm.
\xf{fig:DiagramClass} describes structure of the software based on the classes.
\begin{figure}[!h]
\hrule\vskip4pt
\begin{center}
	{\includegraphics[width=6in]{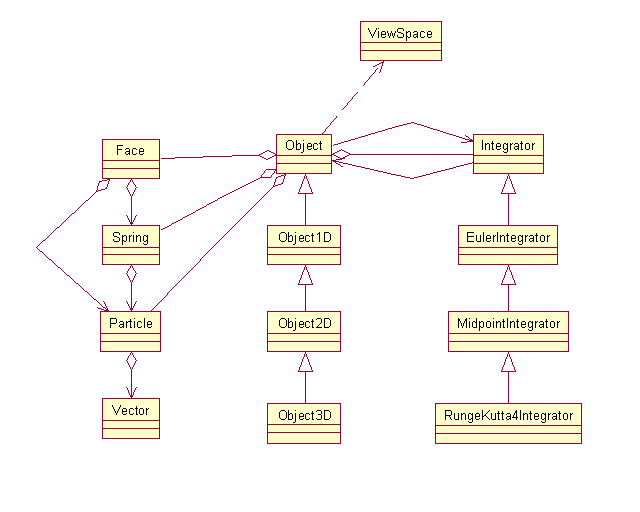}}
\caption{Class Diagram}
\label{fig:DiagramClass}
\end{center}
\hrule\vskip4pt
\end{figure}
\begin{itemize}
\item The three data structures, such as particle, spring, and face compose an elastic object.
\item The elastic object types can be varied by the dimensionality: one-, two-, or three-dimensional. %s, one-dimensional object, two dimensional object, or three dimensional object.
\item The types of integrators are also varied by their complexities, such as Euler, Midpoint, and Runge-Kutta.
\item An ``Object'' instance contains an instance of an ``Integrator''. The relationship between them is aggregation rather than a common composition because when the elastic object is destroyed, the integrator object is not necessary destroyed. The ``Object'' has an aggregation of the ``Integrator'' by containing only a reference or pointer to the ``Integrator''.
\item The classes ``Object'', ``ViewSpace'', and ``Integrator'' are associated to each other based the Model-View-Controller model.
\end{itemize}
Let's have a close view at each model and the related classes with their parameters and member functions.
% \clearpage
\subsection{Design and Implementation of Data Types}
\begin{figure}[!h]
\hrule\vskip4pt
\begin{center}
	{\includegraphics[width=4in]{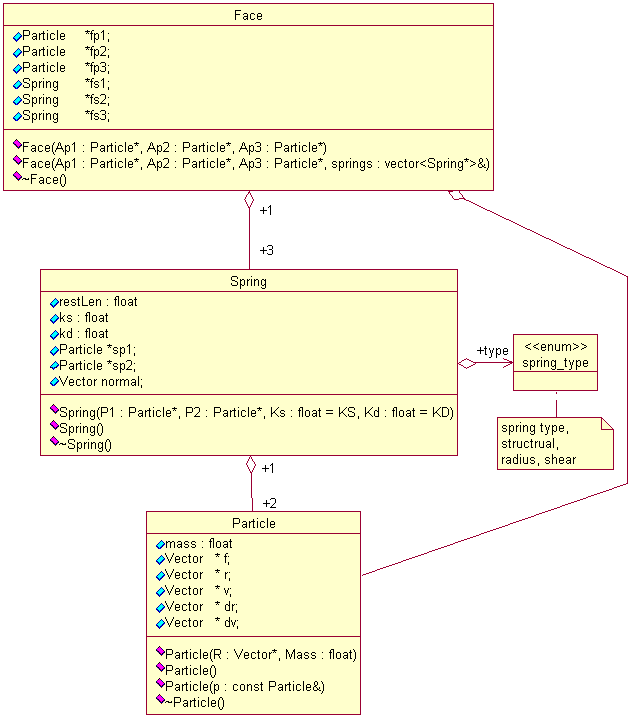}}
\caption{Face-Spring-Particle Class Diagram}
\label{fig:fsp}
\end{center}
\hrule\vskip4pt
\end{figure}

The basic data structure is the object vector, which defines the the scalar value with direction. For the second basic data structure, particle, whose properties, such as position, velocity are made up of the object vector. The next higher data structure is spring, which is defined by two particle objects. Face, which is the highest data structure in this simulation system, is composed of three connected springs.
%\verb"// particle constructor with variables position"
%\verb"Particle (const Particle& p);"
%\verb"Particle(); "%						        // particle constructor without variables"
%\verb"~Particle();"%						        // particle destructor	"
%\verb"void ClearForce(void);"%					// reset forces at a Particle 	 "
\paragraph*{Particle} In \xf{fig:fsp}, the particle class shows that each particle has mass $mass$, position ${\bf r}$, velocity ${\bf v}$, derivative of position ${\bf dr}$, derivative of velocity ${\bf dv}$, and force vector ${\bf f}$. Particle constructor sets up its properties with default values.

%\begin{figure}[!h]\footnotesize
%\hrule\vskip4pt
%\begin{verbatim}
% class Particle
 %{
%    public:
%    float mass;
%    Vector* r;
%    Vector* v;
%    Vector* dr;
%    Vector* dv;
%    Vector* f;
%    Particle (Vector* R, float Mass) : mass(Mass)
%    {
%       r = R;
%      dr = new Vector(0,0,0);
%      dv = new Vector(0,0,0);
%      f  = new Vector(0,0,0);
%      v  = new Vector(0,0,0);
%    }
%}
%\end{verbatim}
%\caption{class Particle}
%\label{fig:cparticle}
%\vskip4pt\hrule
%\end{figure}
\paragraph*{Spring} As shown in \xf{fig:fsp}, the spring class, there are different types of springs to construct the object, such as structural, radius, shear-left, and shear-right springs, declared in the enum type $spring\_type$ and the default spring type is structural. $*sp1$ is the head of the spring and points to a particle; $*sp2$ is the tail of the spring and points to a particle. $restLen$ is the spring length when it is in the resting state. $ks$ is Hooke's spring constant and $kd$ is the spring damping factor. The spring normal vector will be calculated and needed in pressure force calculation.
%\begin{figure}[!h]\footnotesize
%\hrule\vskip4pt
%\begin{verbatim}
% enum spring_type { SPRING_STRUCTURAL, SPRING_RADIUM, SPRING_SHEAR};
% class Spring
% {
%    public:
%    Particle *sp1;
%    Particle *sp2;
%    float restLen;
%    float ks;
%    float kd;
%    Vector normal;
%    Spring(Particle *P1, Particle *P2,float Ks=KS, float Kd=KD);
%    Spring();
%   ~Spring();
%    void setRestLen();
%	}
%\end{verbatim}
%\caption{class Spring}
%\label{fig:cspring}
%\vskip4pt\hrule
%\end{figure}
\paragraph*{Face} In \xf{fig:fsp}, the face class shows that a face contains $*fp1$, $*fp2$, and $*fp3$ point to the first, the second, and the third particles as three of its vertices. It also contains $*fs1$, $*fs2$, and $*fs3$ point to the first, second, and third spring as three of its edges. There are two face constructors. The first one stores the information of three vertices that point to three particles. It represents faces on two-dimensional objects. The faces will only be needed at the display process.
%\begin{figure}[!h]\footnotesize
%\hrule\vskip4pt
%\begin{verbatim}
% class Face
% {
%    public:
%    Particle *fp1;
%    Particle *fp2;
%    Particle *fp3;
%    Spring *fs1;
%    Spring *fs2;
%    Spring *fs3;
%    Face(Particle *Ap1, Particle *Ap2, Particle *Ap3)
%                        :fp1(Ap1), fp2(Ap2), fp3(Ap3);
%    Face(Particle *Ap1, Particle *Ap2, Particle *Ap3,
%                  vector<Spring*> &springs):fp1(Ap1), fp2(Ap2), fp3(Ap3);
%    ~Face();
% }
%\end{verbatim}
%\caption{class Face}
%\label{fig:cface}
%\vskip4pt\hrule
%\end{figure}

\xf{fig:csface1} represents another face constructor along with its algorithm implementation.
It accepts three vertices on each face that point to the three particles, and constructs a spring and stores the spring information into the spring vector. This constructor is called by three-dimensional uniform modeling method. The index of face is the key data structure for subdivision method in subroutine. The constructor initializes the three springs based on the three particles. First spring contains particle $p1$ and $p2$; the second spring contains particle $p2$ and $p3$; the third spring contains particle $p3$ and $p1$.
A special care is taken not to duplicate existing springs (which would result in incorrect behaviour of the model); therefore, we only allow the new and non-existing springs to be saved in the spring vector.
If the first spring already exists with particles $p1$ and $p2$, the new spring $fs1$ will point to the existing spring.
Same method is applied on the second spring $fs2$ and third spring $fs3$. Otherwise, the new spring will be pushed and saved into the spring vector.
Please refer to the actual code for the complete implementation.

\begin{figure}[!h]\footnotesize
\hrule\vskip4pt
\begin{verbatim}
 Face(Particle *Ap1, Particle *Ap2, Particle *Ap3, vector<Spring*> &springs)
     : fp1(Ap1), fp2(Ap2), fp3(Ap3) {
   fs1 = new Spring(Ap1, Ap2); fs2 = new Spring(Ap2, Ap3); fs3 = new Spring(Ap3, Ap1);
   bool a = false, b = false, c = false;
   for(int o = 0; o < springs.size(); o++) {
     if(springs[o]->sp1 == Ap1	&& springs[o]->sp2 == Ap2) {
       delete fs1; fs1 = springs[o]; a = true;
     }
     if(springs[o]->sp1 == Ap2	&& springs[o]->sp2 == Ap3) {
       delete fs2; fs2 = springs[o]; b = true;
     }
     if(springs[o]->sp1 == Ap3	&& springs[o]->sp2 == Ap1) {
       delete fs3; fs3 = springs[o]; c = true;
     }
   }
   if(!a) springs.push_back(fs1);
   if(!b) springs.push_back(fs2);
   if(!c) springs.push_back(fs3);
 }
\end{verbatim}
\caption{Special 3D Uniform Modeling Face Constructor}
\label{fig:csface1}
\vskip4pt\hrule
\end{figure}

% \clearpage
\subsection{Design and Implementation of Components: Model}
\begin{figure}[!h]
\begin{center}
	{ \includegraphics[width=\textwidth]{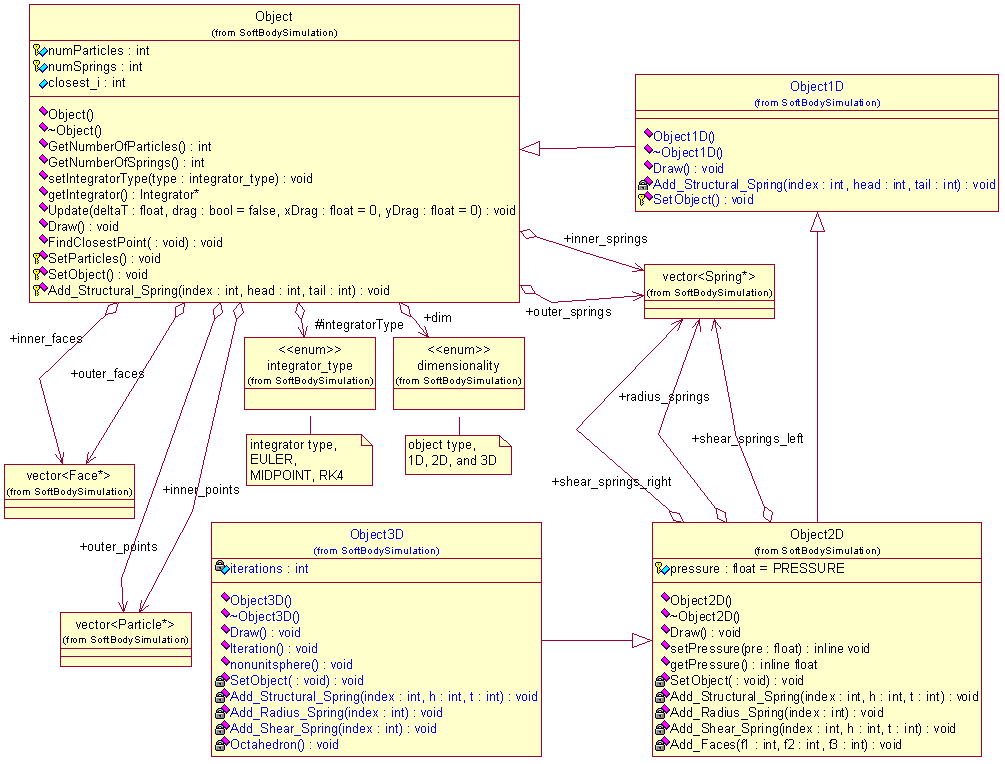}}
\caption{Model Object Class Diagram}
\label{fig:objects}
\end{center}
\end{figure}
The class ``Object'' is the base class for elastic object of any supported dimensionality. %in one dimension, two dimension, or three dimension.
It contains the most common data structure and properties of an elastic object. The geometric complexity is increased according to the dimensions. The ``Object1D'' inherits from the parent class ``Object'', ``Object2D'' inherits from ``Object1D'', and ``Object3D'' inherits from ``Object2D''.
This type of inheritance hierarchy is in place because when each dimensionality is added, the new object type depends on some of the previous implementation and the new things that come with each additional dimension.
For example, 1D object has a notion of structural springs varying in a single dimension;
2D takes the notion of structural springs and augments it with radius and shear springs as well as the notion of pressure
inside an enclosed object; 3D extends 2D by adding the notion of face subdivision and volume making object more
dynamic in terms of run-time number of vertices (to make it more or less smooth depending on the trade off between quality
and performance). All objects share the same
$Update()$/$Draw()$ mechanism, which is used by the OpenGL state machine to update all the vertices of an object
in the Model and reflect the changes in the View by drawing the deformations in real-time.
%\begin{figure}[!h]\footnotesize
%\hrule\vskip4pt
%\begin{verbatim}
% enum dimensionality {DIM1D, DIM2D, DIM3D};
% enum integrator_type {EULER, MIDPOINT, RK4};
% class Integrator;
% class Object
% {
%   protected:
%   int numParticles;
%   int numSprings;
%   Integrator* integrator;
%   integrator_type integratorType;
%   public:
%   vector<Spring*>  inner_springs, outer_springs;
%   vector<Particle*>  inner_points, outer_points;
%   vector<Face*> inner_faces, outer_faces;
%   dimensionality dim;
%   int closest_i ;
%   public:
%   Object();
%   virtual ~Object();
%   int GetNumberOfParticles();
%   int GetNumberOfSprings();
%   void setIntegratorType(integrator_type type);
%   Integrator* getIntegrator();
%   virtual void Update(float, bool = false, float = 0, float = 0);
%   virtual void Draw() = 0;
%   virtual void FindClosestPoint(void) ;
%   protected:
%   virtual void SetObject();
%   virtual void SetParticles();
%   virtual void Add_Structural_Spring(int, int, int);
% }
%\end{verbatim}
%\caption{class Object}
%\label{fig:cobject}
%\vskip4pt\hrule
%\end{figure}
\paragraph*{Object} As shown in \xf{fig:objects}, the object class, an elastic object contains a particle object, a spring object, a face object, and an integrator object.
The data structure varies from inner to outer layers, for example, the pointers to the particles on the inner layer and on the outer layer of the object are saved in different data vectors. $SetObject()$ constructs the geometric shape of the elastic object, which, in turn, constructs the particles $SetParticles()$ and connects the particles by the structural springs via the $Add\_Structural\_Spring()$ call. The enum type $dimensionality$ has one of the values $(DIM1D, DIM2D, DIM3D)$ to determine the object's dimensionality type: 1D, 2D, or 3D; the enum type $integrator\_type$ determines which type of integrator the simulation system uses, Euler, Midpoint, or Runge Kutta Fourth Order integrator. Such design allows extension to add new integrators
and select existing integrators at run-time.
The variable $closest_i$ is the closest point on the outer layer to mouse position and $FindClosestPoint()$ is the function to find such a particle
(used in dragging force application when dragging the object across the simulation window). The function $Update()$
modifies the simulated object's state (either each time point when idle or application of the drag force by the user), and determines the object's
overall forces, velocity, position in the next time step. $Draw()$ visualizes the object after each update.

\begin{figure}[!h]\footnotesize
\hrule\vskip4pt
\begin{verbatim}
 void Idle() {
   object1D.Update(DT, mousedown != 0, xMouse, yMouse);
   object2D.Update(DT, mousedown != 0, xMouse, yMouse);
   object3D.Update(DT, mousedown != 0, xMouse, yMouse);
   glutPostRedisplay();
 }
\end{verbatim}
\caption{$Idle()$ Model Updates}
\label{fig:idle}
\vskip4pt\hrule
\end{figure}

\begin{figure}[!h]\footnotesize
\hrule\vskip4pt
\begin{verbatim}
 void Object::Update(float deltaT, bool drag, float xDrag, float yDrag) {
   if(integrator == NULL) {
    switch(integratorType) {
     case EULER:
       integrator = new EulerIntegrator(*this);
       break;
     case MIDPOINT:
       integrator = new MidpointIntegrator(*this);
       break;
     case RK4:
       integrator = new RungeKutta4Integrator(*this);
       break;
     default:
       assert(false);
       return;
     }
     integrator->setDimension(dim);
   }
   integrator->integrate(deltaT, drag, xDrag, yDrag);
 }
\end{verbatim}
\caption{General $Update()$ Function}
\label{fig:update}
\vskip4pt\hrule
\end{figure}

In the main simulation, the $Idle()$ function shown in \xf{fig:idle}, elastic objects update at every time step $DT$ to tell the the system how the objects behave and the change for their velocity and position. There are four parameters for $Update()$ as shown in \xf{fig:update}, the time step $deltaT$, if there exists user interaction $drag=0$ by default, the mouse position on $x$ and $y$ axises (for dragging upon mouse release) is at 0 by default.
The general algorithm of the $Update()$ presented, illustrates that the most of the actual modifications are based on the dynamically
selected integrator and the dimensionality of the simulation object being integrated. If in the feature a new integrator is added,
this function has to be updated to account for it in the framework.

\paragraph*{1D Object}
%\begin{figure}[!h]\footnotesize
%\hrule\vskip4pt
%\begin{verbatim}
% void Object1D::SetObject()
% {
%   int PosX = 0;
%   int PosY = 0;
%   int i;
%   for(i=0; i<numParticles; i++)
%   {
%     outer_points.push_back( new Particle(new Vector(PosX,PosY,0), MASS));
%     inner_points.push_back( new Particle());
%     PosY += 2;
%   }
%   for(i=0; i<1 ;i++)
%   {
%     Add_Structural_Spring(i, i, (i+1) mod numParticles);
%   }
% }
% void Object1D::Add_Structural_Spring(int index, int head, int tail)
% {
%   outer_springs.push_back( new Spring(outer_points[head],outer_points[tail]));
%   inner_springs.push_back( new Spring());
%   inner_springs[index]->setRestLen();
%   outer_springs[index]->setRestLen();
% }
%\end{verbatim}
%\caption{class Object1D}
%\label{fig:Object1D}
%\vskip4pt\hrule
%\end{figure}

In \xf{fig:objects}, the ``Object1D'' class shows that an one-dimensional object contains two particles and one spring. The type of particles is $outer\_points$ and spring type is structural $outer\_springs$.

\paragraph*{2D Object}

In \xf{fig:objects}, the ``Object2D'' class shows that an two-dimensional object contains inner and outer layers. The type of particles is $inner\_points$ and $outer\_points$. The spring type is structural $inner\_springs$ and $outer\_springs$; moreover, there are another three new types of springs, $radius\_springs$, $shear\_springs\_left$, and $shear\_springs\_right$. The function $Add\_Structural\_Spring()$ models the shape of the inner circle by connecting $inner\_springs$ and the outer circle by connecting the $outer\_springs$ separately. $Add\_Radius\_Spring()$ adds the radius springs with the inner point $i$ and outer point $i$. $Add\_Shear\_Spring()$ adds the left shear springs with inner point $i$ and outer point $i+1$ and the right shear springs with inner point $i+1$ and outer point $i$. The variable $pressure$, which is an additional inner force compared to ``Object1D'', is at each spring along its normal.
%\begin{figure}[!h]\footnotesize
%\hrule\vskip4pt
%\begin{verbatim}
%\end{verbatim}
%\caption{class Object2D}
%\label{fig:Object2D}
%\vskip4pt\hrule
%\end{figure}

\paragraph*{3D Object}
In \xf{fig:objects}, the ``Object3D'' class shows that a three-dimensional object uses similar method as a two-dimensional object by extending the variables into the $z$ axis. However, there are two methods introduced to create a three-dimensional object, such as $nonunitsphere()$ and $SetObject()$, which uses iteration to define an uniform sphere. The base shape for subdivision a sphere is defined in $Octahedron()$ and $Iteration()$ computes the coordinates of the newly generated particles and springs based on the level of detail, the variable $Iterations$.
%\begin{figure}[!h]\footnotesize
%\hrule\vskip4pt
%\begin{verbatim}
%
%\end{verbatim}
%\caption{class Object3D}
%\label{fig:Object3D}
%\vskip4pt\hrule
%\end{figure}

% \clearpage

\subsection{Design and Implementation of Components: Controller}

\begin{figure}[!h]
\hrule\vskip4pt
\begin{center}
	{\includegraphics[width=4in]{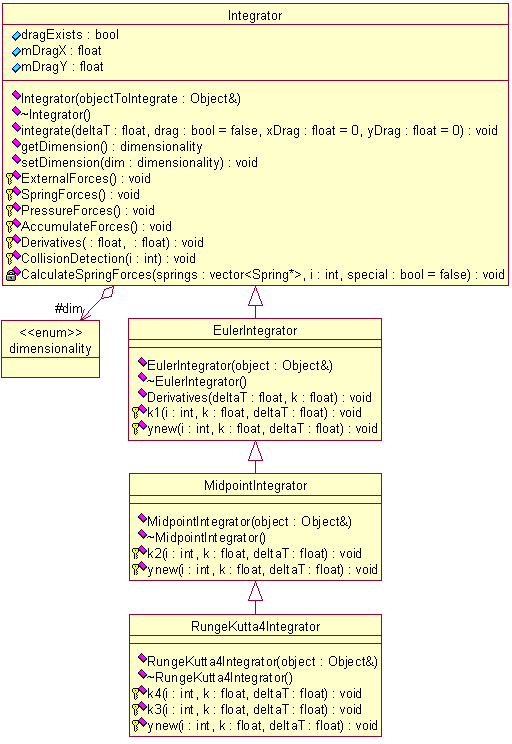}}
\caption{Integrator Framework Class Diagram}
\label{fig:integrators}
\end{center}
\hrule\vskip4pt
\end{figure}

The types of integrators are varied by their complexities, such as Euler, Midpoint, and Runge-Kutta.
The common attributes and methods are defined in the parent class ``Integrator'', as shown in \xf{fig:integrators}.
The subclasses ``EulerIntegrator'', ``MidpointIntegrator'', and ``RungeKuttaIntegrator" inherit the super classes based on the complexity.
The Euler integrator is a basic building block for other integrators which provides the first step of computation of $k_1$ in $k1()$.
Midpoint integrator uses Euler's $k1()$ implementation and provides the 2nd step, $k_2$ implemented
in $k2()$. Finally, the RK4 integrator adds the last two refinement steps $k_3$ (function $k3()$) and $k_4$ (function $k4()$) in addition to what Euler and midpoint have provided. Thus, RK4 implementation depends on the midpoint which, in turn, depends on the Euler
integrator with different parameters.

\begin{figure}[!h]\footnotesize
\hrule\vskip4pt
\begin{verbatim}
void Integrator::integrate(float deltaT, bool drag, float xDrag, float yDrag) {
  dragExists = drag; mDragX = xDrag; mDragY = yDrag;
  AccumulateForces();
  Derivatives(deltaT, 1.0);
}
...
void Integrator::AccumulateForces() {
  ExternalForces();
  SpringForces();

  switch(dim) {
    case DIM1D:
      break;

    case DIM2D:
    case DIM3D:
      PressureForces();
      break;
  }
}
\end{verbatim}
\caption{General $integrate()$ and $AccumulateForces()$ Functions}
\label{fig:integration}
\vskip4pt\hrule
\end{figure}

In \xf{fig:integration} there is a general $integrate()$ function (which is called
from $Object::Update()$) and a general $AccumulateForces()$ function, both of which
play a vital role in the integrator framework in this thesis. They illustrate the
general algorithm of integration applied to the Model's data: first, the effect
of all the forces is accumulated (which includes external forces, such as gravity
and drag, as well as forces induced by springs and pressure); then, the integrator-specific
derivation is performed to each particle of an object. In the general ``Integrator''
the $Derivatives()$ function is pure virtual as is left to be overridden by the ``EulerIntegrator'',
``MidpointIntegrator'', and ``RungeKutta4Integrator'' concrete implementations. It is important
to note that the reverse forces are also accounted at the collision detection at the end
of each $Derivatives()$ implementation. Another note worth mentioning is that the pressure
forces are not applicable in the 1D case as there is no enclosed object, which can hold
pressure in this cases. $ExternalForces()$ checks for the existence of the mouse drag
force (from the user) as well as gravity and sums them up. $SpringForces()$ accumulates
contributions for all spring types (a subject to dimensionality as well, e.g. 1D case
does not have radius or shear springs, only one structural spring).

% \clearpage

\subsection{Simulation Loop Sequence}

The sequence diagram in \xf{fig:DiagramSequence} describes the control-flow of the simulation
sequence and logic of the elastic object simulation system.
The following sequence of steps describes all of the possible states of the elastic object as events occur in greater detail.
There we track the different states how the physical simulation loop works, such as display of the objects,
accumulation of forces, integration of forces, and so on. In other words, this is the main algorithm of
the entire simulation system.

\begin{figure}[!h]
\hrule\vskip4pt
\begin{center}
	{	\includegraphics[width=5.5in]{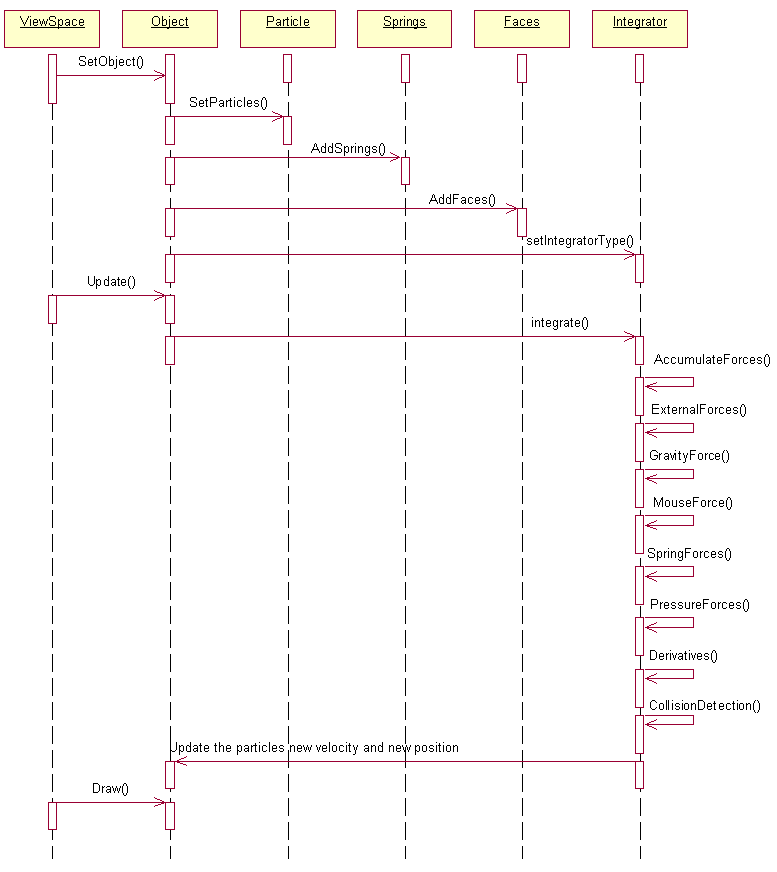}}
\caption{Simulation Loop Sequence Diagram}
\label{fig:DiagramSequence}
\end{center}
\hrule\vskip4pt
\end{figure}

%\begin{figure}[!h]
%\begin{center}
%	{
%	 	 \includegraphics[width=6in]{images/DiagramState.png}}
%\caption{State Diagram}
%\label{fig:DiagramState}
%\end{center}
%\end{figure}

\begin{itemize}
\item Step 1: ``ViewSpace''
%is the start state of the loop when the software runs at the very beginning. It
initializes the virtual world
and provides the user an interactive environment.
It provides the interface to allow user to drag the object, or choose the parameters.
For example, user can choose the object type, one-dimensional, two-dimensional, or three-dimensional.
User can choose the integrator type, Euler, Midpoint, or Runge Kutta 4.
User can set up the springs' stiffness, damping variable, and the pressure.
\item Step 2: $SetObject()$ function creates an elastic object based on the interface variable set from Step 1.
\item Step 3: $SetParticles()$ function sets up the particles' position and their other initial properties, such as mass and velocity.
\item Step 4: $AddSprings()$ function connects particles with springs according to their index.
\item Step 5: $AddFaces()$ connects the springs with faces based on proper index. This step will be ignored if the object is one-dimensional.
\item Step 6: $SetIntegratorType()$ function tells the Controller which integrator users select through the interface.
\item Step 7: $Update()$ updates the integrator's time step.
\item Step 8: $Integrate()$ contains two functions, $AccumulateForces()$ and $Derivatives()$. It is based on all the object geometric information modeled and all the forces information accumulated, to integrate over the time step to get new object position and orientation.
\item Step 9: $AccumulateForces()$ state is to sum up the forces accumulated on each particle.
\item Step 10: $GravityForce()$ is to accumulate gravity force based on the particles' masses.
\item Step 11: $MouseForce()$ is the external force from the interface when user interacts with the object. It will be added or subtracted from the particles depends on the force's direction.
\item Step 12: $SpringForce()$ is to accumulate internal force of the particles connected by springs.
\item Step 13: $PressureForce()$ is to accumulate the internal pressure acted on the particles. For one-dimensional object, this state is omitted.
\item Step 14: $Derivatives()$ does the real derivative computation of acceleration and velocity in order to get new velocity and position of elastic objects based on the integrator type defined by users.  
\item Step 15: $CollisionForce()$ is to check if the object is out of boundaries after the integration state. If the new position is outside of the boundary, then it will be corrected and reset on the edge of the boundary. Moreover, the new collision force will be added to the object.
\item Step 16: $Draw()$ displays the object with new position, velocity, and deformed shape.
\end{itemize}

\chapter{Experimental Results}\index{Experimental Results}

In this chapter, the one-dimensional, two-dimensional, and three-dimensional objects are illustrated at different animation sequences, with different simulation parameters, and by simulation with different numerical integration methods.
 
\section{Animation Sequence}

The screenshots in this section present the animation sequence of the one-dimensional, two-dimensional, and three-dimensional objects when they are at the initial state, colliding with floor, bouncing back from the floor, responding to user's external dragging, and at the resting state.

\subsection{1D}

This simulation shows two masses connected with one spring. The one-dimensional object moves in a three-dimensional environment, which consists of ceiling, walls, and floor. Users can drag the mass with the mouse to change the object's position and direction. \xf{fig:rOneD1} presents the initial state of the object; \xf{fig:rOneD2} shows the object collides with the floor when it drops with gravity force; \xf{fig:rOneD3} displays the collision response of the object based on the penalty method; \xf{fig:rOneD4} shows the moment when users drag the object; \xf{fig:rOneD5} shows how the object reacts on the external impact, such as mouse dragging force or bouncing force with walls; \xf{fig:rOneD6} displays the object resting on the floor after a while when there is no interaction from the user. 

\begin{figure*}[h]
\hrule\vskip4pt
\begin{center}
	\subfigure[The initial state]
	{\label{fig:rOneD1}
	 \includegraphics[width=1.8in]{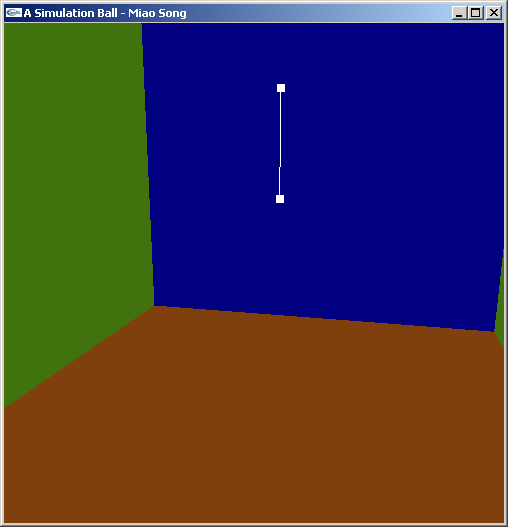}}
	 \hspace{.1in}
	\subfigure[Collide with floor]
  {\label{fig:rOneD2}
	 \includegraphics[width=1.8in]{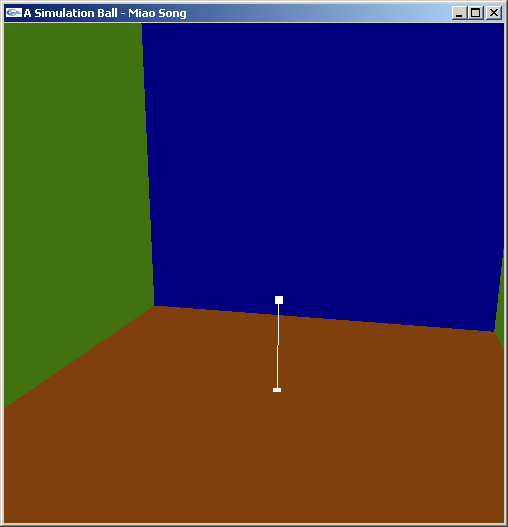}}
	  \hspace{.1in}
	\subfigure[Bounce back from the floor]
  {\label{fig:rOneD3}
	 \includegraphics[width=1.8in]{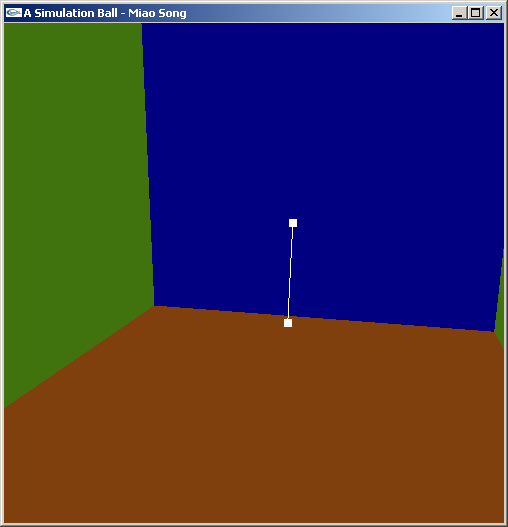}}
	  \hspace{.1in}
	\subfigure[Drag the object]
  {\label{fig:rOneD4}
	 \includegraphics[width=1.8in]{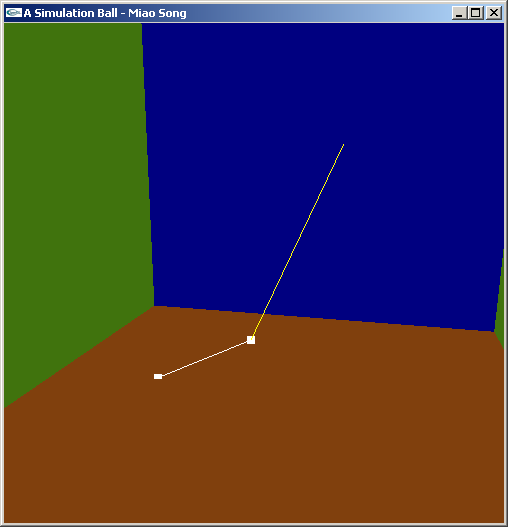}}
	 	  \hspace{.1in}
	\subfigure[Response to compact]
  {\label{fig:rOneD5}
	 \includegraphics[width=1.8in]{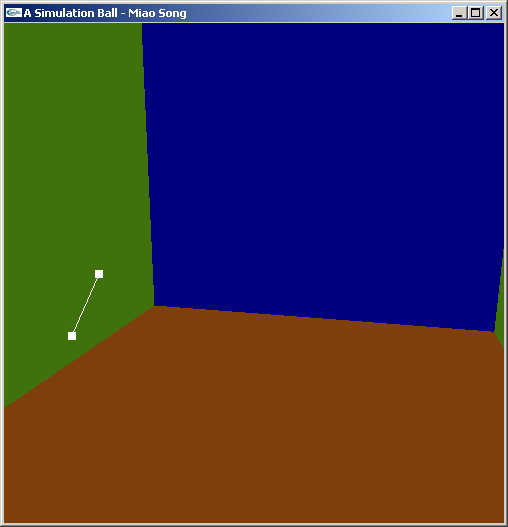}}
	 	  \hspace{.1in}
	\subfigure[The resting state]
  {\label{fig:rOneD6}
	 \includegraphics[width=1.8in]{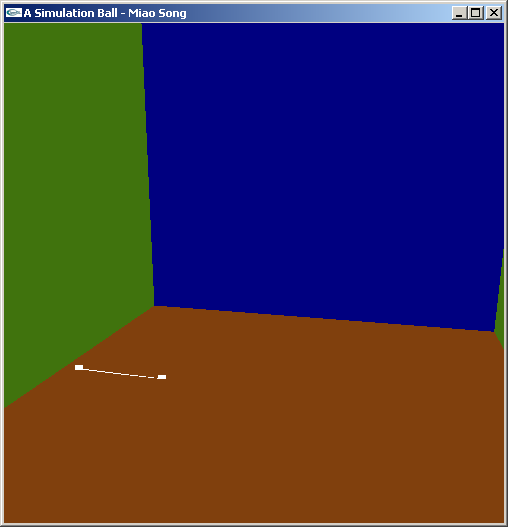}}
\caption{Animation Sequence of One Dimensional Elastic Object}
\end{center}
\hrule\vskip4pt
\end{figure*}

\subsection{2D}

The simulation as shown in \xf{fig:rTwoD1} through \xf{fig:rTwoD6} is how a two-dimensional object moves in a three-dimensional environment. This two-layer object consists of 10 particles and 10 structural springs on both inner and outer circles. Moreover, it contains 10 radius springs, 10 shear left springs, and 10 shear right springs between the inner and outer layers. If a two-dimensional object with only one layer, or the object has no pressure force within, the spring's stiffness has to be a larger value than without, then the object will not collapse.
However, as shown in \xf{fig:rTwoD2}, if the spring stiffness is small enough, the object does not collapse, neither overlap with the layers because of the stability of the two-layer structure.

\begin{figure*}[h]
\hrule\vskip4pt
\begin{center}
	\subfigure[The initial state]
	{\label{fig:rTwoD1}
	 \includegraphics[width=1.8in]{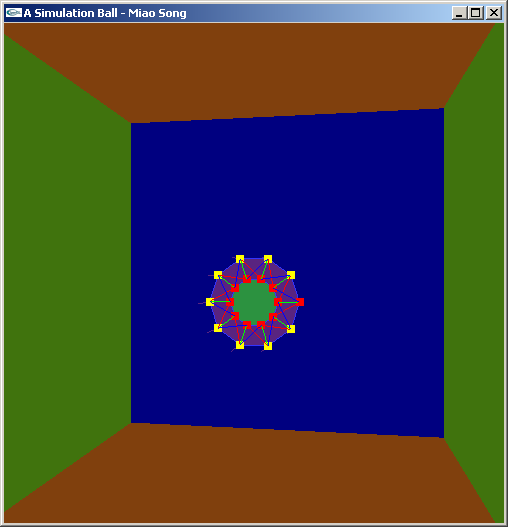}}
	 \hspace{.1in}
	\subfigure[Collide with floor]
  {\label{fig:rTwoD2}
	 \includegraphics[width=1.8in]{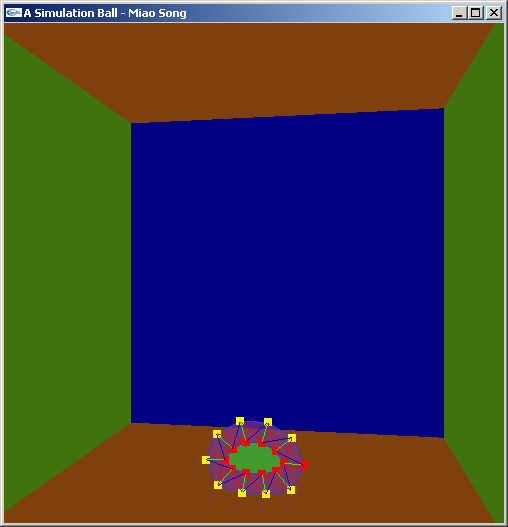}}
	  \hspace{.1in}
	\subfigure[Bounce back from the floor]
  {\label{fig:rTwoD3}
	 \includegraphics[width=1.8in]{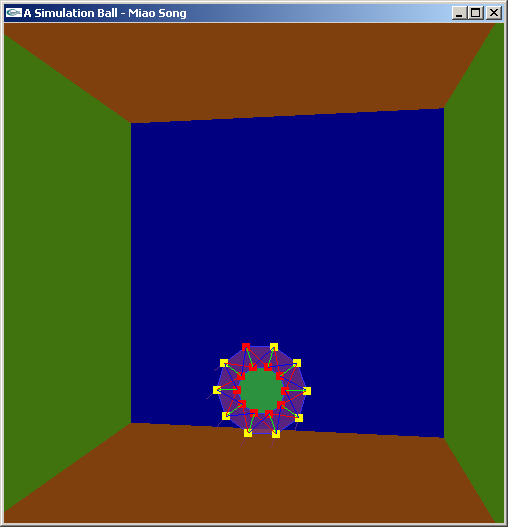}}
	  \hspace{.1in}
	\subfigure[Drag the object]
  {\label{fig:rTwoD4}
	 \includegraphics[width=1.8in]{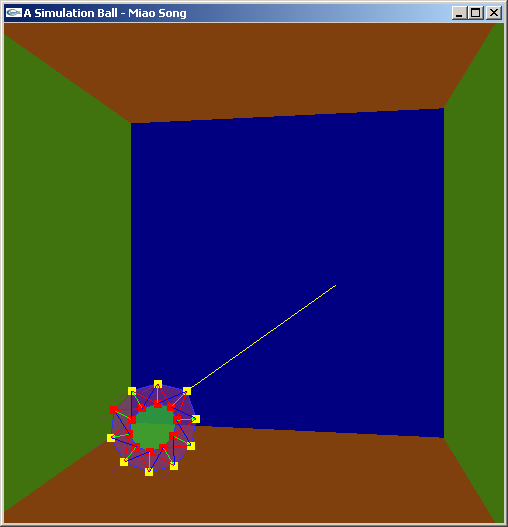}}
	 \hspace{.1in}
	\subfigure[Response to compact]
  {\label{fig:rTwoD5}
	 \includegraphics[width=1.8in]{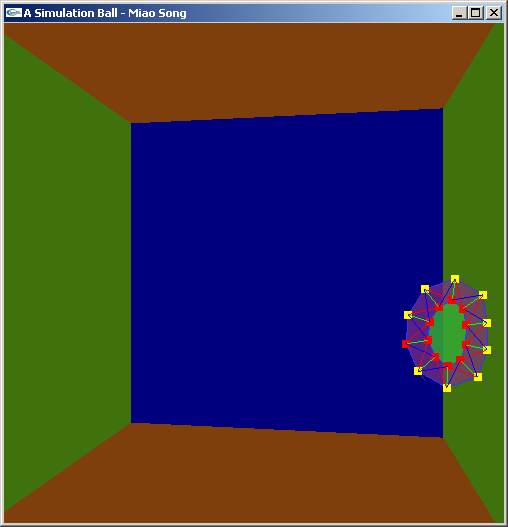}}
	 	  \hspace{.1in}
	\subfigure[The resting state]
  {\label{fig:rTwoD6}
	 \includegraphics[width=1.8in]{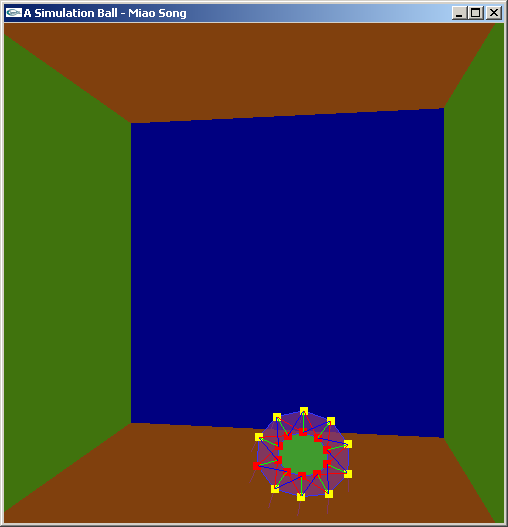}}
\caption{Animation Sequence of Two Dimensional Elastic Object}
\end{center}
\hrule\vskip4pt
\end{figure*}

\subsection{3D}

The simulation as shown in \xf{fig:rThreeD1} through \xf{fig:rThreeD6} is how a three-dimensional uniform facet object moves in a three-dimensional environment. This two-layer object, which is generated by subdividing an octahedron once, consists of 12 particles, 36 structural springs, and 32 faces, on both inner and outer spheres. Moreover, the object also contains 36 radius springs, 36 shear left springs, and 36 shear right springs between the inner and outer layers. Just like in two dimensions, the two-layer structure gives the three-dimensional sphere more stability.

\begin{figure*}[h]
\hrule\vskip4pt
\begin{center}
	\subfigure[The initial state]
	{\label{fig:rThreeD1}
	 \includegraphics[width=1.8in]{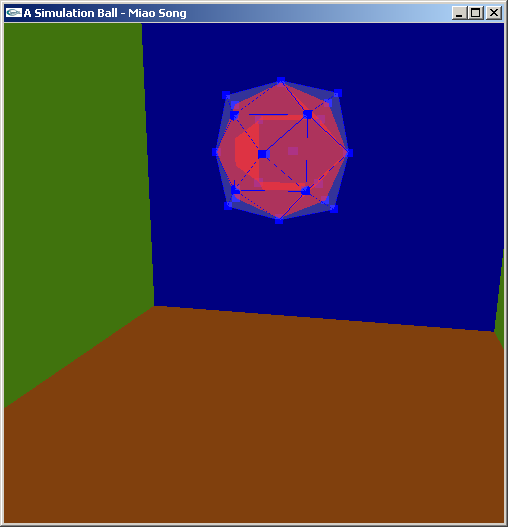}}
	 \hspace{.1in}
	\subfigure[Collide with floor]
  {\label{fig:rThreeD2}
	 \includegraphics[width=1.8in]{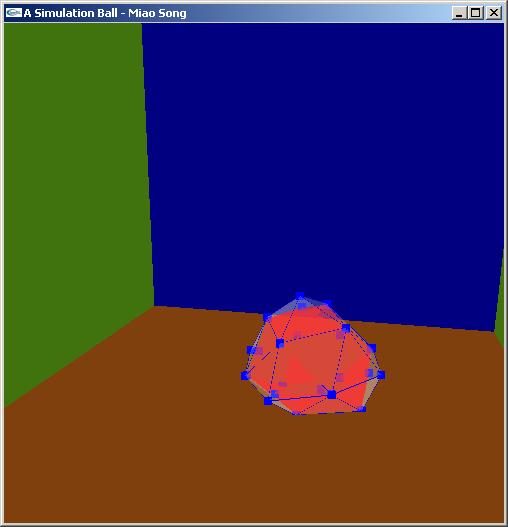}}
	  \hspace{.1in}
	\subfigure[Bounce back from the floor]
  {\label{fig:rThreeD3}
	 \includegraphics[width=1.8in]{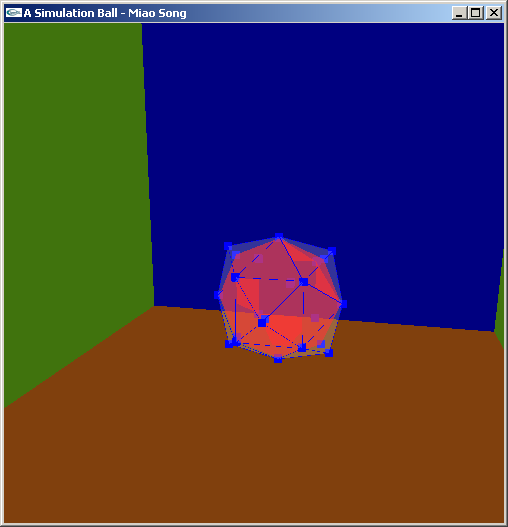}}
	  \hspace{.1in}
	\subfigure[Drag the object]
  {\label{fig:rThreeD4}
	 \includegraphics[width=1.8in]{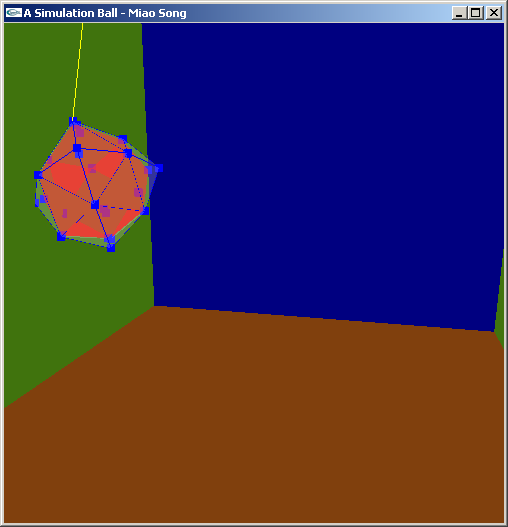}}
	 	  \hspace{.1in}
	\subfigure[Response to compact]
  {\label{fig:rThreeD5}
	 \includegraphics[width=1.8in]{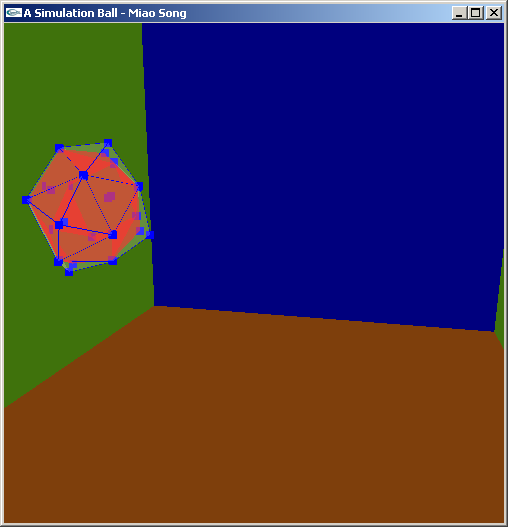}}
	  \hspace{.1in}
	\subfigure[The resting state]
  {\label{fig:rThreeD6}
	 \includegraphics[width=1.8in]{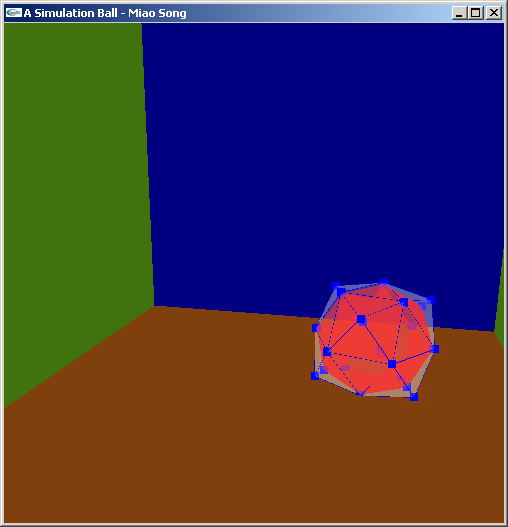}}
\caption{Animation Sequence of Three Dimensional Elastic Object}
\end{center}
\hrule\vskip4pt
\end{figure*}

\section{Simulation Parameters}

The parameters in the simulation such as mass, spring stiffness, and friction (damping) can be changed.
One can drag the object mass with a mouse to change its position.
Effects of different simulation parameters are discussed. 

\subsection{Summary of the Adjustable Parameters}

The parameters that influence the behavior of the simulated environment
are summarized below, with their default values. Most initial and default
values were based on the 2D case from \cite{mm041}; otherwise, the values
are empirical and are partially dependent on the hardware the simulation
is executing on.

\begin{itemize}
\item
KS = 800.0f where KS is structural spring stiffness constant. The larger
this value is, the less elastic the object is and it is more resistant to the inner pressure and
deformation. The lesser this value is the more object is deformable and a subject to break up
if the inner pressure force is high.

\item
KD = 15.0f where KD is structural spring damping constant, opposite to the spring retraction force. It denotes how fast the object is to resist its motion.

\item
RKS = 700.0f where RKS is radius and shear spring stiffness constant, similar to KS, but for
radius and shear springs as opposed to the structural springs.

\item
RKD = 50.0f where RKD is radius and shear spring damping constant, similar to KD, but for radius
and shear springs.

\item
MKS = 150.0f where MKS is the spring stiffness constant of the spring connected with the mouse and the approximate nearest particle on the object.
This constitutes the elasticity of the ``drag'' spring connected to the mouse: the lesser the value is, the more elastic it is, and the harder it is
to drag the object as a result.

\item
MKD = 25.0f where MKD is the damping constant of the spring connect with the mouse and the approximate nearest point on the object.
%,that helps the drag spring to retract back. The larger it is, the faster the retraction process is.

\item
PRESSURE = 20.0f where PRESSURE is gas constant used in the ideal gas equation mentioned earlier to
determine the pressure force inside the enclosed object. If this constant is too high, and the combined
spring stiffness for all the spring types is low enough, the object can ``blow up''.

\item
MASS = 1.0f where MASS is the mass for each particle. The object can be made heavier or lighter if this
value is larger or smaller respectively, in order to experiment with the gravity effects. Naturally,
the heavier objects will be more difficult to drag upwards in the simulation environment. Conversely,
the smaller-mass object can be dragged around with less effort given the rest of the parameters remain
constant.

\end{itemize}

\subsection{Stability vs. Time Step}

First, the figures in this section (\xf{fig:r003euler}, \xf{fig:r003midpoint}, and \xf{fig:r003rk4})
show the stability of the three integrators. We consider the integration time step parameter in these
scenarios only, assuming all the other parameters (discussed later) are not change for the described simulations.
As shown in those figures, when the time step is small, such as $DT = 0.003$\footnote{This is an empirical value; dependent on the performance of the hardware.}, three of the integrators
behave well and the object does not ``blow up''. 
However, when one increases the time step by a factor of 10 to $DT = 0.03$, the midpoint (see \xf{fig:r03midpoint})
and RK4 (see \xf{fig:r03rk4}) integrators are still stable and the object integrated with Euler integrator ``blows up'' as in \xf{fig:r03euler}. 
Furthermore, when the time step is increased 10-fold more to $DT = 0.3$, only the object integrated with RK4 (see \xf{fig:r3rk4})
is stable and another two objects integrated with Euler (\xf{fig:r3euler}) and Midpoint (\xf{fig:r3midpoint}) methods ``blow up''.

\begin{figure*}[h]
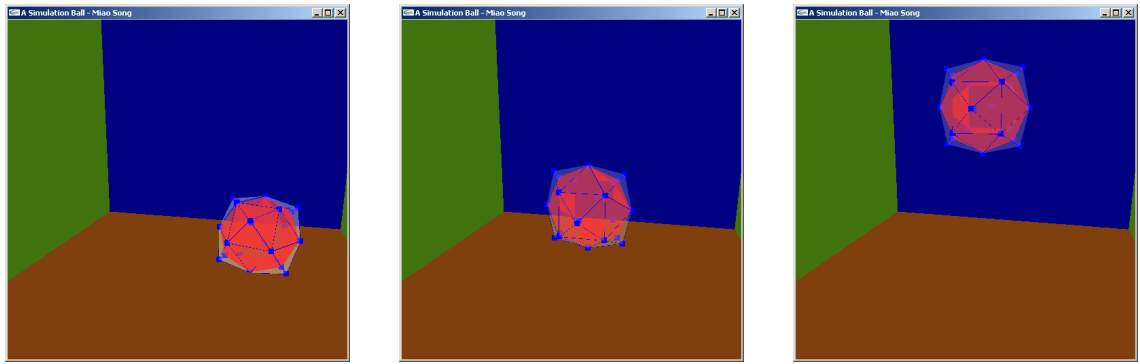

\hrule\vskip4pt
\begin{center}
	\subfigure[The object integrated with Euler Method]
	{\label{fig:r003euler}
	 \includegraphics[width=1.8in]{images/result/4d6.png}}
	 \hspace{.1in}
	\subfigure[The object integrated with Midpoint Method]
  {\label{fig:r003midpoint}
	 \includegraphics[width=1.8in]{images/result/4d3.png}}
	  \hspace{.1in}
	\subfigure[The object integrated with RK4]
  {\label{fig:r003rk4}
	 \includegraphics[width=1.8in]{images/result/4d1.png}}	
	 \caption{Elastic Object at Timestep = 0.003}
\end{center}
\hrule\vskip4pt
\end{figure*}

\begin{figure*}[h]
\hrule\vskip4pt
\begin{center}
	\subfigure[The object integrated with Euler Method]
	{\label{fig:r03euler}
	 \includegraphics[width=1.8in]{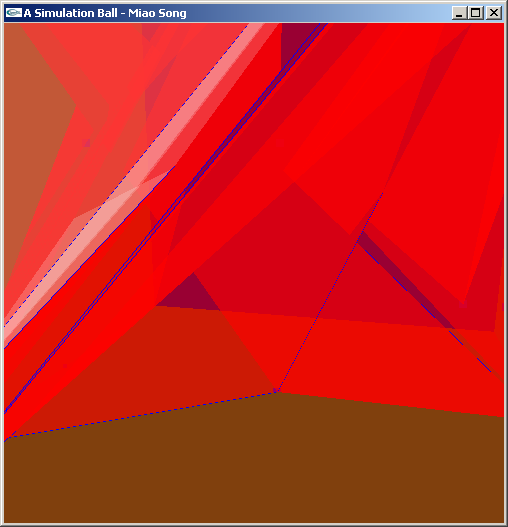}}
	 \hspace{.1in}
	\subfigure[The object integrated with Midpoint Method]
  {\label{fig:r03midpoint}
	 \includegraphics[width=1.8in]{images/result/4d3.png}}
	  \hspace{.1in}
	\subfigure[The object integrated with RK4]
  {\label{fig:r03rk4}
	 \includegraphics[width=1.8in]{images/result/4d1.png}}	
	 \caption{Elastic Object at Timestep = 0.03}
\end{center}
\hrule\vskip4pt
\end{figure*}

\begin{figure*}[h]
\hrule\vskip4pt
\begin{center}
	\subfigure[The object integrated with Euler Method]
	{\label{fig:r3euler}
	 \includegraphics[width=1.8in]{images/result/brokeneuler.png}}
	 \hspace{.1in}
	\subfigure[The object integrated with Midpoint Method]
  {\label{fig:r3midpoint}
	 \includegraphics[width=1.8in]{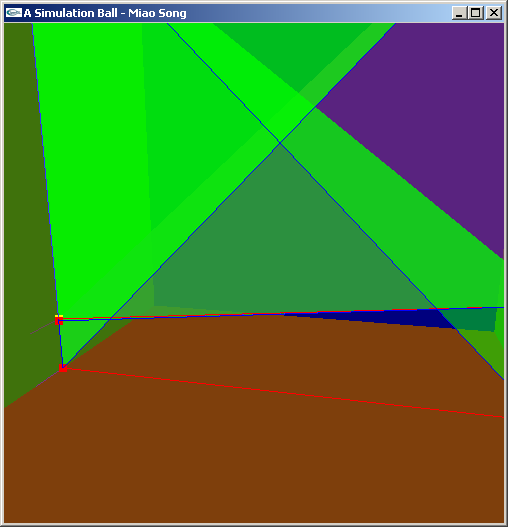}}
	  \hspace{.1in}
	\subfigure[The object integrated with RK4 Method]
  {\label{fig:r3rk4}
	 \includegraphics[width=1.8in]{images/result/4d1.png}}	
	 \caption{Elastic Object at Timestep = 0.3}
\end{center}
\hrule\vskip4pt
\end{figure*}

\subsection{Efficiency and Accuracy}

The more computational effort is required, the less efficient
algorithm is. Likewise, the more accurate algorithm is, the
more computation effort it requires, the less efficient it is.
Thus, in our simulation system the most efficient and least
accurate integration method is Euler's, followed by Midpoint
(about twice as more accurate and slower), followed by
RK4 (four times slower than Euler's and the most accurate
of the three). This can be illustrated in 
\xf{fig:r003euler}, \xf{fig:r003midpoint}, and \xf{fig:r003rk4}
running concurrently with the same time step of $0.003$, where
one can see the simulation with Euler's method reaches the floor
fastest and RK4 slowest. Of course, the efficiency of the simulation
and the accuracy of the shape and movement depends on the amount
of particles (and as a result, all kinds of springs) in the object.

%\begin{figure*}[h]
%\hrule\vskip4pt
%\begin{center}
%	\subfigure[Euler Integrator]
%	{\label{fig:re}
%	 \includegraphics[width=1.8in]{images/result/euler.png}}
%	 \hspace{.1in}
%	\subfigure[Midpoint Integrator]
%  {\label{fig:rm}
%	 \includegraphics[width=1.8in]{images/result/midpoint.png}}
%	  \hspace{.1in}
%	\subfigure[RK4 Integrator]
%  {\label{fig:rr}
%	 \includegraphics[width=1.8in]{images/result/rk4.png}}
%\caption{A Three-dimensional Object Simulated with Three Different Integrators}
%\end{center}
%\hrule\vskip4pt
%\end{figure*}

\section{Computational Errors}

This section briefly summarizes the error accumulated in the
application of the described algorithms and their effects.

\subsection{Collision Detection}

We have applied the Penalty Method in our simulation system.
This simple but inaccurate algorithm causes the object to ``stick'' on the collision
surface when dragging the object at the same time and it may become difficult to drag
the object away for a period of time.

\subsection{Subdivision Method}

The spherical shape is not perfect round because the number of springs associated to each
particle is not uniform. If one wants more quality subdivision has to be done in more than
one subdivision operation, but the simulation may rapidly become very slow as the number
of particles grow requiring a much greater computational effort, which is suitable only
for the high-end hardware if one wishes to do it in real-time. In \xf{fig:rshape} is an
example of the two iterations of the subdivision.

\begin{figure}[h]
\hrule\vskip4pt
\begin{center}
 {  \includegraphics[width=2in]{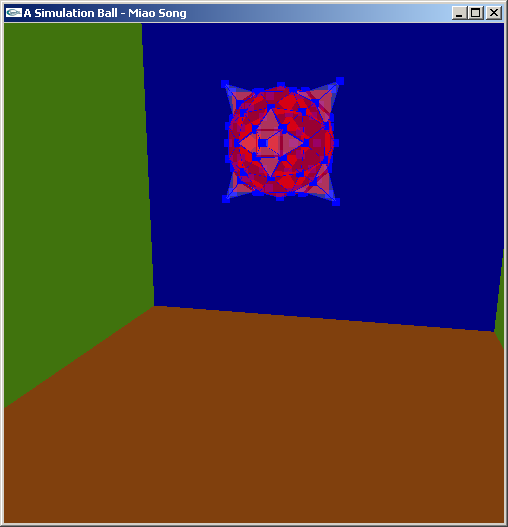}}
\end{center}
\hrule\vskip4pt
\caption{Second Subdivision Iteration}
\label{fig:rshape}
\end{figure}

\chapter{Conclusion and Future Work}
\parskip        1.5ex plus 1ex minus 0.5ex
\parindent      0em
\label{chapt:conclusion}
\index{conclusion}

This chapter describes our contribution based on the existing elastic model and analyzes the possible development and related work in the future. 
\section{Contribution}
The new model, two-layer elastic object with uniform-surfaces is a simple, efficient approach to imitate the liquid effects of elastic object, such as human's tissue and soft body. Since the modeling and structure of the tissue kind elastic object is closer to real tissue than an one layer object, the level of realism has been increased. The images in this chapter are screenshots from the elastic simulation system we have developed. The modeling method and the density setting provides significant improvements on the conflicts of accuracy and interactivity on previous models. The realism of the results, such as liquid motion and inertia effects are also enhanced. 

\paragraph*{Procedural Modeling} 
We have applied the procedural modeling method with particle system to model elastic objects. From simple one-dimensional to most complicated three-dimensional object, we introduced the modeling method for different dimensional objects and related physics knowledge gradually. In the elastic object simulation system, each particle has its local coordinate which is easy to be computed at every time step. Moreover, this modeling method can efficiently control the level of detail as required by graphics artists and computer hardware available. As shown in \xf{fig:2dparticle} and \xf{fig:3dparticle},  this modeling method also most approximately approaches the ideal equal faces; therefore, the edges(springs) on the faces and the forces on each particle are approximately to be equal at initial state in order to minimize the computation error caused by the object geometry. 
%The initial shape can be detailed in different grades upon requirements. The subdivision of the surfaces by particles does not need manual pre-calculation. 
%Also, the two-layer elastic object is more accurate than single layer to imitate real tissue simulation with multi layers. 

\begin{figure*}[h]
\begin{center}
	\subfigure[Two-dimensional Object]
	{\label{fig:2dparticle}
	 \includegraphics[width=2in]{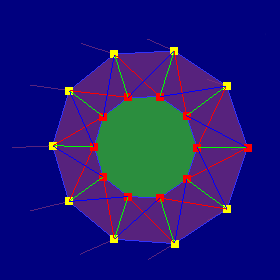}}
	 \hspace{.6in}
	 	\subfigure[Three-dimensional Object]
	{\label{fig:3dparticle}
	 \includegraphics[width=2in]{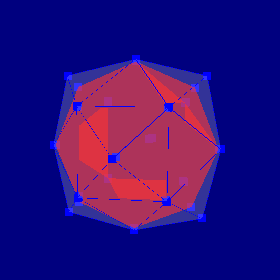}}
\caption{Uniform Shape Modeling}
\end{center}
\end{figure*}

\paragraph*{Density} 
As shown in \xf {fig:2dgravity} and \xf{fig:3dgravity}, the density is defined only for each particle on the elastic surface and the internal density is represented by air pressure physics equation. The weights of particles on inner and outer layer can be set differently. For example, a balloon half filled with liquid, the bottom is heavier than the top part because the density is at the bottom is liquid and top part is air. The weights on inner layer can be set much heavier than outer layer. This special feature gives us flexibilities to imitate different material effects with such simple model.  %The bottom particles on the object are not necessary to have the same gravity force as the particles on the top part of the object. Also, the density of the inner layer, the outer layer, and the layer between them can be defined differently as the material of fat, liquid, air, and so on. The self modifiable density is useful for self deformation effects and describes liquidity effects.
%The Disney animators discovered that mass is very important to achieve realistic animation\cite{TR81}.       
\begin{figure*}[h]
\begin{center}
\subfigure[Two-dimensional Object]
	{\label{fig:2dgravity}
	 \includegraphics[width=2in]{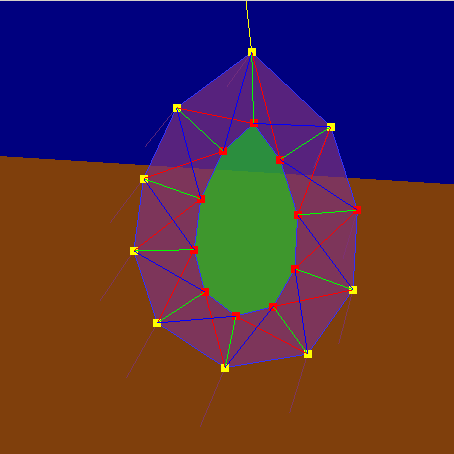}}
	 \hspace{.6in}
	 	\subfigure[Three-dimensional Object]
	{\label{fig:3dgravity}
	 \includegraphics[width=2in]{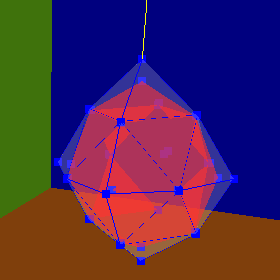}}
\caption{Non-Uniform Density}
\end{center}
\end{figure*}

\paragraph*{Inertia} 
Inertia effect is a unique effect in two layer-elastic simulation system, which can not be achieved with one-layer object. \xf{fig:2dinertia} and \xf{fig:3dinertia} show the inertial movement of a two-dimensional and three-dimensional elastic object. In \xf{fig:2dinertia}, the inner layer and the outer layer have the opposite internal force drive them along axis x. Since the two layers are connected by springs, the inner particles and outer particles have an extra force applied on them, interactive force between inner and outer particles. And their movement, position, and acceleration will be computed according to the contribution of this extra interactive force. This interactive force does not exist in a single layer object. \xf{fig:3dinertia} displays the moment when the elastic object drops down onto the ground. The outer and inner particles will fall with the object based on their gravity and springs force. Here, the inertia for inner particle and outer particle are dependent not only on the force from their own motion, the force from the neighbors on the same layer, but also from the interaction on the other layer. This simulation system is more accurate to describe the inertia property happened in the liquid object.% and make the object looks more wiggling.  

\begin{figure*}[h]
\begin{center}
\subfigure[Two-dimensional Object]
	{\label{fig:2dinertia}
	 \includegraphics[width=2in]{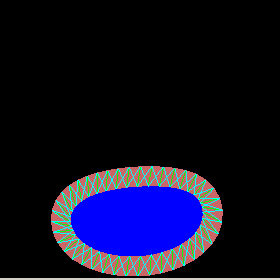}}
	 \hspace{.6in}
	 	\subfigure[Three-dimensional Object]
	{\label{fig:3dinertia}
	 \includegraphics[width=2in]{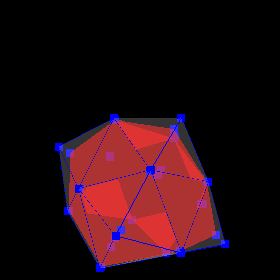}}
\caption{Liquid Motion and Inertia}
\end{center}
\end{figure*}

\paragraph*{Stability} 
The two layered system is stable. Even without the internal pressure force, the shape will not collapse because the two layers are connected by different types of springs. The simulation system works well even with the very inaccurate Euler integrator at large time step, which will result shape collapse or blow up on a one-layer object with the same set of values. We have also implemented the higher level integrators, such as Midpoint and Runga Kutta 4. %The object behaves more smooth and liquidity with these advanced integrators. 

\paragraph*{Re-usability}
The design of this simulation system is based on well-known software design pattern. It decomposes the novel concepts into concrete small components. The functions and classes are easy to be plugged and adapted into other program.
 
This elastic simulation model simplifies the physical modeling method with a group of masses and springs. Also, the simulation is computed in real time based on the numerical integration of the physical laws of dynamics. %It provides both the most efficient and realistic properties. 
%\paragraph*{Integrator Implementation}
\section{Conclusion}

We have developed a one-dimensional elastic object, a two-layer two-dimensional elastic object, and extend it into three-dimensions. These models are all physically based, making use of results from gravity and pressure forces and are implemented with three types of integrations: Euler, Midpoint, and Runge Kutta Fourth Order. The procedural uniform surface generation algorithm provides a convenient mechanism for collision detection. It can generate convincing behaviors when the objects collide with rigid floors or walls because all the particles are checked in every update cycle. Moreover, the rendering is fast because graphics software and hardware renders triangular facets very efficiently.

\section{Future Work}

\paragraph*{Character Animation} The functionality development of elastic simulation modeling for 3D software design and implementation has emerged as a new challenge in computer graphics. One of the existing software with the elastic modeling functionality is Maya, which provides shape deformation, especially facial animation, for a group of objects. It is more convenient than traditional frame animation. However, the elastic object movement is not attached to skeleton animation. Furthermore, this elastic simulation is not in real time.   

A possible future work that can be done based on the elastic simulation is to define a skeleton system and to map the mesh body onto it. The different parts of the body can be defined as the different freedom of deformable based on the elasticity. For example, the mesh is less elastic on the arms, legs; the mesh is more elastic on the areas that consist fats, like breast, belly. The weight of the elastic property of the muscles can be mapped and dynamically set according to the skeleton. The system can be integrated into advanced animation software as a Plug-in.      

%maybe can demo the inner mass more than outer, compare one layer object with only mass on surface. the two layer can MONI the object with mass inside of fluid as liquid object implementation.

%\paragraph*{Computer Game} A realistic and convenient dynamic elastic deformable method can be implemented for more special effects.   

%\paragraph*{Medical} The current system is just a quick simple and less expensive calculation two layered structure with the pressure inside. However, it MONI and achieves reality of elastic object movements convincely. In the future, the modeling of the object can be even more close to real organs. 
%Diagram can show the structures, like several balls, or can be called cells, in as the inner of the object, connected to the outer layer of the object.

\paragraph*{Collision Detection} between soft objects is a complex phenomenon, which has not been widely developed in physics. In our current system, we are using the penalty methods \cite{MJ88}, which do not generate the contact surface between the interacting objects. This method uses the amount of inter-penetration for computing a force which pushes the objects apart instead. Even though the result is fair enough based on estimation, in reality, the contact surfaces should be generated rather than local inter-penetrations. Especially, if we want to use computer animation to imitate organ surgery and help surgeon practice as if interact with real objects, the penalty method is no longer appropriate. There must be a more accurate algorithm to define the collision between rigid body and soft body, or soft body to soft body. Our software should be able to describe other soft body deformation, such as fractures.   

%more automs inside of the object

%Auto set the gravity with different weights for the particles at the bottom of the object.

%tissue density to describe liquidity effect, for example in a elastic bag half filled with water... the bottom is heavy

%demo obj file format loader with face index...

%graphics needed here to show three cells in side of object

%\paragraph*{cutting} 

%%
%% Bibliography
%%
\clearpage
\thispagestyle{empty}
\addcontentsline{toc}{chapter}{Bibliography}
\nocite{*}
\label{pg:bib}
\bibliographystyle{alpha}
\bibliography{thesis}
%\addcontentsline{toc}{chapter}{Bibliography}
%\label{chapt:bibliography}
%\bibliography{thesis}
%\bibliographystyle{alpha}
%\bibliographystyle{named}
%\bibliographystyle{unsrt}

% EOF

%\include{otherfeatures}
%\include{index}
%\include{dictionary}
%%%%%%%%%%%%%%%%%%%%%%%%%%%%%%%%%%%%%%%%%%%%%%%%%%%%%%%%%%%%%%%%%%%%
\end{document}